\documentclass[aps,prl,superscriptaddress,twocolumn]{revtex4-1}

\usepackage{graphicx}
\usepackage{dcolumn}
\usepackage{amssymb}
\usepackage{amsmath}
\usepackage{bbold}
\usepackage{bm}
\usepackage{color}
\usepackage{physics}
\usepackage{verbatim}

\usepackage{soul}

\usepackage[colorlinks,urlcolor=blue,citecolor=blue,linkcolor=blue]{hyperref}
\usepackage{comment}
\usepackage{cleveref}
\usepackage{mathtools}
\usepackage{mathrsfs}

\usepackage[utf8]{inputenc}
\usepackage[T1]{fontenc}

\renewcommand{\k}{\mathbf{k}}
\newcommand{\q}{\mathbf{q}}
\newcommand{\p}{\mathbf{p}}

\begin{document}

\title{Quantum Droplets of Light in Semiconductor Microcavities}

\author{Matteo Caldara}
\affiliation{International School for Advanced Studies (SISSA), via Bonomea 265, 34136 Trieste, Italy}
\affiliation{School of Physics and Astronomy, Monash University, Victoria 3800, Australia}

\author{Olivier Bleu}
\affiliation{School of Physics and Astronomy, Monash University, Victoria 3800, Australia}
\affiliation{Instit\"ut f\"ur Theoretische Physik, Heidelberg University, 69120 Heidelberg, Germany}

\author{Francesca Maria Marchetti}
\affiliation{Departamento de F\'isica Te\'orica de la Materia Condensada, Universidad Aut\'onoma de Madrid, Madrid 28049, Spain}
\affiliation{Condensed Matter Physics Center (IFIMAC), Universidad Aut\'onoma de Madrid, Madrid 28049, Spain}

\author{Jesper Levinsen}
\affiliation{School of Physics and Astronomy, Monash University, Victoria 3800, Australia}

\author{Meera M. Parish}
\affiliation{School of Physics and Astronomy, Monash University, Victoria 3800, Australia}

\date{\today}

\begin{abstract}
Quantum droplets are dilute self-bound configurations of bosons that result from the balance between a mean-field attraction and a repulsion induced by quantum fluctuations. Such droplets have been successfully realized in cold atomic gases and represent a signature of their quantum nature. Here, we predict the existence of a similar droplet phase in a solid-state system, involving polaritons formed from the strong coupling between excitons (bound electron-hole pairs) and photons in a semiconductor microcavity. 
We consider a spin mixture of exciton-polaritons near a biexciton Feshbach resonance, which allows one to tune the interspecies interactions to be attractive and comparable in magnitude to the intraspecies repulsion. We find that self-bound quantum droplets are achievable for realistic parameters in atomically thin semiconductors, and that they can be detected via their excitation spectrum and spatial profile. This exotic phase could potentially lead to polariton condensation at lower thresholds and it opens an alternative avenue to achieve the long-sought quantum polaritonic regime. 
\end{abstract}

\maketitle

Classical liquids arise from the balance between two competing interatomic forces---a short-range repulsion and a long-range induced-dipole-dipole attraction~\cite{VanDerWaals1873}---and may form dense droplets held together by surface tension. Similar configurations also appear in the quantum realm in the form of dilute nanodroplets of liquid Helium~\cite{Barranco2006}, the prototypical  strongly correlated quantum liquid~\cite{Leggett2008}. 
Recently, a novel ultradilute self-bound quantum droplet was predicted to exist in ultracold mixtures of bosonic atoms  with tunable interactions~\cite{Petrov2015} (see also \cite{Bulgac2002}). This arises due to a delicate competition between a net mean-field (MF) attraction---occuring when the interspecies attraction overcomes the intraspecies repulsion---and repulsive Lee-Huang-Yang (LHY) quantum fluctuations~\cite{Lee1957,Larsen1963}. 
Remarkably, shortly after this was proposed, 
quantum droplets were experimentally realized in homonuclear~\cite{Cabrera2018,Semeghini2018} and heteronuclear~\cite{DErrico2019} binary mixtures, as well as in dipolar condensates~\cite{FerrierBarbut2016, Chomaz2016}. This has led to intense theoretical and experimental efforts~\cite{Petrov2025}, for instance on lowering the dimensionality~\cite{Petrov2016,Cheiney2018,Lavoine2021b}, analyzing rotational~\cite{Kartashov2018,Caldara2022} and collisional properties~\cite{Ferioli2019, Cavicchioli2025}, considering new stabilizing mechanisms~\cite{Cappellaro2017, Lavoine2021,Chiquillo2025,Mixa2025}, and even different mixtures~\cite{Wang2020, Xu2022}. While quantum droplets have proved to be an extremely active research field, the focus has thus far been limited to ultracold atomic gases. 

In this Letter, we predict the existence of self-bound quantum droplets in a solid-state system involving light and matter. Specifically, we consider exciton-polaritons, which are hybrid light-matter quasiparticles that emerge from the strong coupling between excitons (electron-hole bound states) and photons in a two-dimensional (2D) semiconductor microcavity~\cite{Deng2010,Shelykh2010,Carusotto2013}. 
Due to their bosonic nature and small effective mass, exciton-polaritons display remarkable collective coherent phenomena like Bose-Einstein condensation~\cite{Kasprzak2006,Deng2010,Balili2007}, superfluidity~\cite{Amo2009NatPhys,Lerario2017} and quantum vortices~\cite{Lagoudakis2008,Lagoudakis2009,Sanvitto2010}. 
Over the last two decades, most experiments have been in the semiclassical regime where the collective effects can be accurately explained by a MF theory of non-linear classical waves~\cite{Deng2010,Carusotto2013}.
However, recently, there has been a growing effort to access the regime of strongly correlated polaritons, which is necessary for a range of quantum photonic applications in scalable semiconductor systems~\cite{Gerace2009,Ballarini2013,Sanvitto2016,Gerace2019,Liew2023}.
Common proposals in this direction involve enhancing polariton-polariton repulsion (e.g., by exploiting 
dipolaritons~\cite{Rivera2015,Arora2017,Calman2018,Horng2018,Niehues2019} or Rydberg polaritons~\cite{Gu_NatComm2021,Orfanakis2022,Makhonin_LSA2024}), or strongly confining the cavity  photon mode~\cite{MunozMatutano2019,Delteil2019,Kuriakose2022}. Here, we propose to instead \emph{enhance} quantum fluctuations by effectively \emph{weakening} the MF effects. 

\begin{figure}[tp]
    \centering
    \begin{minipage}{0.48\textwidth}
    \centering
    \includegraphics[width=\textwidth]{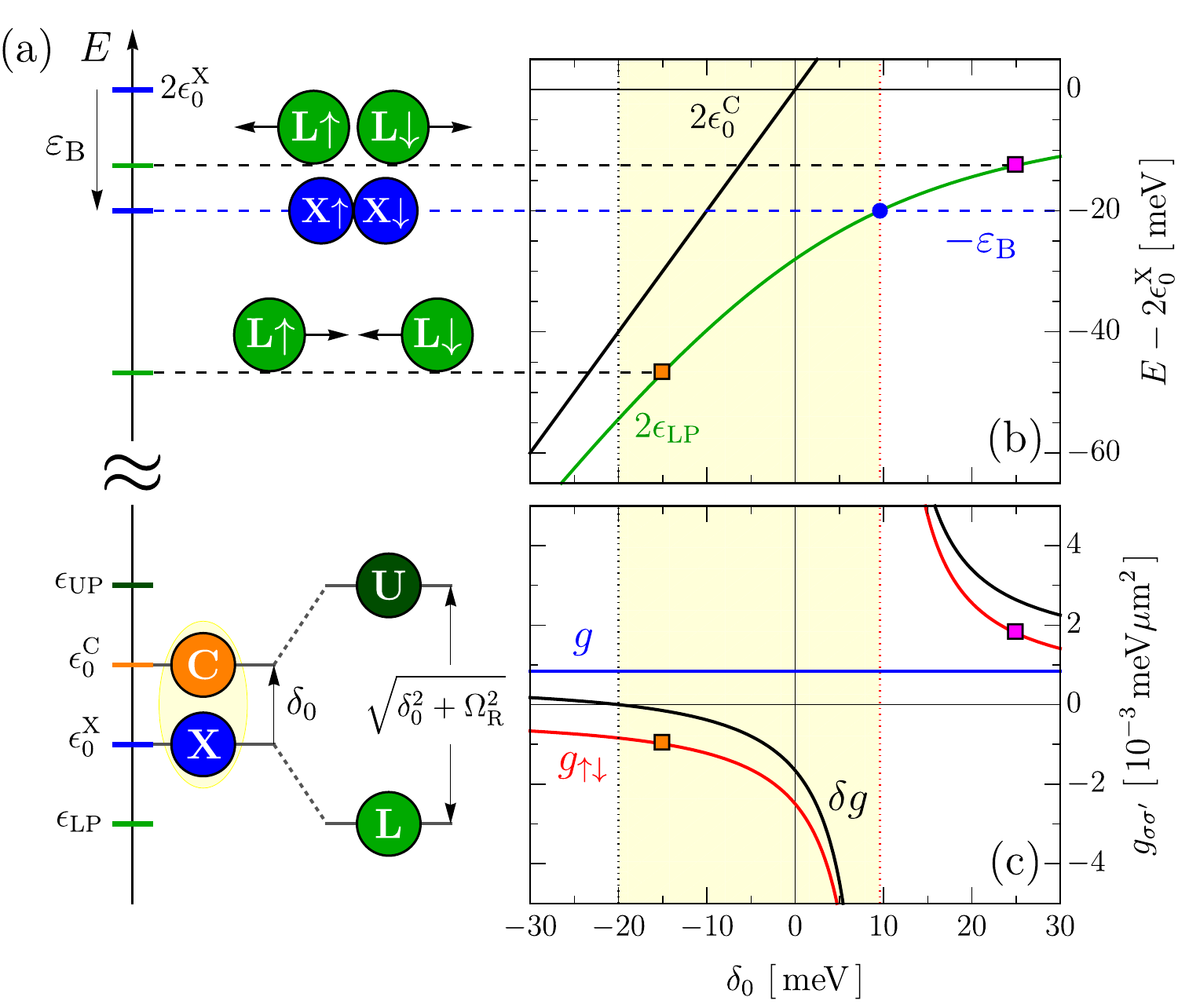}
    \end{minipage} 
    \caption{(a) Schematic illustration of the energy scales for one (bottom) and two (top) particles. The hybridization of an exciton (X) and a photon (C) yields upper (U) and lower (L) polariton quasiparticles. Two $\uparrow$ and $\downarrow$ lower polaritons repel (attract) each other when their energy $2 \epsilon_{\rm LP}$ is above (below) the energy $-\varepsilon_{\rm B}$ of the biexciton (X$\uparrow$-X$\downarrow$). 
    (b) Two-lower-polariton (green) and two-photon (black) energies as a function of detuning $\delta_0$. 
    The blue point denotes the Feshbach resonance that occurs when $2 \epsilon_{\rm LP} = -\varepsilon_{\rm B}$. 
    (c) Detuning dependence of the intra- (blue) and inter-species (red) exciton-exciton interaction coefficients. The magenta (orange) square is an example of repulsive (attractive) interspecies interactions.
    The black curve shows $\delta g = g + g_{\uparrow \downarrow}$, and the yellow region denotes the necessary condition $\delta g < 0$ for the appearance of quantum droplets.
    We use parameters inspired by recent experiments with MoSe$_2$ monolayers: $\Omega_{\rm R} = 28$ meV, $\varepsilon_{\rm X} = 470$ meV~\cite{Dufferwiel2015}, $\varepsilon_{\rm B} = 20$ meV~\cite{Hao2017}, and exciton mass $m_{\rm X} = 1.14 m_0$~\cite{Kylanpaa2015} (with $m_0$ the bare electron mass).
    }
    \label{fig:Fig1}
\end{figure}

Key to our proposal is the ability to tune the interactions between polaritons, similarly to the case of ultracold atomic gases~\cite{FerrierBarbut2016, Chomaz2016, Cabrera2018, Semeghini2018}. Specifically, we take advantage of the polariton 
Feshbach resonance~\cite{Saba_PRL2000,Borri2000,Wen_NJP2013,Wouters2007,Carusotto2010,Bleu2020}, which arises when
the energy of two opposite-spin polaritons matches that of a biexciton bound state~\cite{Miller_PRB1982,Lovering_PRL1992,Borri_SST2003}. As depicted in Fig.~\ref{fig:Fig1}, this leads to a strong enhancement of interactions and a sudden change in their sign, thus enabling us to enter a regime where quantum fluctuations dominate. The Feshbach resonance has recently been observed both in quantum well (QW) semiconductors~\cite{TakemuraNatPhys2014,Takemura2017,NavadehToupchi2019} and in transition metal dichalcogenide (TMD) monolayers \cite{BingTan2023, Zhao2024}, and it has also received significant attention in the context of  many-body effects such as Bose polarons~\cite{Levinsen2019, BastarracheaMagnani2019,BingTan2023,Choo2024}, photon dimers~\cite{CamachoGuardian2021}, and novel collective paired phases~\cite{Marchetti2014,Vermileya2024}.

Using a Bogoliubov approximation combined with a bosonic pairing approach~\cite{Nozieres1982,HuPRL2020,HuPRA2020}, we demonstrate that the quantum droplet phase is robust over a wide range of experimentally accessible parameter regimes and can exist whenever mean-field attraction is present. We obtain several experimental signatures of this novel quantum polariton phase such as the excitation spectrum, the saturation density, and the spatial profile. In particular, we predict that the droplet will occur for parameters that are within reach of current experiments on semiconductor microcavities. 
\paragraph{Model.---} We consider a 2D spin mixture of exciton-polaritons described by the Hamiltonian $\hat{H} = \hat{H}_0 + \hat{V}$ written in the photon-exciton basis (setting $\hbar$ and the system area to 1):
\begin{subequations}
\begin{align}
    \label{eq:H0}
    \hat{H}_0 =& \sum_{\k \sigma} 
    \begin{pmatrix}
        \hat{c}_{\k \sigma}^\dagger & \hat{x}_{\k \sigma}^\dagger 
    \end{pmatrix}
    \begin{pmatrix}
        \epsilon_{\k}^{\rm C}-\mu_{\sigma} & \Omega_{\rm R}/2 \\
        \Omega_{\rm R}/2 & \epsilon_{\k}^{\rm X}-\mu_{\sigma}
    \end{pmatrix}
    \begin{pmatrix}
        \hat{c}_{\k \sigma} \\
        \hat{x}_{\k \sigma}
    \end{pmatrix}, \\
    \label{eq:V}
    \hat{V} =& \sum_{\sigma \sigma'} \sum_{\k \k' \q} \frac{v_{\sigma \sigma'}}{2} \hat{x}_{\k+\q \sigma}^\dagger \hat{x}_{\k'-\q \sigma'}^\dagger \hat{x}_{\k' \sigma'} \hat{x}_{\k \sigma}.    
\end{align}
\end{subequations}
Here, $\hat{x}_{\k \sigma}^\dagger$ ($\hat{c}_{\k \sigma}^\dagger$) creates an exciton (cavity photon) with in-plane momentum $\k$, spin (polarization) $\sigma = \{ \uparrow, \downarrow \}$ and dispersions $\epsilon_{\k}^{\rm X}= k^2/2m_{\rm X}$ and $\epsilon_{\k}^{\rm C}=k^2/2m_{\rm C}+\delta_0$, respectively, where the typical photon to exciton mass ratio is $m_{\rm C}/m_{\rm X} = 10^{-5}$. 
We measure energies with respect to the $\k=0$ exciton, with $\delta_0$ the photon-exciton detuning, and
$\mu_\sigma$ the chemical potentials for each spin.
Equation~\eqref{eq:H0} treats the excitons as structureless bosons which is valid as long as the light-matter coupling is much smaller than the exciton binding energy, i.e., $\Omega_{\rm R} \ll \varepsilon_{\rm X}$, as is the case in microcavities with either monolayer TMDs or a few III-V quantum wells. 

The non-interacting Hamiltonian $\hat{H}_0$ is diagonalized by the transformations $\hat{L}_{\k \sigma} = C_{\k} \hat{c}_{\k \sigma} + X_{\k} \hat{x}_{\k \sigma}$, $\hat{U}_{\k \sigma} = X_{\k} \hat{c}_{\k \sigma} - C_{\k} \hat{x}_{\k \sigma}$, where $\hat{L}_{\k \sigma}$ and $\hat{U}_{\k \sigma}$ are the lower (LP) and upper (UP) polariton quasiparticles with dispersions $E_{\rm UP/LP}(\k)=\frac{1}{2}\left( \epsilon_{\k}^{\rm C} + \epsilon_{\k}^{\rm X} \pm \sqrt{\left( \epsilon_{\k}^{\rm C} - \epsilon_{\k}^{\rm X} \right)^2 + \Omega_{\rm R}^2} \right)$, while $X_{\k}$/$C_{\k}$ are the Hopfield coefficients which quantify the exciton/photon fraction inside each quasiparticle (their explicit expressions can be found in the Supplemental Material (SM)~\cite{SM}). In the following, we denote the polariton energies at zero momentum as $\epsilon_{\rm{UP}/\rm{LP}}$. 

Polaritons inherit interactions from their exciton component, which have underlying spin-dependent interaction strengths $v_{\sigma \sigma'}$, as in Eq.~\eqref{eq:V}. 
Since exciton interactions are short-ranged, they can be described by an effective interaction strength $g_{\sigma\sigma'}$, which is obtained from the Born series expansion~\cite{SM}
\begin{align} \label{eq:Born0}
    g_{\sigma\sigma'} = v_{\sigma\sigma'} \left(1 + v_{\sigma\sigma'} \sum_\k\frac{1}{2 \epsilon_{\rm LP} - 2 \epsilon_{\k}^{\rm X}} + \ldots \right) \, ,
\end{align}
where the collision energy in the scattering is set by the two interacting lower polaritons~\cite{Bleu2020}. 
For the repulsive intraspecies interactions $g_{\uparrow \uparrow} = g_{\downarrow \downarrow} \equiv g$, it is sufficient to expand up to second order and use the standard Born approximation~\cite{Ciuti1998,Tassone1999} for the underlying interactions $v_{\uparrow \uparrow} = v_{\downarrow \downarrow} \equiv v$. We take $g = 4 \pi /m_{\rm X}$ since this approximately corresponds to the values for both TMDs and QW semiconductors (within a $\sim$10\% uncertainty~\cite{deLaFuente2025}). 

On the other hand, the underlying interspecies interactions are attractive, i.e., $v_{\uparrow \downarrow} = v_{\downarrow \uparrow} <0$, and they lead to a $\uparrow \downarrow$ biexciton bound state which requires the infinite Born series in Eq.~\eqref{eq:Born0}. Importantly, in the regime $\Omega_{\rm R} \ll \varepsilon_{\rm X}$ the effective interaction $g_{\uparrow \downarrow}$ takes a universal low-energy form in the vicinity of the biexciton bound state with binding energy $\varepsilon_{\rm B}$~\cite{Bleu2020}:
\begin{equation}
    \label{eq:inter_cont_int_ren}
    \frac{1}{g_{\uparrow \downarrow}} = \frac{1}{v_{\uparrow \downarrow}} + \sum_{\k} \frac{1}{2 \epsilon_{\k}^{\rm X}-2\epsilon_{\rm LP}} = \frac{m_{\rm X}}{4 \pi} \ln \left( \frac{\varepsilon_{\rm B}}{2 |\epsilon_{\rm LP}|} \right).
\end{equation}
In contrast to 2D ultracold gases, $g_{\uparrow\downarrow}$ depends on the light-matter coupling instead of the system density. 

As shown in Fig.~\ref{fig:Fig1},  
the interspecies interaction can be tuned 
from attractive ($g_{\uparrow \downarrow}<0$) to repulsive ($g_{\uparrow \downarrow}>0$) by varying the photon detuning $\delta_0$ across the biexciton Feshbach resonance $\epsilon_{\rm LP}=-\varepsilon_{\rm B}/2$~\footnote{Note that the biexciton resonance essentially coincides with the bipolariton resonance due to the smallness of the cavity photon mass~\cite{Borri_SST2003}.}. For concreteness, we use typical experimental parameters for TMD monolayers.
While a collapse instability of the polariton condensate is predicted at the MF level~\cite{Vladimirova2010} when $\delta g \equiv g + g_{\uparrow \downarrow} < 0$ [yellow region in Figs.~\ref{fig:Fig1}(b-c)], we show here that beyond-MF quantum fluctuations can stabilize the system into a self-bound droplet phase. Note that since the Born approximation represents an upper bound on $g$~\cite{Li2021}, in reality we expect the yellow region to extend to larger negative detunings.
The precise value of the lower critical detuning can also depend on the finite range of the interactions.

\paragraph{Bogoliubov theory with bosonic pairing.---} We now turn to the many-body system and consider the thermodynamic potential $\Omega = \langle \hat{H} \rangle_0$, where $\langle \dots \rangle_0$ denotes an expectation value over the Bose-Einstein condensate (BEC) of LPs. 
To capture the biexciton Feshbach resonance, we introduce the pairing field $\Phi = - v_{\uparrow \downarrow} \sum_{\p} \langle \hat{x}_{\p \uparrow} \hat{x}_{-\p \downarrow} \rangle$ and we perform a MF decoupling of the singlet sector of the interaction term. Thus, Eq.~\eqref{eq:V} becomes
\begin{align}
    \hat{V} =& \sum_{\k \k' \q, \sigma} \frac{v}{2} \hat{x}_{\k+\q \sigma}^\dagger \hat{x}_{\k'-\q \sigma}^\dagger \hat{x}_{\k' \sigma} \hat{x}_{\k \sigma} \notag \\
    &- \frac{|\Phi|^2}{v_{\uparrow \downarrow}} - \sum_{\k} \left( \Phi \hat{x}_{\k \uparrow}^\dagger \hat{x}_{-\k \downarrow}^\dagger + \Phi^* \hat{x}_{-\k \downarrow} \hat{x}_{\k \uparrow} \right).
\end{align}
A similar bosonic pairing approach was first considered in the context of attractive single-component Bose liquids~\cite{Nozieres1982}, and has already been used to model quantum droplets in 3D Bose mixtures~\cite{HuPRL2020, HuPRA2020}.

The macroscopic condensation of polaritons into the zero-momentum state can be described by a zero-temperature semiclassical approximation of the exciton and photon operators, i.e., $\hat{x}_{\mathbf{0} \sigma} \sim \alpha_{\sigma}$ and $\hat{c}_{\mathbf{0} \sigma} \sim \lambda_{\sigma}$.
Since the system Hamiltonian $\hat{H}$ is endowed with a $U(1) \times U(1)$ symmetry, i.e., it is invariant under a global  transformation $\alpha_{\sigma}(\lambda_{\sigma}) \mapsto e^{i \theta_{\sigma}} \alpha_{\sigma}(\lambda_{\sigma})$ and $\Phi \mapsto e^{i ( \theta_{\uparrow} + \theta_{\downarrow})} \Phi$, we choose $\alpha_{\sigma}, \lambda_{\sigma}, \Phi \in \mathbb{R}$ without any loss of generality. 
While a general derivation can be found in the SM~\cite{SM}, here we focus on the experimentally relevant case of a symmetric spin mixture ($\alpha_{\sigma}= \alpha/\sqrt{2}$, $\lambda_{\sigma}= \lambda/\sqrt{2}$ and $\mu_{\sigma} = \mu$) where the thermodynamic potential at the mean-field level reads:
\begin{equation}
    \label{eq:OmMFgeneral}
    \Omega_{\rm MF} = - \frac{\Phi^2}{g_{\uparrow \downarrow}} - (\Phi + \mu) \alpha^2 + (\delta_0 - \mu) \lambda^2 + \Omega_{\rm R} \lambda \alpha + \frac{g}{4} \alpha^4,
\end{equation}
in terms of $g$ and $g_{\uparrow\downarrow}$.
Since photons are non-interacting, the photon densities enter $\Omega$ only through its MF part, so the stationary condition $\partial_{\lambda}\Omega_{\rm MF} = 0$ relates them to the exciton densities as $\lambda = - \Omega_{\rm R} \alpha/[2(\delta_0-\mu)]$. This relation, which in the non-interacting case determines the total occupation of the LP branch~\cite{SM}, can be inserted into Eq.~\eqref{eq:OmMFgeneral} to give
\begin{equation}
    \Omega_{\rm MF}(\alpha,\Phi,\mu) = - \frac{\Phi^2}{g_{\uparrow \downarrow}} - \left( \Phi + \mu + \frac{\Omega_{\rm R}^2}{4 (\delta_0 - \mu)}\right) n_0 + \frac{g}{4} n_0^2,
    \label{eq:OmMF}
\end{equation}
where $n_0=\alpha^2$ is the exciton density of the polariton condensate.
$\Omega_{\rm MF}$ explicitly depends on the chemical potential, whose MF value $\mu_{\rm MF}$ is obtained by minimizing $\Omega_{\rm MF}$ with respect to $\alpha$~\cite{SM}.  
\begin{figure}[tp]
    \begin{minipage}{0.475\textwidth}
    \centering
    \includegraphics[width=\textwidth]{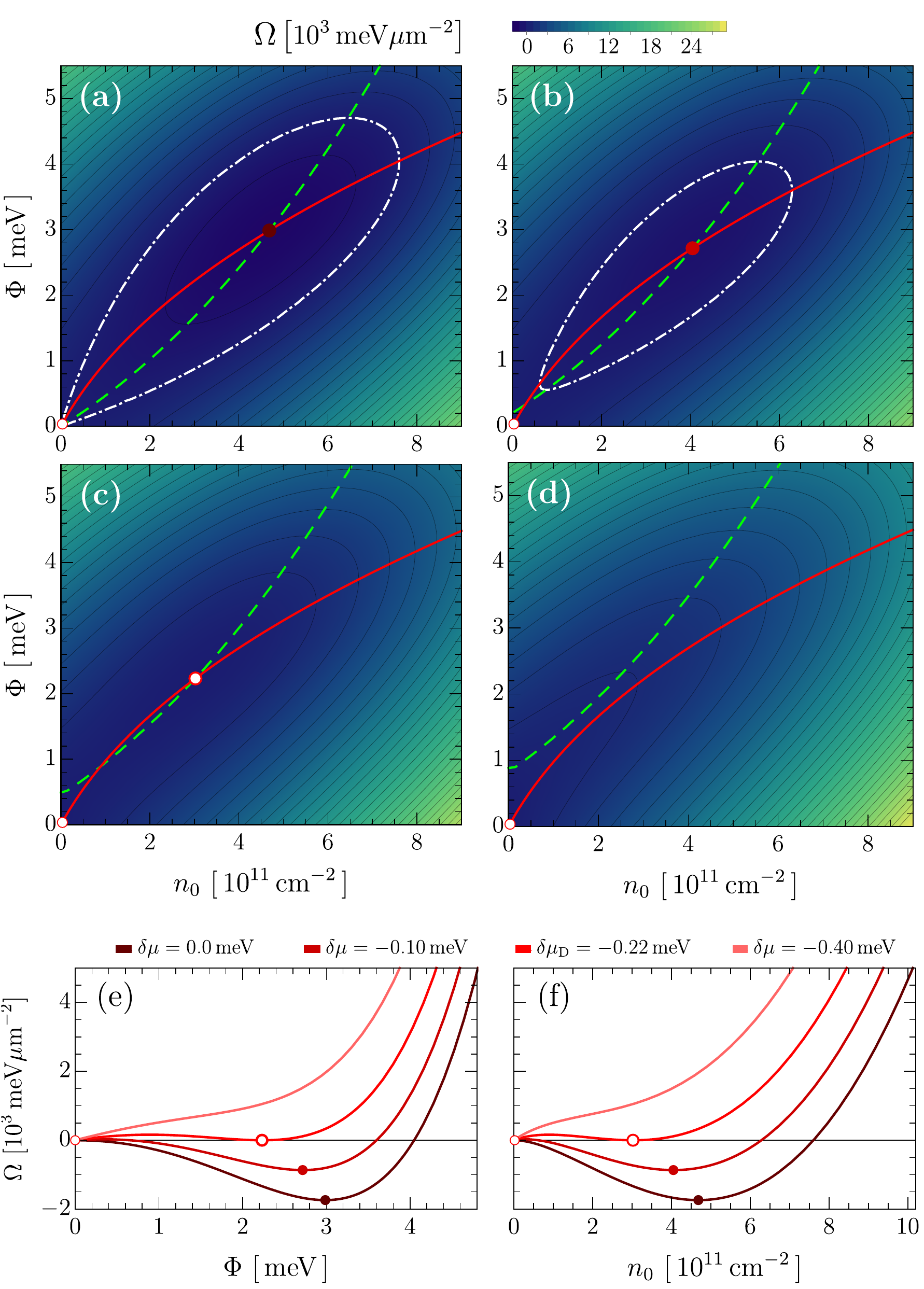}
    \end{minipage}
    \caption{(a-d) Thermodynamic potential as a function of exciton density and pairing field for 4 values of the chemical potential $\delta \mu = \mu-\epsilon_{\rm LP}$: $0$ (a), $-0.10$~meV (b), $-0.22$~meV (c), $-0.40$~meV (d). 
    The white dot-dashed contour denotes the zero pressure line, i.e., the zeros of $\Omega$, while the green dashed and red solid curves correspond to the conditions $\partial_{\alpha}\Omega = 0$ and $\partial_{\Phi}\Omega = 0$, respectively. The filled dots in panels (a) and (b) are the stationary points satisfying both conditions. The empty dot in panel (c) indicates the quantum droplet which appears as a first order transition, signaling phase coexistence with the vacuum (white-red point). Thermodynamic potential as a function of pairing field (e) and exciton density (f) for the same values of $\delta\mu$, obtained by evaluating $\Omega$ along the red solid curve of the corresponding panels (a-d).  
    The detuning is fixed at $\delta_0 = 0$, while all the other system parameters are the same as in Fig.~\ref{fig:Fig1}.}
    \label{fig:Fig2}
\end{figure}

The next contribution is the LHY term coming from quantum fluctuations around the condensate. This is evaluated at $\mu = \mu_{\rm MF}$ and reads
\begin{align}
    \label{eq:OmLHY}
    \Omega_{\rm LHY}(\alpha,\Phi) = \frac{1}{2} &\sum_{\k \ne 0} 
    \bigg[ E_{+, \k}^a +  E_{-, \k}^a + E_{+, \k}^b +  E_{-, \k}^b \notag \\
    &- 2 \left(
    A_{\k} + B_{\k} \right)
    + \frac{C_+^2+C_-^2}{2 \epsilon_{\k}^{\rm X} + 2 |\epsilon_{\rm LP}|} \bigg].
\end{align}
Here, $E_{\pm, \k}^{a/b}$ are the four Bogoliubov branches of the system and we have introduced $A_{\k} = \epsilon_{\k}^{\rm C} - \mu_{\rm MF}$, $B_{\k} =  \epsilon_{\k}^{\rm X} - \mu_{\rm MF} + g n_0$ and $C_{\pm} = gn_0/2 \pm \Phi$. 
The spectrum of excitations is obtained from the diagonalization of the Bogoliubov Hamiltonian, and analytical expressions of the four branches are found for a symmetric mixture. In the limit of small photon mass $m_{\rm C} \ll m_{\rm X}$, the Bogoliubov spectrum simplifies and the momentum integral in Eq.~\eqref{eq:OmLHY} can be computed analytically~\cite{SM}. Notice that the validity of the Bogoliubov theory improves by moving further from the biexciton resonance, i.e., for negative values of detuning (cf. Fig.~\ref{fig:Fig1}).

\begin{figure}[tp]
    \centering
    \begin{minipage}{0.48\textwidth}
    \centering
    \includegraphics[width=\textwidth]{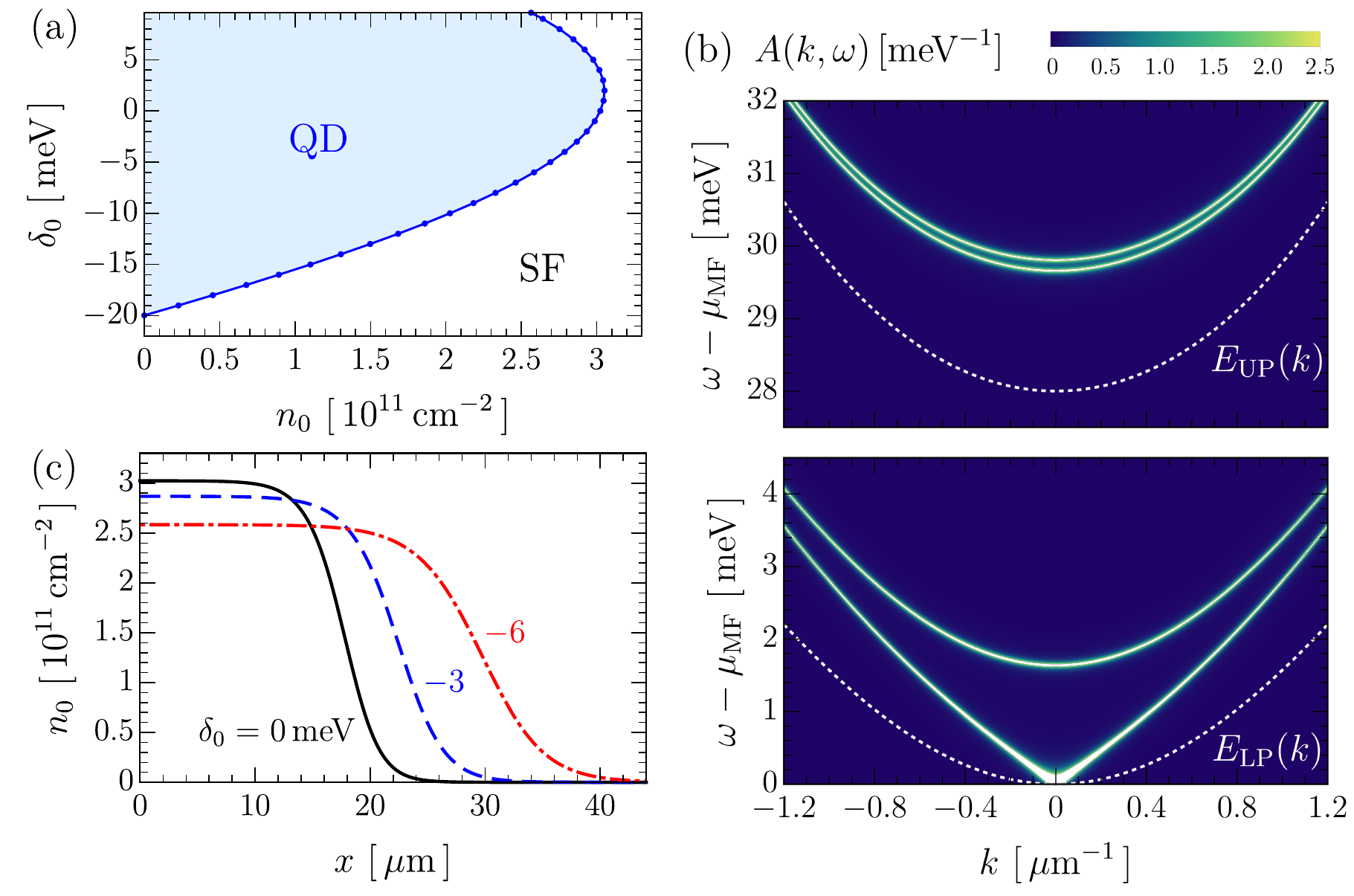}
    \end{minipage} 
    \caption{(a) Phase diagram of detuning versus exciton density in the polariton condensate. The blue solid line bounds the region (QD) where quantum droplets can form, and it corresponds to the saturation density of the quantum droplet in phase coexistence with the vacuum. Outside this region, the system is in a superfluid  miscible phase (SF). Quantum droplets only occur in the detuning range $-19.97$ meV $<\delta_0<9.6$ meV, where $\delta g\le 0$ for our parameters.   
    (b) Absorption spectrum, featuring the four Bogoliubov branches $E_{-,\k}^{a/b}$ (bottom panel) and $E_{+,\k}^{a/b}$ (top panel), which includes a linear gapless mode. The LP and UP dispersions are plotted as white dashed lines. 
    The cavity linewidth has been fixed to $\Gamma=0.025$~meV. 
    (c) Droplet density profiles, obtained from a Gross-Pitaevskii (GP)-like approach (see text), for $\delta_0 = 0.0$ (black, solid line), $-3.0$ (blue dashed) and $-6.0$ meV (red dash-dotted). System parameters are as in Fig.~\ref{fig:Fig1}.}
    \label{fig:Fig3}
\end{figure}

\paragraph{Droplet phase.---} The existence of self-bound quantum droplets relies on the coexistence between vacuum (zero density) and a finite-density phase at zero pressure $P$. 
Since the total thermodynamic potential $\Omega = \Omega_{\rm MF} + \Omega_{\rm LHY}=-P$ depends on the exciton density $n_0$ (via $\alpha$), the pairing field $\Phi$ and the chemical potential $\mu$, the quantum droplet state $\text{QD} \equiv (n_{0\text{D}}, \Phi_{\rm D}, \mu_{\rm D})$ is thus the simultaneous solution of the three conditions  
\begin{align}
    \label{eq:QDconds}
    \text{(i)} \, \Omega|_\text{QD} &= 0, & \text{(ii)} \, \partial_{\alpha} \Omega|_\text{QD} &= 0, & \text{(iii)} \, \partial_{\Phi} \Omega|_\text{QD} &= 0,
\end{align}
at non-zero $n_0.$
To determine the existence of the droplet phase, we plot in Fig.~\ref{fig:Fig2}(a-d) the thermodynamic potential in the $n_0$-$\Phi$ plane at fixed chemical potential for the specific case of zero detuning---these results are extended to the full range of the quantum droplet phase ($\delta g < 0$) in Fig.~\ref{fig:Fig3}. Here, we parametrize $\mu$ as a  deviation from its non-interacting value, i.e., $\mu=\epsilon_{\rm LP}+\delta \mu$. 

Figure~\ref{fig:Fig2}(a) shows the case $\delta \mu = 0$ where the intersections between the green ($\partial_{\alpha} \Omega = 0$) and red ($\partial_{\Phi} \Omega = 0$) curves yield two stationary points: the vacuum, and a global minimum at finite density and positive pressure. This is apparent in Figs.~\ref{fig:Fig2}(e-f) which show the behavior of $\Omega$ along the red curve as a function of the pairing field and the exciton density. Taking instead $\delta \mu=-0.10$~meV in Fig.~\ref{fig:Fig2}(b), the line of zeros shrinks and the vacuum becomes an isolated zero.   
As a result, the corresponding curves in Fig.~\ref{fig:Fig2}(e-f) display a saddle point and a global minimum at negative and positive pressures, respectively. 

For the critical value $\delta \mu_{\rm D} \simeq -0.22$ meV in Fig.~\ref{fig:Fig2}(c), the region of positive pressure disappears as the line of zeros contracts to a single point with $n_{0\rm D} = 3.0 \times 10^{11}\,\text{cm}^{-2}$ and $\Phi_{\rm D} = 2.2 \, \text{meV}$.
This corresponds to phase separation between the vacuum and a polariton condensate at a particular density---the so-called saturation density---thus yielding a quantum droplet of exciton-polaritons. 
On the other hand, in Fig.~\ref{fig:Fig2}(d) where $\delta \mu < \delta \mu_{\rm D}$, the vacuum is the only stable phase. 

As depicted in Fig.~\ref{fig:Fig3}(a), we find that the quantum droplet exists with a well-defined saturation density across the full range of cavity detunings where $\delta g \le 0$ [Fig.~\ref{fig:Fig1}(c)].  
Above this density, the system forms the usual superfluid phase without any instability. 
However, if polaritons are pumped such that the exciton density $n_0$ lies within the quantum-droplet region of the phase diagram in Fig.~\ref{fig:Fig3}(a), then the system increases its density to form a quantum droplet surrounded by vacuum. This suggests the tantalizing possibility of achieving polariton lasing at a lower pump power.

At $\delta_0 \simeq -20$ meV, the saturation density hits zero and only the miscible superfluid phase exists, while above the Feshbach resonance at $\delta_0 > 9.6$ meV, the ground state is unstable towards forming biexcitons. 
Note that the non-monotonic behavior of the phase boundary in Fig.~\ref{fig:Fig3}(a) can be ascribed to the presence of the pairing field since a similar behavior was observed in previous studies of pairing in quantum mixtures~\cite{Marchetti2008, Marchetti2014,Vermileya2024}.
Phase diagrams as a function of either the pairing field or the chemical potential can be found in~\cite{SM}. 
\paragraph{Excitation spectrum.---}
Another key observable of the droplet phase is the excitation spectrum which can be accessed experimentally in polariton systems~\cite{Pieczarka2020,Ballarini2020,Pieczarka2022,Claude2023,Frerot2023}. We plot in Fig.~\ref{fig:Fig3}(b) the excitation absorption spectrum, which is proportional to the
photon spectral function
$A(\k,\omega) = - \frac{1}{\pi} {\rm Im} \left[ G_\text{C}(\k,\omega+ i \Gamma) \right]$~\cite{Cwik2016},
where the photon Green's function is defined in~\cite{SM} and $\Gamma$ is the cavity photon linewidth. 
We observe that the lowest Bogoliubov branch is linear and gapless, i.e., $E^a_{-,\k =0}=0$, as a consequence of the $U(1)$ symmetry breaking of the system. The pairing field in our theory leads to a finite energy gap in the next branch $E^b_{-,\k}$, which  one expects will be removed once further fluctuations are included.  
The spectrum in Fig.~\ref{fig:Fig3}(b) also shows two additional branches which lie higher in energy, as they result from the splitting of the UP branch.
A clear experimental signature of the polariton quantum droplet phase is the behavior of the speed of sound of the gapless mode, since it remains constant for particle numbers below the saturation density but increases in the miscible superfluid phase.

\paragraph{Droplet profiles.---}
Finally, we estimate the size and spatial profile of polariton quantum droplets using a Gross-Pitaevskii (GP) approach within a slab-geometry configuration~\cite{Stringari1987,Cikojevic2021}. Here we assume a sizable condensate with a uniform density along the $y$ direction and a liquid-vacuum interface along the $x$-axis (see~\cite{SM} for details), thus yielding an upper bound for the droplet size. 
As a result, we find that the spatial profile of the exciton density $n_0(x)$ is governed by the equation
\begin{equation}
    \label{eq:den_prof}
    \frac{\hbar^2}{8 M_*} \left[ - 2 \frac{n_0''(x)}{n_0(x)} + \left( \frac{n_0'(x)}{n_0(x)} \right)^2 \right] + \frac{\partial \Omega}{\partial n_0}[n_0(x)] = 0,
\end{equation}
where the light-matter coupling is accounted for by the effective mass $M_*^{-1} = m_{\rm X}^{-1} +\Omega_R^2/[4 m_{\rm C} (\delta_0-\mu)^2]$, which is linked to the LP effective mass via $M_* \simeq X_0^2 M_{\rm LP}$~\cite{Carusotto2013}. Here we have eliminated the dependence on the pairing field using the stationary condition~(\ref{eq:QDconds}.iii). 
Note that the interplay between the inter- and intraspecies interactions is encoded within the density derivative of the thermodynamic potential which thus results in an effective ``trapping potential''~\cite{SM}.
Figure~\ref{fig:Fig3}(c) shows how the density profiles have a flat top, and that the saturation density and droplet radius decreases and increases, respectively, as $|\delta g|$ becomes smaller (i.e., considering more negative cavity detuning). Overall, we find droplet radii of order $\sim 10$ $\mu$m, which is compatible with ongoing experiments.

\paragraph{Conclusions.---}
In this work, we have shown that exciton-polaritons can host quantum droplets, the ultimate liquid-like behavior of a quantum fluid of light. 
Our findings complement previous studies of liquid light in the context of nonlinear optics with competing nonlinearities~\cite{Michinel2002, Michinel2006, Novoa2009, Wu2013}, as well as quasi-equilibrium photon fluids in plasma and semiconductor environments~\cite{Figueiredo2024}.
The advantage of polariton droplets over their cold-atom counterparts is that they do not suffer from three-body recombination, due to the lack of deep two-body bound states~\cite{Braaten2006}, and there is in principle a greater range of particle numbers that can be accessed. Furthermore, the Bogoliubov excitation spectrum is more readily probed in polariton condensates, thus providing a deeper insight into the broken symmetries and associated Goldstone modes of a Bose mixture. Specifically, there is currently a lack of consensus between different perturbative methods~\cite{Petrov2015,HuPRL2020} about the dispersions of the spin and density modes. 
The driven-dissipative nature of the polariton system is not expected to change our results qualitatively for long polariton lifetimes~\cite{Marchetti2014, Sun2017, Ballarini2020} and these effects can also be straightforwardly incorporated in our GP analysis, e.g., through a nonresonant pumping scheme \cite{Keeling2008}.

Our results suggest that quantum droplets of light are achievable for realistic ranges of parameters in current TMD experiments.  
Key signs of the droplet phase include an ultra-low-threshold for condensation, the absence of a blueshift with varying pump power, an unchanging Bogoliubov spectrum, and a condensate size of $\sim$ 10 $\mu$m.  
The observation of quantum droplets would provide an unmistakable signature of the elusive quantum nature of exciton-polariton quasiparticles, thus opening promising avenues into the long-standing challenge of quantum polaritonics~\cite{Gerace2019,Liew2023}.

\medskip
\acknowledgments
We gratefully acknowledge fruitful discussions with Dmitry Petrov, and we thank Cesar Cabrera for helpful feedback on the manuscript. We also thank Humberto Michinel for making us aware of Refs.~\cite{Michinel2002, Michinel2006, Novoa2009, Wu2013}.
MC warmly acknowledges hospitality at the School of Physics and Astronomy of Monash University, where the present work started. 
OB, JL and MMP acknowledge support from the Australian Research Council (ARC) Centre of Excellence in Future Low-Energy Electronics Technologies (CE170100039), and from ARC Discovery Projects DP240100569 and DP250103746. OB also acknowledges support from the Deutsche Forschungsgemeinschaft (DFG) via the Collaborative Research Centre SFB 1225 ISOQUANT (Project-ID No. 273811115). MMP is also supported through an ARC Future Fellowship FT200100619. FMM acknowledges financial support from the Spanish Ministry of Science, Innovation and Universities through the ``Maria de Maetzu'' Programme for Units of Excellence in R\&D (CEX2023-001316-M) and from the Ministry of Science, Innovation and Universities MCIN/AEI/10.13039/501100011033, FEDER UE,  projects No.~PID2020-113415RB-C22 (2DEnLight) and No.~PID2023-150420NB-C31 (Q).

\medskip
The data that support the findings of this article are openly available~\cite{dataav}.

\newpage

\onecolumngrid
\clearpage
\begin{center}
\textbf{\large Supplemental Material:
			Quantum Droplets of Light in Semiconductor Microcavities}\\
\vspace{4mm}
{Matteo~Caldara$^{1,2}$, Olivier~Bleu$^{2,3}$, Francesca~M.~Marchetti$^{4,5}$, Jesper~Levinsen$^{2}$, and Meera~M.~Parish$^{2}$}\\
\vspace{2mm}
{\em \small
$^1$ International School for Advanced Studies (SISSA), via Bonomea 265, 34136 Trieste, Italy\\
$^2$School of Physics and Astronomy, Monash University, Victoria 3800, Australia\\
$^3$Instit\"ut f\"ur Theoretische Physik, Heidelberg University, 69120 Heidelberg, Germany \\
$^4$Departamento de F\'isica Te\'orica de la Materia Condensada, Universidad Aut\'onoma de Madrid, Madrid 28049, Spain\\
$^5$Condensed Matter Physics Center (IFIMAC), Universidad Aut\'onoma de Madrid, Madrid 28049, Spain\\
}\end{center}
\setcounter{equation}{0}
\setcounter{figure}{0}
\setcounter{table}{0}
\setcounter{page}{1}
\makeatletter
\renewcommand{\theequation}{S\arabic{equation}}
\renewcommand{\thefigure}{S\arabic{figure}}

We present here the derivation of the Bogoliubov Hamiltonian for the general case of a spin mixture of exciton-polaritons. Then, we will provide a detailed calculation of the Bogoliubov spectra and the Lee-Huang-Yang (LHY) correction for a symmetric mixture, which is the case presented in the main text. Finally, we will derive the effective Gross-Pitaevskii equation which describes the spatial profile of the droplet density. 

\section{Bogoliubov theory with bosonic pairing}
Let us consider a two-dimensional mixture of exciton-polaritons in spin states $\sigma=\{\uparrow, \downarrow\}$, which is modeled by the Hamiltonian $\hat{H} = \hat{H}_0 + \hat{V}$, where
\begin{subequations}
\label{eq:ex_ph_ham_sup}
\begin{align}
    \label{eq:ex_ph_ham_sup_H0}
    \hat{H}_0 &= \sum_{\k,\sigma} 
    \begin{pmatrix}
        \hat{c}_{\k \sigma}^\dagger & \hat{x}_{\k \sigma}^\dagger
    \end{pmatrix}
    \begin{pmatrix}
        \epsilon^{\rm C}_{\k} - \mu_{\sigma} & \Omega_{\rm R}/2 \\
        \Omega_{\rm R}/2 &  \epsilon^{\rm X}_{\k} - \mu_{\sigma}
    \end{pmatrix}
    \begin{pmatrix}
        \hat{c}_{\k \sigma} \\
        \hat{x}_{\k \sigma}
    \end{pmatrix},
    \\
    \label{eq:ex_ph_ham_sup_V}
    \hat{V} &=
    \sum_{\substack{\k, \k', \q \\ \sigma, \sigma'}} \frac{v_{\sigma \sigma'}}{2 \mathcal{A}} \hat{x}_{\k+\q \sigma}^\dagger \hat{x}_{\k'-\q \sigma'}^\dagger \hat{x}_{\k' \sigma'} \hat{x}_{\k \sigma}.
\end{align}
\end{subequations}
Here, $\hat{x}_{\k \sigma}^\dagger$ creates an exciton with in-plane momentum $\k$, spin $\sigma$ and kinetic energy $ \epsilon^{\rm X}_{\k} = \hbar^2 k^2/2m_{\rm X}$, while $\hat{c}_{\k \sigma}^\dagger$ creates a photon with momentum $\k$, polarization $\sigma$ and kinetic energy $ \epsilon^{\rm C}_{\k} = \hbar^2k^2/2m_{\rm C}+\delta_0$. The detuning $\delta_0$ is the energy of the cavity photon relative to the exciton energy at $\k=0$, while $\Omega_{\rm R}$ is the Rabi coupling between the exciton and the cavity photon. We denote the system area as $\mathcal{A}$.

Formally, our system is reminiscent of the scenario where two bosonic species of atoms are Rabi coupled, which has also been shown to feature droplet formation~\cite{Cappellaro2017, Lavoine2021,Chiquillo2025}. However, we emphasize that there are several crucial differences. In our work, the two matter components are coupled separately to two photon components, rather than being coupled to each other. This leads to a system with four distinct components (two matter and two light, with two Rabi couplings), rather than two components with a single Rabi coupling. Furthermore, the light-matter coupling fundamentally alters the system as, in the absence of many-body effects, it leads to new part-light, part-matter exciton-polariton quasiparticles with a non-parabolic dispersion, something that cannot occur in ultracold atomic gases.

In writing Eq.~\eqref{eq:ex_ph_ham_sup_H0} we have applied the rotating wave approximation and considered the excitons as structureless bosons. 
The former requires the light-matter coupling to be much smaller than the band gap energy, while the latter that the light-matter coupling is much smaller than the exciton binding energy, i.e., $\Omega_{\rm R} \ll \varepsilon_{\rm X}$. Both conditions are well satisfied in monolayer TMDs and devices with a few QWs.

The second term, Eq.~\eqref{eq:ex_ph_ham_sup_V}, contains the exciton-exciton interactions. These arise from the van der Waals force between two neutral composite particles (i.e., excitons), and are hence of short range. In particular, the relevant energy scales for scattering in our problem (e.g., the light-matter coupling $\Omega_\mathrm{R}$ or the mean-field energy shifts) are  much smaller than the energy scale set by the range of the exciton-exciton interaction potential, i.e., $\varepsilon_\mathrm{X}$, and hence we can simply approximate the interaction potential by a low-energy $s$-wave interaction constant ($v_{\uparrow \uparrow} = v_{\downarrow \downarrow} = v$ or $v_{\uparrow\downarrow} = v_{\downarrow \uparrow}$). However, as is well known in Bogoliubov theory~\cite{LandauLifshitzStat2}, some care has to be taken in this limit. We now carefully go through these considerations. 

\subsection{Interaction strength and polariton T-matrix}
To relate the bare interaction coefficients $v$ and $v_{\uparrow\downarrow}$ to the polariton interaction constants, we carry out a similar calculation as in Ref.~\cite{Bleu2020}, where the polariton-polariton T-matrix has been derived exactly.  

The T-matrix operator is given by the Born series expansion
\begin{equation}
    \label{eq:Born}
    \hat{T} = \hat{V} + \hat{V} \frac{1}{E - \hat{H}_0 + i 0^+} \hat{V} + \dots ,
\end{equation}
where $E$ is the collision energy and $i 0^+$ is an infinitesimal imaginary part which shifts the poles slightly into the lower imaginary half plane.
We consider the scattering between two lower polaritons with zero total momentum, hence we evaluate the matrix elements of the above equation over the two-particle states
\begin{equation}
    \ket{L_{\sigma}, L_{\sigma'}, \p} = \hat{L}_{-\p \sigma}^\dagger \hat{L}_{\p \sigma'}^\dagger \ket{0}.
\end{equation}
With the on-shell condition $|\p|=|\p'|$, the two terms of the Born series~\eqref{eq:Born} can be shown to read:
\begin{subequations}
    \begin{align}
        \bra{L_{\sigma}, L_{\sigma'}, \p'} \hat{V} \ket{L_{\sigma}, L_{\sigma'}, \p} &= (1 + \delta_{\sigma \sigma'}) |X_{\p}|^4 v_{\sigma \sigma'} \, ,\\
        \bra{L_{\sigma}, L_{\sigma'}, \p'} \hat{V} \frac{1}{E-\hat{H}_0} \hat{V} \ket{L_{\sigma}, L_{\sigma'}, \p} &= (1 + \delta_{\sigma \sigma'}) |X_{\p}|^4 v_{\sigma \sigma'}^2 \Pi(E) \, ,
    \end{align}
\end{subequations}
where the normalization factor $1+\delta_{\sigma \sigma'}$ accounts for scattering between identical particles, $X_{\p}$ is the Hopfield coefficient and $\Pi(E)$ is the one-loop polarization bubble. Each additional term in the Born series will simply contain an extra factor $v_{\sigma \sigma'} \Pi(E).$  Exploiting the large difference between photon and exciton masses, $m_{\rm C} \ll m_{\rm X}$, the integrand in the polarization bubble is dominated by large momenta up to the momentum cut-off $\Lambda$, leading to the approximation
\begin{equation}
    \Pi(E) \simeq - \frac{1}{\mathcal{A}} \sum_{\k}^{\Lambda} \frac{1}{2 \epsilon_{\k}^{\rm X} - E}.
\end{equation}
Using these results, the interaction strength for lower polaritons is given by
\begin{align}
    T_{\sigma \sigma'}(p) &\equiv \frac{ \bra{L_{\sigma}, L_{\sigma'}, \p'} \hat{T}(2 E_{\rm LP}(\p)) \ket{L_{\sigma}, L_{\sigma'}, \p}}{1+\delta_{\sigma \sigma'}} \notag \\
    &= |X_{\p}|^4 v_{\sigma \sigma'} \left( 1 - \frac{v_{\sigma \sigma'}}{\mathcal{A}} \sum_{\k}^{\Lambda} \frac{1}{2 \epsilon_{\k}^{\rm X} - 2 E_{\rm LP}(\p)}+\dots \right).
\end{align}
We then write the regularized interaction strengths between excitons from the polariton T-matrix at zero momentum as
\begin{equation}
    \label{eq:renorm_final}
    g_{\sigma \sigma'} = \frac{T_{\sigma \sigma'}(0)}{|X_0|^4}=v_{\sigma \sigma'} \left( 1 - \frac{v_{\sigma \sigma'}}{\mathcal{A}} \sum_{\k}^{\Lambda} \frac{1}{2 \epsilon_{\k}^{\rm X} + 2 |\epsilon_{\rm LP}|} + \dots \right),
\end{equation}
where we divide by the Hopfield coefficient as is usual for the relation between exciton and polariton interactions and we used $E_{\rm LP}(p=0)=\epsilon_{\rm LP}<0$. 
In order to obtain cutoff-independent results, a distinction between the two spin channels is in order. 

For the intraspecies repulsion, which we assume to be spin-independent ($v_{\sigma \sigma}=v$), it is sufficient to retain the first two terms in the Born series expansion~\eqref{eq:renorm_final} and take the Born approximation result for the bare interaction strength, with $g_{\sigma \sigma} = g = 4 \pi \hbar^2/m_{\rm X}$~\cite{Tassone1999}. Sending $\Lambda \to \infty$, we can approximate the bare interaction as $v \simeq g$ in the second-order correction of Eq.~\eqref{eq:renorm_final} and obtain:
\begin{equation}
    \label{eq:intra_sup}
    v = g \left( 1 + \frac{g}{\mathcal{A}} \sum_{\k} \frac{1}{2 \epsilon_{\k}^{\rm X} + 2 \left|\epsilon_{\rm LP}\right|} \right).
\end{equation}

The interspecies interaction requires a more careful discussion because the underlying attraction $v_{\uparrow \downarrow} < 0$ leads to a biexciton bound state. This prevents any perturbative treatment, and the infinite Born series in Eq.~\eqref{eq:renorm_final} is required.
As is standard in 2D quantum gases (see Ref.~\cite{Levinsen2015}), the underlying bare coupling constant and the momentum cutoff $\Lambda$~\cite{Levinsen2019} are related to the biexciton binding energy $\varepsilon_{\rm B}$ through
\begin{equation}
    \label{eq:v_Lambda}
    \frac{1}{v_{\uparrow \downarrow}} = - \frac{1}{\mathcal{A}}\sum_{\k}^{\Lambda} \frac{1}{2 \epsilon_{\k}^{\rm X}+\varepsilon_{\rm B}}. 
\end{equation}
In the limit $\Lambda \to \infty$, it yields the universal low-energy form 
\begin{equation}
    \label{eq:inter_sup}
    \frac{1}{v_{\uparrow \downarrow}} + \frac{1}{\mathcal{A}} \sum_{\k} \frac{1}{2 \epsilon_{\k}^{\rm X} + 2 \left|\epsilon_{\rm LP}\right|}  = \frac{m_{\rm X}}{4 \pi \hbar^2}\ln \left( \frac{\varepsilon_{\rm B}}{2 |\epsilon_{\rm LP}|} \right) = \frac{1}{g_{\uparrow \downarrow}},
\end{equation}
where in the last step, we resummed the full geometric series appearing in Eq.~\eqref{eq:renorm_final}. This result for the polariton-polariton T-matrix $T_{\uparrow \downarrow}=|X_0|^4g_{\uparrow\downarrow}$ was derived in Ref.~\cite{Bleu2020}.

\subsection{Bogoliubov approximation}
The Bogoliubov approximation consists of separating both the excitonic and photonic operators into a homogeneous condensate contribution at $\k=0$ and small-amplitude excitations on top of it. To this end, we introduce the classical condensate modes $\hat{x}_{\mathbf{0} \sigma} = \alpha_{\sigma}$ and $\hat{c}_{\mathbf{0} \sigma} = \lambda_{\sigma}$ for excitons and photons, respectively, and we expand the Hamiltonian~\eqref{eq:ex_ph_ham_sup} in powers of $\hat{x}_{\k \ne 0, \sigma}$, $\hat{c}_{\k \ne 0, \sigma}$ up to second order.
The light-matter Hamiltonian becomes
\begin{align}
    \label{eq:H0Bog_sup}
    \hat{H}_0 \simeq \sum_{\sigma} \left[ -\mu_{\sigma} \left|\alpha_{\sigma}\right|^2 + \left( \delta_0 - \mu_{\sigma} \right) \left|\lambda_{\sigma}\right|^2 + \frac{\Omega_{\rm R}}{2} \left( \alpha^*_{\sigma} \lambda_{\sigma} + \alpha_{\sigma} \lambda^*_{\sigma} \right) \right] - \sum_{\k \sigma}{\vphantom{\sum}}' \left( \frac{\epsilon_{\k}^{\rm X}+\epsilon_{\k}^{\rm C}}{2} - \mu_{\sigma} \right) + \frac{1}{2} \sum_{\k}{\vphantom{\sum}}' \hat{\Psi}_{\k}^\dagger \mathcal{H}_0(\k) \hat{\Psi}_{\k},
\end{align}
where the primed sum runs over $\k\ne0$ and we have introduced the spinor
\begin{equation}
    \hat{\Psi}_{\k}
    =
    \begin{pmatrix}
        \hat{c}_{\k \uparrow} & \hat{c}_{\k \downarrow} & \hat{c}_{-\k \uparrow}^\dagger & \hat{c}_{-\k \downarrow}^\dagger & \hat{x}_{\k \uparrow} & \hat{x}_{\k \downarrow} & \hat{x}_{-\k \uparrow}^\dagger & \hat{x}_{-\k \downarrow}^\dagger
    \end{pmatrix}^T
\end{equation}
and the $8 \times 8$ matrix
\begin{equation}
    \mathcal{H}_0(\k)
    =
    \begin{psmallmatrix}
    \epsilon^{\rm C}_{\k} - \mu_{\uparrow} & 0 & 0 & 0 & \Omega_{\rm R}/2 & 0 & 0 & 0 \\
    0 & \epsilon^{\rm C}_{\k} - \mu_{\downarrow} & 0 & 0 & 0 & \Omega_{\rm R}/2 & 0 & 0 \\
    0 & 0 & \epsilon^{\rm C}_{\k} - \mu_{\uparrow} & 0 & 0 & 0 & \Omega_{\rm R}/2 & 0 \\
    0 & 0 & 0 & \epsilon^{\rm C}_{\k} - \mu_{\downarrow} & 0 & 0 & 0 & \Omega_{\rm R}/2 \\
    \Omega_{\rm R}/2 & 0 & 0 & 0 & \epsilon^{\rm X}_{\k} - \mu_{\uparrow} & 0 & 0 & 0 \\
    0 & \Omega_{\rm R}/2 & 0 & 0 & 0 & \epsilon^{\rm X}_{\k} - \mu_{\downarrow} & 0 & 0 \\
    0 & 0 & \Omega_{\rm R}/2 & 0 & 0 & 0 & \epsilon^{\rm X}_{\k} - \mu_{\uparrow} & 0 \\
    0 & 0 & 0 & \Omega_{\rm R}/2 & 0 & 0 & 0 & \epsilon^{\rm X}_{\k} - \mu_{\downarrow} 
    \end{psmallmatrix}.
\end{equation}
We then treat the two spin contributions to the interacting term $\hat{V} = \hat{V}_{\sigma \sigma} + \hat{V}_{\uparrow \downarrow}$ on different footings. Introducing the exciton density of the polariton condensate as $n_{0\sigma} = \left| \alpha_{\sigma} \right|^2/\mathcal{A}$, the intraspecies interaction becomes:
\begin{align}
    \label{eq:V_intra_sup}
    \hat{V}_{\sigma \sigma} &= \sum_{\substack{\k, \k', \q \\ \sigma}} \frac{v}{2 \mathcal{A}} \hat{x}_{\k+\q \sigma}^\dagger \hat{x}_{\k'-\q \sigma}^\dagger \hat{x}_{\k' \sigma} \hat{x}_{\k \sigma} \notag \\
    &\simeq \mathcal{A} \sum_{\sigma} \frac{v}{2} n_{0\sigma}^2 - \sum_{\k \sigma} {\vphantom{\sum}}' v n_{0\sigma} + \frac{1}{2}\sum_{\k} {\vphantom{\sum}}'
    \begin{psmallmatrix}
        \hat{x}_{\k \uparrow}^\dagger & \hat{x}_{\k \downarrow}^\dagger & \hat{x}_{-\k \uparrow} & \hat{x}_{-\k \downarrow}
    \end{psmallmatrix}
    \begin{psmallmatrix}
        2 v n_{0 \uparrow} & 0 & v n_{0 \uparrow} & 0 \\
        0 & 2 v n_{0 \downarrow} & 0 & v n_{0 \downarrow} \\
        v n_{0 \uparrow} & 0 & 2 v n_{0 \uparrow} & 0 \\
        0 & v n_{0 \downarrow} & 0 & 2 v n_{0 \downarrow}
    \end{psmallmatrix}
    \begin{psmallmatrix}
        \hat{x}_{\k \uparrow} \\
        \hat{x}_{\k \downarrow} \\
        \hat{x}_{-\k \uparrow}^\dagger \\
        \hat{x}_{-\k \downarrow}^\dagger
    \end{psmallmatrix}.
\end{align}
For the attractive interspecies interaction, we first perform a mean-field decoupling with a pairing field at the saddle-point level
\begin{equation}
    \label{eq:Phi_sup}
    \Phi = - \frac{v_{\uparrow \downarrow}}{\mathcal{A}} \sum_{\textbf{p}} \langle \hat{x}_{\textbf{p} \uparrow} \hat{x}_{-\textbf{p} \downarrow} \rangle, 
\end{equation}
and then we realize a Bogoliubov expansion in the excitonic operators which gives:
\begin{align}
    \label{eq:V_inter_sup}
    \hat{V}_{\uparrow \downarrow} &= \sum_{\substack{\k, \k', \textbf{q} \\ \sigma}} 
    \frac{v_{\uparrow \downarrow}}{2 \mathcal{A}} \hat{x}_{\k+\textbf{q} \sigma}^\dagger \hat{x}_{\k'-\textbf{q} \bar{\sigma}}^\dagger \hat{x}_{\k' \bar{\sigma}} \hat{x}_{\k \sigma} \notag \\
    &\simeq - \frac{1}{2} \sum_{\sigma} \left( \mathcal{A} \frac{\left|\Phi\right|^2}{v_{\uparrow \downarrow}} + \Phi \alpha^*_{\sigma} \alpha^*_{\bar{\sigma}} + \Phi^* \alpha_{\sigma} \alpha_{\bar{\sigma}} \right)
    +
    \frac{1}{2}\sum_{\k} {\vphantom{\sum}}'
    \begin{psmallmatrix}
        \hat{x}_{\k \uparrow}^\dagger & \hat{x}_{\k \downarrow}^\dagger & \hat{x}_{-\k \uparrow} & \hat{x}_{-\k \downarrow}
    \end{psmallmatrix}
    \begin{psmallmatrix}
       0 & 0 & 0 & - \Phi \\
       0 & 0 & - \Phi & 0 \\
       0 & - \Phi^* &  0 & 0 \\
       - \Phi^* & 0 & 0 & 0
    \end{psmallmatrix}
    \begin{psmallmatrix}
        \hat{x}_{\k \uparrow} \\
        \hat{x}_{\k \downarrow} \\
        \hat{x}_{-\k \uparrow}^\dagger \\
        \hat{x}_{-\k \downarrow}^\dagger
    \end{psmallmatrix}.
\end{align}

The initial quartic Hamiltonian~(\ref{eq:ex_ph_ham_sup}) has been recast into the following quadratic form:
\begin{align}
    \label{eq:H_quadratic_sup}
    \hat{H} \simeq& 
    \sum_{\sigma} \left[ - \frac{\mathcal{A}}{2} \frac{\left|\Phi\right|^2}{v_{\uparrow \downarrow}} - \frac{1}{2} \left( \Phi \alpha^*_{\sigma} \alpha^*_{\bar{\sigma}} + \Phi^* \alpha_{\sigma} \alpha_{\bar{\sigma}} \right) - \mu_{\sigma} \left|\alpha_{\sigma}\right|^2 + \left( \delta_0 - \mu_{\sigma} \right) \left|\lambda_{\sigma}\right|^2 + \frac{\Omega_{\rm R}}{2} \left( \alpha_{\sigma} \lambda^*_{\sigma} + \alpha^*_{\sigma} \lambda_{\sigma} \right) + \frac{v}{2} \frac{\left|\alpha_{\sigma}\right|^4}{\mathcal{A}} \right] \notag \\
    &- \frac{1}{2}\sum_{\k \sigma}{\vphantom{\sum}}' \left(  \epsilon_{\k}^{\rm C} - \mu_{\sigma} + \epsilon_{\k}^{\rm X} - \mu_{\sigma} + 2 g n_{0\sigma} - \frac{\Phi^2 + (g n_{0\sigma})^2}{2 \epsilon_{\k}^{\rm X} + 2 \left|\epsilon_{\rm LP}\right|}
    \right) + \frac{1}{2} \sum_{\k}{\vphantom{\sum}}' \hat{\Psi}_{\k}^\dagger \mathcal{H}(\k) \hat{\Psi}_{\k}, 
\end{align}
where we used Eqs.~\eqref{eq:intra_sup} and~\eqref{eq:inter_sup} to replace the bare interactions in the condensate term. The explicit form of the full Hamiltonian matrix of fluctuations is
\begin{equation}
    \mathcal{H}(\k) = 
    \begin{psmallmatrix}
     \epsilon_{\k}^{\rm C}-\mu_{\uparrow} & 0 & 0 & 0 & \Omega_{\rm R}/2 & 0 & 0 & 0 \\
     0 & \epsilon_{\k}^{\rm C}-\mu_{\downarrow} & 0 & 0 & 0 & \Omega_{\rm R}/2 & 0 & 0 \\
     0 & 0 & \epsilon_{\k}^{\rm C}-\mu_{\uparrow} & 0 & 0 & 0 & \Omega_{\rm R}/2 & 0 \\
     0 & 0 & 0 & \epsilon_{\k}^{\rm C}-\mu_{\downarrow} & 0 & 0 & 0 & \Omega_{\rm R}/2 \\
     \Omega_{\rm R}/2 & 0 & 0 & 0 & \epsilon^{\rm X}_{\k} - \mu_{\uparrow} + 2 g n_{0 \uparrow} & 0 & g n_{0 \uparrow} & - \Phi \\
     0 & \Omega_{\rm R}/2 & 0 & 0 & 0 & \epsilon^{\rm X}_{\k} - \mu_{\downarrow} + 2 g n_{0 \downarrow} & - \Phi & g n_{0 \downarrow} \\
     0 & 0 & \Omega_{\rm R}/2 & 0 & g n_{0 \uparrow} & - \Phi^* &  \epsilon^{\rm X}_{\k} - \mu_{\uparrow} + 2 g n_{0 \uparrow} & 0 \\
     0 & 0 & 0 & \Omega_{\rm R}/2 & - \Phi^* &  g n_{0 \downarrow} &  0 &  \epsilon^{\rm X}_{\k} - \mu_{\downarrow} + 2 g n_{0 \downarrow}
    \end{psmallmatrix}.
\end{equation}

\subsection{Mean-field}
The first line of Eq.~\eqref{eq:H_quadratic_sup} is the mean-field free energy $F_{\rm MF}$. We notice that quantum fluctuations only depend on the densities of excitons, $n_{0\sigma}$, and not on the photonic fraction $\lambda_{\sigma}$. As a result, the minimization of the total free energy with respect to the photon densities yields the simple relation
\begin{equation}
    \label{eq:lambda_alpha_sup}
    \partial_{\lambda_{\sigma}} F_{\rm MF} = 0 \qquad \rightarrow \qquad \lambda_{\sigma} = -\frac{\Omega_{\rm R}}{2 \left( \delta_0 - \mu_{\sigma} \right)} \alpha_{\sigma}.
\end{equation}
If we recall the expressions for the lower-polariton energy and the Hopfield coefficients at zero momentum
\begin{equation}
    \label{eq:Elp_Hopfs}
    \epsilon_{\rm LP} = \frac{1}{2} \left( \delta_0 - \sqrt{\delta_0^2 + \Omega_{\rm R}^2} \right), \qquad \left|C_0\right|^2 = \frac{1}{2} \left( 1 - \frac{\delta_0}{\sqrt{\delta_0^2 + \Omega_{\rm R}^2}} \right), \qquad \left|X_0\right|^2 = 1 - \left|C_0\right|^2,
\end{equation}
we see that in the non-interacting case where $\mu_{\sigma} = \epsilon_{\rm LP}$, the above relation reads $\lambda_{\sigma} = (C_0/X_0) \alpha_{\sigma}$. The physical meaning looks clear if we consider the rotation into the polariton basis, where we restrict to the condensate part at $\k=0$:
\begin{equation}
    \label{eq:simpleRels}
    \begin{pmatrix}
        U_{0 \sigma} \\
        L_{0 \sigma}
    \end{pmatrix}
    =
    \begin{pmatrix}
        C_0 & -X_0 \\
        X_0 & C_0
    \end{pmatrix}
    \begin{pmatrix}
        \alpha_{\sigma} \\
        \lambda_{\sigma}
    \end{pmatrix}
    = 
    \begin{pmatrix}
        0 \\
        \alpha_{\sigma}/X_0
    \end{pmatrix}.
\end{equation}
The optimal value of $\lambda_{\sigma}$ is such that only the lower-polariton condensate is populated, with density $|L_{0 \sigma}|^2 = |\alpha_{\sigma}|^2/X_0^2 = |\lambda_{\sigma}|^2/C_0^2$. In the weakly-interacting regime, the chemical potential deviates from the non-interacting value $\epsilon_{\rm LP}$ and the LP condensate is depleted due to quantum fluctuations. This is precisely the scenario we analyze in the present work. 

Using Eq.~\eqref{eq:lambda_alpha_sup}, one obtains the mean-field thermodynamic potential $\Omega_{\rm MF} = F_{\rm MF}/\mathcal{A}$:
\begin{equation}
    \label{eq:OmMF_sup}
    \Omega_{\rm MF}
    =
    - \frac{\left|\Phi\right|^2}{g_{\uparrow \downarrow}} 
    + 
    \sum_{\sigma} \left[ - \frac{1}{2} \left( \Phi \frac{\alpha^*_{\sigma} \alpha^*_{\bar{\sigma}}}{\mathcal{A}} + \Phi^* \frac{\alpha_{\sigma} \alpha_{\bar{\sigma}}}{\mathcal{A}} \right)
    - \left( \mu_{\sigma} + \frac{\Omega_{\rm R}^2}{4 \left( \delta_0 - \mu_{\sigma} \right)} \right) n_{0\sigma} + \frac{g}{2} n_{0\sigma}^2 \right]. 
\end{equation}

\subsection{LHY quantum fluctuations}
In order to analyze the beyond mean-field Hamiltonian in the second line of Eq.~\eqref{eq:H_quadratic_sup}, one has to consider the matrix (see Ref.~\cite{Petrov2025})
\begin{equation}
    \label{eq:HTilde_sup}
    \tilde{\mathcal{H}}(\k) = 
    \begin{psmallmatrix}
     \epsilon_{\k}^{\rm C}-\mu_{\uparrow} & 0 & 0 & 0 & \Omega_{\rm R}/2 & 0 & 0 & 0 \\
     0 & \epsilon_{\k}^{\rm C}-\mu_{\downarrow} & 0 & 0 & 0 & \Omega_{\rm R}/2 & 0 & 0 \\
     0 & 0 & -\epsilon_{\k}^{\rm C}+\mu_{\uparrow} & 0 & 0 & 0 & -\Omega_{\rm R}/2 & 0 \\
     0 & 0 & 0 & -\epsilon_{\k}^{\rm C}+\mu_{\downarrow} & 0 & 0 & 0 & -\Omega_{\rm R}/2 \\
     \Omega_{\rm R}/2 & 0 & 0 & 0 & \epsilon^{\rm X}_{\k} - \mu_{\uparrow} + 2 g n_{0 \uparrow} & 0 & g n_{0 \uparrow} & - \Phi \\
     0 & \Omega_{\rm R}/2 & 0 & 0 & 0 & \epsilon^{\rm X}_{\k} - \mu_{\downarrow} + 2 g n_{0 \downarrow} & - \Phi & g n_{0 \downarrow} \\
     0 & 0 & -\Omega_{\rm R}/2 & 0 & - g n_{0 \uparrow} & \Phi^* &  - \epsilon^{\rm X}_{\k} + \mu_{\uparrow} - 2 g n_{0 \uparrow} & 0 \\
     0 & 0 & 0 & -\Omega_{\rm R}/2 & \Phi^* &  - g n_{0 \downarrow} &  0 &  - \epsilon^{\rm X}_{\k} + \mu_{\downarrow} - 2 g n_{0 \downarrow}
    \end{psmallmatrix},
\end{equation}
which has the eight eigenvalues $\pm E^{a}_{\pm,\k}$, $\pm E^{b}_{\pm,\k}$. 
The Hamiltonian matrix $\mathcal{H}(\k)$ is diagonalized by the Bogoliubov transformation $\hat{\Psi}_{\k} = \mathcal{R} \hat{\phi}_{\k}$, with 
$\hat{\phi}_{\k} = \begin{pmatrix} \hat{a}_{\k, +} & \hat{a}_{\k, -} & \hat{b}_{\k, +} & \hat{b}_{\k, -} & \hat{a}_{-\k, +}^\dagger & \hat{a}_{-\k, -}^\dagger & \hat{b}_{-\k, +}^\dagger & \hat{b}_{-\k, -}^\dagger \end{pmatrix}^T$.
The columns of the matrix $\mathcal{R}$ are (properly normalized) eigenvectors of $\tilde{\mathcal{H}}(\k)$ corresponding, respectively, to the eigenvalues $E^{a}_{+, \k}$, $E^{a}_{-, \k}$, $E^{b}_{+, \k}$, $E^{b}_{-, \k}$, $-E^{a}_{+, \k}$, $-E^{a}_{-, \k}$, $-E^{b}_{+, \k}$ and $-E^{b}_{-, \k}$. Applying this transformation, one obtains
\begin{align*}
    &
    \hat{\Psi}_{\k}^\dagger
     \left(\mathcal{R}^\dagger \right)^{-1} \, \mathcal{R}^\dagger
     \mathcal{H}(\k)
     \mathcal{R} \, \mathcal{R}^{-1}
     \hat{\Psi}_{\k}
     \\
     &=
     \begin{psmallmatrix}
        \hat{a}_{\k, +}^\dagger & \hat{a}_{\k, -}^\dagger & \hat{b}_{\k, +}^\dagger & \hat{b}_{\k, -}^\dagger & \hat{a}_{-\k, +} & \hat{a}_{-\k, -} & \hat{b}_{-\k, +} & \hat{b}_{-\k, -}
     \end{psmallmatrix}
     \begin{psmallmatrix}
       E_{+, \k}^a & & & & & & & \\
        & E_{-, \k}^a & & & & & & \\
        & & E_{+, \k}^b & & & & & \\
        & & & E_{-, \k}^b & & & & \\
        & & & & E_{+, \k}^a & & & \\
        & & & & & E_{-, \k}^a & & \\
        & & & & & & E_{+, \k}^b & \\
        & & & & & & & E_{-, \k}^b 
     \end{psmallmatrix} 
     \begin{psmallmatrix}
        \hat{a}_{\k, +} \\
        \hat{a}_{\k, -} \\
        \hat{b}_{\k, +} \\
        \hat{b}_{\k, -} \\
        \hat{a}_{-\k, +}^\dagger \\
        \hat{a}_{-\k, -}^\dagger \\
        \hat{b}_{-\k, +}^\dagger \\
        \hat{b}_{-\k, -}^\dagger
     \end{psmallmatrix}
     \\
     &= 2 \left( E_{+, \k}^a  \hat{a}_{\k, +}^\dagger \hat{a}_{\k, +}  +  E_{-, \k}^a  \hat{a}_{\k, -}^\dagger \hat{a}_{\k, -} + E_{+, \k}^b  \hat{b}_{\k, +}^\dagger \hat{b}_{\k, +}  +  E_{-, \k}^b  \hat{b}_{\k, -}^\dagger \hat{b}_{\k, -}  \right)   
     +
     E_{+, \k}^a +  E_{-, \k}^a + E_{+, \k}^b +  E_{-, \k}^b, 
\end{align*}
which leads to the quadratic Hamiltonian:
\begin{equation}
    \hat{H}_2
    =
    \mathcal{A} \, \Omega_{\rm LHY} + \sum_{\k, s=\pm}{\vphantom{\sum}}' 
    \left(
    E_{s, \k}^a  \hat{a}_{\k, s}^\dagger \hat{a}_{\k, s}
    +
    E_{s, \k}^b  \hat{b}_{\k, s}^\dagger \hat{b}_{\k, s}
    \right).
\end{equation}
The above equation shows that the zero-point energy of the Bogoliubov modes corresponds to the LHY contribution to the thermodynamic potential:
\begin{equation}
    \label{eq:OmLHY_sup}
    \Omega_{\rm LHY} = \frac{1}{2 \mathcal{A}} \sum_{\k}{\vphantom{\sum}}' 
    \left[ 
    E_{+, \k}^a +  E_{-, \k}^a + E_{+, \k}^b +  E_{-, \k}^b
    - \sum_{\sigma} \left(
    \epsilon_{\k}^{\rm C} - \mu_{\sigma}
    +
    \epsilon_{\k}^{\rm X} - \mu_{\sigma} + 2 g n_{0\sigma} - \frac{\Phi^2 + (g n_{0\sigma})^2}{2 \epsilon_{\k}^{\rm X} + 2 \left|\epsilon_{\rm LP}\right|} \right)
    \right].
\end{equation}
The knowledge of the Bogoliubov excitation spectrum is essential to compute $\Omega_{\rm LHY}$. In the following section we present analytical results for an unpolarized spin mixture in the limit of small photon mass.

\section{Symmetric spin mixture}
We derive here results for the specific case of a symmetric mixture, where $n_{0 \uparrow} = n_{0 \downarrow} = n_0/2$, with $n_0 = |\alpha|^2/\mathcal{A}$ the total exciton density, and $\mu_{\uparrow} = \mu_{\downarrow} = \mu$.  Without loss of generality, we can take $\alpha, \Phi \in \mathbb{R}$, such that the mean-field thermodynamic potential~\eqref{eq:OmMF_sup} reads
\begin{equation}
    \label{eq:OmMFsym}
    \Omega_{\rm MF}(\alpha, \Phi, \mu)
    =
    - \frac{\Phi^2}{ g_{\uparrow \downarrow}} 
    -
    \left( \Phi + \mu + \frac{\Omega_{\rm R}^2}{4 \left( \delta_0 - \mu \right)}\right) n_0
    +
    \frac{g}{4} n_0^2 \, .
\end{equation}
Solving the stationary condition $\partial_{\alpha} \Omega_{\rm MF} = 0$ with respect to the chemical potential, one gets the two solutions 
\begin{equation}
    \label{eq:muMF}
    \mu_{\pm} = \frac{1}{2} \left( \delta_0 + \frac{g}{2} n_0 - \Phi \pm \sqrt{\left( \delta_0 - \frac{g}{2} n_0 + \Phi \right)^2 + \Omega_{\rm R}^2}  \right).
\end{equation}
The above equation provides an intuitive understanding of the effects of interactions, which are encoded in the combination $g n_0/2-\Phi$. To see this, we first use the approximate relation between the pairing field and the interspecies interaction strength $\Phi \sim - g_{\uparrow \downarrow} n_0/2$ (coming from the minimization of $\Omega_{\rm MF}$ with respect to $\Phi$) to rewrite the combination between interactions as $\delta g (n_0/2)$, with $\delta g = g + g_{\uparrow \downarrow}$. In the droplet phase where $\delta g \lesssim 0$, we can expand Eq.~\eqref{eq:muMF} at first order in $\delta g$ to get
\begin{equation}
    \mu_{\pm} \simeq 
    \begin{cases}
        \epsilon_{\rm UP} + C_0^2 \delta g \frac{n_0}{2} \\
        \epsilon_{\rm LP} + X_0^2 \delta g \frac{n_0}{2}
    \end{cases},
\end{equation}
thus showing the shift in the chemical potential with respect to the non-interacting case.
Since we are specifically considering a spin mixture of lower polaritons, in the following we take $\mu_{\rm MF} \equiv \mu_{-}$ as the mean-field chemical potential. As usual in Bogoliubov theory, either with Petrov's approach~\cite{Petrov2015} or within the bosonic pairing theory~\cite{HuPRL2020}, the excitation spectrum and the LHY correction are derived fixing the chemical potential to its mean-field value into the second line of the Hamiltonian~\eqref{eq:H_quadratic_sup} which describes quantum fluctuations.
After introducing the short-hand notations
\begin{align}
    \label{eq:shorthand}
    A_{\k} &= \epsilon_{\k}^{\rm C}-\mu_{\rm MF}, &
    B_{\k} &= \epsilon_{\k}^{\rm X} -\mu_{\rm MF} + g n_0, &
    C_{\pm} &= \frac{g}{2} n_0 \pm \Phi,
\end{align}
the Bogoliubov spectrum obtained from the diagonalization of the matrix~\eqref{eq:HTilde_sup} consists of the following four branches:
\begin{equation}
    \label{eq:spectrum_sup}
    \begin{aligned}
        {E_{\pm, \k}^{a}}^2 &= 
        \frac{A_{\k}^2+B_{\k}^2-C_{+}^2}{2} + \left( \frac{\Omega_{\rm R}}{2} \right)^2
        \pm
        \sqrt{
        \left(
        \frac{A_{\k}^2-B_{\k}^2+C_{+}^2}{2}
        \right)^2
        +
        \left[ \left( A_{\k}+B_{\k} \right)^2 - C_{+}^2 \right] \left( \frac{\Omega_{\rm R}}{2} \right)^2
        } \, ,
        \\
        {E_{\pm, \k}^{b}}^2 &=
        \frac{A_{\k}^2+B_{\k}^2-C_{-}^2}{2} + \left( \frac{\Omega_{\rm R}}{2} \right)^2
        \pm
        \sqrt{
        \left(
        \frac{A_{\k}^2-B_{\k}^2+C_{-}^2}{2}
        \right)^2
        +
        \left[ \left( A_{\k}+B_{\k} \right)^2 - C_{-}^2 \right] \left( \frac{\Omega_{\rm R}}{2} \right)^2
        } \, .
    \end{aligned}
\end{equation}
Written in terms of these parameters, the beyond mean-field thermodynamic potential~\eqref{eq:OmLHY_sup} for an unpolarized mixture is given by:
\begin{equation}
    \label{eq:OmLHYsym}
    \Omega_{\rm LHY}(\alpha, \Phi) = \frac{1}{2 \mathcal{A}} \sum_{\k}{\vphantom{\sum}}' 
    \left[ 
    E_{+,\k}^a +  E_{-,\k}^a + E_{+,\k}^b +  E_{-,\k}^b
    - 2 A_{\k} - 2 B_{\k} + \frac{C_{+}^2 + C_{-}^2}{2 \epsilon_{\k}^{\rm X} + 2 \left|\epsilon_{\rm LP}\right|}
    \right].
\end{equation}

We notice that the condition $E_{-,\k=0}^{a}=0$ yields:
\begin{align}
    0 &= \left[ A_0(B_0-C_+) - \frac{\Omega_{\rm R}^2}{4} \right] \left[ A_0(B_0+C_+) - \frac{\Omega_{\rm R}^2}{4} \right] \notag \\
    &= (\delta_0-\mu_{\rm MF})^2 \left[ -\left( \mu_{\rm MF} + \Phi + \frac{\Omega_{\rm R}^2}{4 (\delta_0-\mu_{\rm MF})} \right) +g\frac{n_0}{2} \right] \left[ -\left( \mu_{\rm MF} - \Phi + \frac{\Omega_{\rm R}^2}{4 (\delta_0-\mu_{\rm MF})} \right) +\frac{3}{2}g n_0 \right]. 
\end{align}
The first square bracket is precisely the stationary condition $\partial_{\alpha} \Omega_{\rm MF} = 0$ which is satisfied when $\mu = \mu_{\rm MF}$. Therefore, we are guaranteed that the lowest branch of the excitation spectrum is gapless, as expected from Goldstone theorem due to the breaking of the continuous $U(1)$ symmetry of the \emph{density} field $\alpha$. The branch $E^b_{-,\k}$, instead, displays a gap at $\k=0$ as a result of a finite pairing field $\Phi>0$.The remaining two branches $E^{a/b}_{+,\k}$ are further higher in energy, as shown in Fig.~3(b) of the main text.
It is instructive to investigate the following limiting cases:
\begin{itemize}
    \item in the absence of light-matter coupling, which is recovered by setting $A_{\k} = \Omega_{\rm R} = 0$, we get $\mu_{\rm MF} = g n_0/2 - \Phi$ and the spectrum reduces to the two solutions
    \begin{align}
        E_{\pm,\k}^{a} \, &\mapsto \, E_{-}(\k) = \sqrt{\epsilon_{\k}^{\rm X} \left( \epsilon_{\k}^{\rm X} + gn_0 + 2 \Phi \right)} \, , \notag \\
        E_{\pm,\k}^{b} \, &\mapsto \, E_{+}(\k) = \sqrt{\left( \epsilon_{\k}^{\rm X} + gn_0 \right) \left( \epsilon_{\k}^{\rm X} + 2 \Phi \right)} \, , \notag 
    \end{align}
    which are precisely the gapless and gapped branches derived in Refs.~\cite{HuPRL2020,HuPRA2020} in the context of bosonic pairing theory applied to a Bose-Bose mixture. Ref.~\cite{HuPRA2020}, in particular, presents a detailed comparison with the spectrum derived by Petrov~\cite{Petrov2015} relying on the standard Bogoliubov approximation;
    \item in the absence of interactions, $g = \Phi = 0$, the chemical potential is just $\mu_{\rm MF}=\epsilon_{\rm LP}$ and we obtain the lower and upper polariton branches $E_{\rm UP/LP}(\k)$ shifted by the energy of the lower polariton at zero momentum:
    \begin{align}
        E_{-,\k}^{a/b} \, &\mapsto \, E_{-}(\k) = \frac{1}{2} \left( A_{\k} + B_{\k} - \sqrt{ (A_{\k} - B_{\k})^2 + \Omega_{\rm R}^2 } \right) = -\epsilon_{\rm LP} + E_{\rm LP}(\k),  \notag \\
        E_{+,\k}^{a/b} \, &\mapsto \, E_{+}(\k) = \frac{1}{2} \left( A_{\k} + B_{\k} + \sqrt{ (A_{\k} - B_{\k})^2 + \Omega_{\rm R}^2 } \right) = -\epsilon_{\rm LP} + E_{\rm UP}(\k); \notag 
    \end{align}
    \item in the absence of interspecies interaction, $\Phi=0$, we can define $C_{\pm} = C = g n_0/2$ and $\mathcal{K}_{\k \pm} \equiv (A_{\k}^2 \pm B_{\k}^2 \mp C^2)/2$ so that we easily recover the quasiparticle energy spectrum in Ref.~\cite{Hu2022}:
    \begin{equation}
        E_{\pm,\k}^{a/b} \, \mapsto \, E_{\pm}(\k) = \mathcal{K}_{\k +} + \left( \frac{\Omega_{\rm R}}{2} \right)^2
        \pm
        \sqrt{
        \mathcal{K}_{\k -}
        +
        \left[ \left( A_{\k}+B_{\k} \right)^2 - C^2 \right] \left( \frac{\Omega_{\rm R}}{2} \right)^2
        }. \notag
    \end{equation}
\end{itemize}

\subsection{Limit of small photon mass}
An analytical expression for $\Omega_{\rm LHY}$ can be derived in the limit of a very small photon mass, $m_{\rm C}/m_{\rm X} \ll 1$. A similar approach has been followed in Ref.~\cite{Hu2022} for 2D exciton-polaritons with repulsive interaction. In the limit $m_{\rm X}/m_{\rm C} \to \infty$, the function $A_{\k}$ is the dominant term for any $\k \ne 0$ and one can approximate the Bogoliubov branches~\eqref{eq:spectrum_sup} as
\begin{align}
    {E_{+,\k}^{a/b}}^2 &\simeq A_{\k}^2 + \left( \frac{\Omega_{\rm R}}{2} \right)^2 \frac{A_{\k} \left( A_{\k}+B_{\k} \right)}{A_{\k}^2-B_{\k}^2+C_{\pm}^2} \, , \notag \\
    {E_{-,\k}^{a/b}}^2 &\simeq B_{\k}^2-C_{\pm}^2 - \left( \frac{\Omega_{\rm R}}{2} \right)^2 \frac{\left( A_{\k}+B_{\k} \right)B_{\k} - C_{\pm}^2}{A_{\k}^2-B_{\k}^2+C_{\pm}^2} \, . \notag
\end{align}
At the leading order, the dispersion relations of the photon field and exciton field for each species are effectively decoupled
\begin{align}
    E_{+,\k}^{a/b} =& A_{\k} + \mathcal{O}\left( m_{\rm C}/m_{\rm X} \right), &
    E_{-,\k}^{a/b} =& \sqrt{B_{\k}^2-C_{\pm}^2} + \mathcal{O}\left( m_{\rm C}/m_{\rm X} \right),
\end{align}
which yields the following form for the fluctuation contribution to the thermodynamic potential:
\begin{equation}
    \Omega_{\rm LHY}(\alpha,\Phi) = \frac{1}{2 \mathcal{A}} \sum_{\k, s=\pm}{\vphantom{\sum}}' \left[ \sqrt{B_{\k}^2-C_{s}^2} - B_{\k} + \frac{C_{s}^2}{\hbar^2{\k}^2/m_{\rm X} + 2 |\epsilon_{\rm LP}|} \right].
\end{equation}
Following the discussion in Ref.~\cite{Hu2022}, it is then convenient to split the momentum summation into two distinct bits as
\begin{equation}
    \label{eq:sums}
    \frac{1}{\mathcal{A}}\sum_{\k}{\vphantom{\sum}}' 
        \left( 
        \sqrt{B_{\k}^2-C_{s}^2} - B_{\k}
        +
        \frac{C_{s}^2}{2 B_{\k}}
        \right) 
    +
     \frac{1}{\mathcal{A}}\sum_{\k}{\vphantom{\sum}}' 
        \left( 
        \frac{ C_{s}^2}{ \hbar^2 \k^2/m_{\rm X} + 2 |\epsilon_{\rm LP}|} 
        -
        \frac{C_{s}^2}{2 B_{\k}}
        \right). 
\end{equation}
Defining the dummy variable $y=B_{\k}/C_s-1$, the leftmost sum is evaluated as:
\begin{align}
    &\frac{1}{2 \pi} \int_0^{\infty} dk \, k \, \left( 
    \sqrt{B_{\k}^2-C_{s}^2} - B_{\k}
    +
    \frac{C_{s}^2}{2 B_{\k}}
    \right) \notag \\
    &=\frac{m_{\rm X} C_s^2}{2 \pi \hbar^2} \,\int_{B_0/C_s-1}^{\infty} dy \left[ \sqrt{y(y+2)} - (y+1) + \frac{1}{2y + 2} \right] \notag \\
    &=\frac{m_{\rm X} C_s^2}{4 \pi \hbar^2} \left[ (y+1) \sqrt{y(y+2)} - 2 {\rm arcsinh} \left( \sqrt{\frac{y}{2}} \right) - y^2 - 2 y + \ln\left( 2 + 2 y \right) \right]_{B_0/C_s-1}^{\infty} \notag \\
    &=\frac{m_{\rm X}}{4 \pi \hbar^2} \left[ B_0 \left( B_0 - \sqrt{B_0^2 - C_s^2} \right) + C_s^2 \ln\left( \frac{B_0 + \sqrt{B_0^2 - C_s^2}}{2 \sqrt{e} \, B_0} \right) \right],
\end{align}
where the identity ${\rm arcsinh}x = \ln \left( x + \sqrt{x^2 + 1} \right) $ has been used. Then, the rightmost sum in Eq.~\eqref{eq:sums} gives:
\begin{equation}
    \frac{C_s^2}{2 \pi} \int_0^{\infty} dk \left( 
    \frac{ k}{\hbar^2 k^2/m_{\rm X} + 2 |\epsilon_{\rm LP}|} 
    -
    \frac{k}{2 B_{\k}}
    \right) = \frac{m_{\rm X} C_s^2}{4 \pi \hbar^2} \ln \left( \frac{B_{0}}{|\epsilon_{\rm LP}|} \right).
\end{equation}
Recalling the definitions of $B_0$ and $C_{\pm}$ in Eq.~\eqref{eq:shorthand}, the total thermodynamic potential $\Omega = \Omega_{\rm MF} + \Omega_{\rm LHY}$ takes the analytical form:
\begin{align}
    \label{eq:Omega_Bogoliubov}
    \Omega(\alpha, \Phi, \mu) =& - \frac{\Phi^2}{ g_{\uparrow \downarrow}} - \left( \Phi + \mu + \frac{\Omega_{\rm R}^2}{4 \left( \delta_0 - \mu \right)}\right) n_0 + \frac{g}{4} n_0^2 \notag \\
    &+ \frac{m_{\rm X}}{8 \pi \hbar^2} \sum_{s=\pm} \left[ B_{0} \left( B_{0} - \sqrt{B_{0}^2 - C_s^2} \right) + C_s^2 \ln\left( \frac{B_{0} + \sqrt{B_{0}^2 - C_s^2}}{2 \sqrt{e} \, |\epsilon_{\rm LP}|} \right) \right].
\end{align}

\section{Green's function}
The previously derived Bogoliubov theory is equivalent to a path-integral formalism at the gaussian level. Here, we briefly present this analogy and we compute the Green's function of the system which enters the photon spectral function in the main text. The polariton spin mixture can be described in real space by the Hamiltonian density
\begin{equation}
    \begin{aligned}
        \mathscr{H}(\mathbf{r}) = \sum_{\sigma} \bigg[&
        C_{\sigma}^\dagger(\mathbf{r}) \left( - \frac{\hbar^2 \bm{\nabla}^2}{2m_\text{C}} + \delta_0 - \mu_{\sigma} \right) C_{\sigma}(\mathbf{r}) + X_{\sigma}^\dagger(\mathbf{r}) \left( - \frac{\hbar^2 \bm{\nabla}^2}{2m_\text{X}} - \mu_{\sigma} \right) X_{\sigma}(\mathbf{r})  \\
        &+ \frac{\Omega_{\rm R}}{2} \left( C_{\sigma}^\dagger(\mathbf{r}) X_{\sigma}(\mathbf{r}) + X_{\sigma}^\dagger(\mathbf{r}) C_{\sigma}(\mathbf{r}) \right) 
        \bigg] + \sum_{\sigma, \sigma'} \frac{v_{\sigma \sigma'}}{2} X_{\sigma}^\dagger(\mathbf{r}) X_{\sigma'}^\dagger(\mathbf{r}) X_{\sigma'}(\mathbf{r}) X_{\sigma}(\mathbf{r}),
    \end{aligned}
\end{equation}
where $C_{\sigma}(\mathbf{r})$/$X_{\sigma}(\mathbf{r})$ are the annihiliation field operators of a photon/exciton of species $\sigma=\uparrow, \downarrow$.
Within a general formalism at finite temperature ($\beta = 1/k_{\rm B} T$), the action of the system is given by:
\begin{equation}
    \begin{aligned}
        \mathcal{S} =& \int d\mathbf{r} \int_0^{\beta \hbar} d\tau \, \left\{ \sum_{\sigma } \left[  C_{\sigma}^\dagger(\mathbf{r}, \tau) \hbar \partial_{\tau} C_{\sigma}(\mathbf{r},\tau) +  X_{\sigma}^\dagger(\mathbf{r}, \tau) \hbar \partial_{\tau} X_{\sigma}(\mathbf{r},\tau) \right] + \mathscr{H}(\mathbf{r},\tau) \right\}.
    \end{aligned}
\end{equation}
We perform a Hubbard-Stratonovich transformation to decouple the interspecies interactions by explicitly introducing a bosonic pairing field at the saddle-point level, i.e. $\Phi(\mathbf{r}) \equiv \Phi$. After doing so, the action becomes:
\begin{equation}
    \begin{aligned}
        \mathcal{S} =& \int d\mathbf{r} \int_0^{\beta \hbar} d\tau \, \left\{ 
        \sum_{\sigma} \frac{\Omega_{\rm R}}{2} \left( C_{\sigma}^\dagger(\mathbf{r},\tau) X_{\sigma}(\mathbf{r},\tau) + X_{\sigma}^\dagger(\mathbf{r},\tau) C_{\sigma}(\mathbf{r},\tau) \right)
        \right. \\
        & + \sum_{\sigma} \left[ C_{\sigma}^\dagger(\mathbf{r}, \tau) \left( \hbar \partial_{\tau} - \frac{\hbar^2 \bm{\nabla}^2}{2m_\text{C}} + \delta_0 - \mu_{\sigma} \right) C_{\sigma}(\mathbf{r},\tau) + X_{\sigma}^\dagger(\mathbf{r}, \tau) \left( \hbar \partial_{\tau} - \frac{\hbar^2 \bm{\nabla}^2}{2m_\text{X}} - \mu_{\sigma} \right) X_{\sigma}(\mathbf{r},\tau) \right]   \\
        & \left. - \frac{\Phi^2}{v_{\uparrow \downarrow}} - \Phi \left[ X_\uparrow(\mathbf{r},\tau) X_\downarrow(\mathbf{r},\tau) + X_\downarrow^\dagger(\mathbf{r},\tau) X_\uparrow^\dagger(\mathbf{r},\tau) \right] + \sum_{\sigma} \frac{v}{2} X_{\sigma}^\dagger(\mathbf{r},\tau) X_{\sigma}^\dagger(\mathbf{r},\tau) X_{\sigma}(\mathbf{r},\tau) X_{\sigma}(\mathbf{r},\tau)
        \right\}.
    \end{aligned}
\end{equation}
Finally, we perform the usual Bogoliubov approximation for the photon and exciton operators 
\begin{align}
    C_{\sigma}(\mathbf{r}) \, &\mapsto \,  \mathcal{A}^{-1/2} \lambda_{\sigma} + C_{\sigma}(\mathbf{r}), &
    X_{\sigma}(\mathbf{r}) \, &\mapsto \,  \mathcal{A}^{-1/2} \alpha_{\sigma} + X_{\sigma}(\mathbf{r}),
    \label{eq:flucts}
\end{align}
which allows to decompose the total action into  
$
 \mathcal{S} = \mathcal{S}_{\rm MF} + \mathcal{S}_{\rm B}.
$
The mean-field term 
$
\mathcal{S}_{\rm MF} = \hbar \beta \mathcal{A} \, \Omega_{\rm MF}
$
is proportional to the the thermodynamic potential given in Eq.~\eqref{eq:OmMFsym} for the case of a symmetric mixture. Using the short-hand notation $(\mathbf{r}, \tau) \equiv r$, $\int d\mathbf{r} \int_0^{\beta \hbar} d\tau \equiv \int dr$ and introducing the Nambu spinor
\begin{equation}
    \Psi(r) = 
    \begin{pmatrix}
    C_\uparrow(r) & C_\downarrow(r) & C_\uparrow^\dagger(r) & C_\downarrow^\dagger(r) &
    X_\uparrow(r) & X_\downarrow(r) & X_\uparrow^\dagger(r) & X_\downarrow^\dagger(r)  
    \end{pmatrix}
    ^T,
\end{equation}
the contribution to the action at the gaussian level takes the compact form 
\begin{equation}
    \mathcal{S}_{\rm B} = \frac{1}{2} \int dr \int dr' \Psi^\dagger(r) \left[ - \mathcal{G}^{-1}(r, r') \right] \Psi(r').
\end{equation}
In order to analyze the gaussian kernel, it is convenient to perform a Fourier transform and work in momentum space $(i \omega_m, \mathbf{k})$, where $\omega_m=2\pi m/\beta$ ($m \in \mathbb{Z}$) is a bosonic Matsubara frequency. The results at zero temperature can then be obtained through the analytic continuation $-\hbar \partial_{\tau} \rightarrow i \omega_m \rightarrow \omega + i 0^+$. The (inverse) Green's function of the system is given by
\begin{equation}
\label{eq:gf_full}
    \mathcal{G}^{-1}(\k,\omega) = 
    \begin{psmallmatrix}
     \omega - A_{\k} & 0 & 0 & 0 & -\Omega_{\rm R}/2 & 0 & 0 & 0 \\
     0 & \omega - A_{\k} & 0 & 0 & 0 & -\Omega_{\rm R}/2 & 0 & 0 \\
     0 & 0 & -\omega - A_{\k} & 0 & 0 & 0 & -\Omega_{\rm R}/2 & 0 \\
     0 & 0 & 0 & -\omega - A_{\k} & 0 & 0 & 0 & -\Omega_{\rm R}/2 \\
     -\Omega_{\rm R}/2 & 0 & 0 & 0 & \omega - B_{\k} & 0 & -g n_0 & \Phi \\
     0 & -\Omega_{\rm R}/2 & 0 & 0 & 0 &  \omega - B_{\k} & \Phi & -g n_0 \\
     0 & 0 & -\Omega_{\rm R}/2 & 0 & -g n_0 & \Phi &  - \omega - B_{\k} & 0 \\
     0 & 0 & 0 & -\Omega_{\rm R}/2 & \Phi & - g n_0 &  0 & - \omega - B_{\k}
    \end{psmallmatrix},
\end{equation}
where we have restricted to the symmetric mixtures of lower polaritons that we analyze in the main text. Notice that the poles of the Green's function, given by the condition $\text{det}\left[ \mathcal{G}^{-1}(\k,\omega \rightarrow E_{\k}) \right]=0$, coincide with the eigenvalues of the Bogoliubov matrix $\tilde{\mathcal{H}}(\k)$ defined in Eq.~\eqref{eq:HTilde_sup}.

The first element of the matrix obtained from inverting Eq.~\eqref{eq:gf_full}, 
\begin{equation}
    \label{eq:gf_phot}
    G_\text{C}(\k,\omega) \equiv \left[ \mathcal{G}(\k,\omega) \right]_{11},
\end{equation}
is the photon Green's function which enters the spectral function $A(\k, \omega)$ discussed in the main text. 

\section{Droplet phase and cavity detuning}
Here we briefly present the dependence of the quantum droplet phase on the cavity detuning, which is the experimental knob to tune the interactions and then drive the system into the MF collapsing state. 

\begin{figure}[h]
    \centering
    \begin{minipage}{0.99\textwidth}
    \centering
    \includegraphics[width=\textwidth]{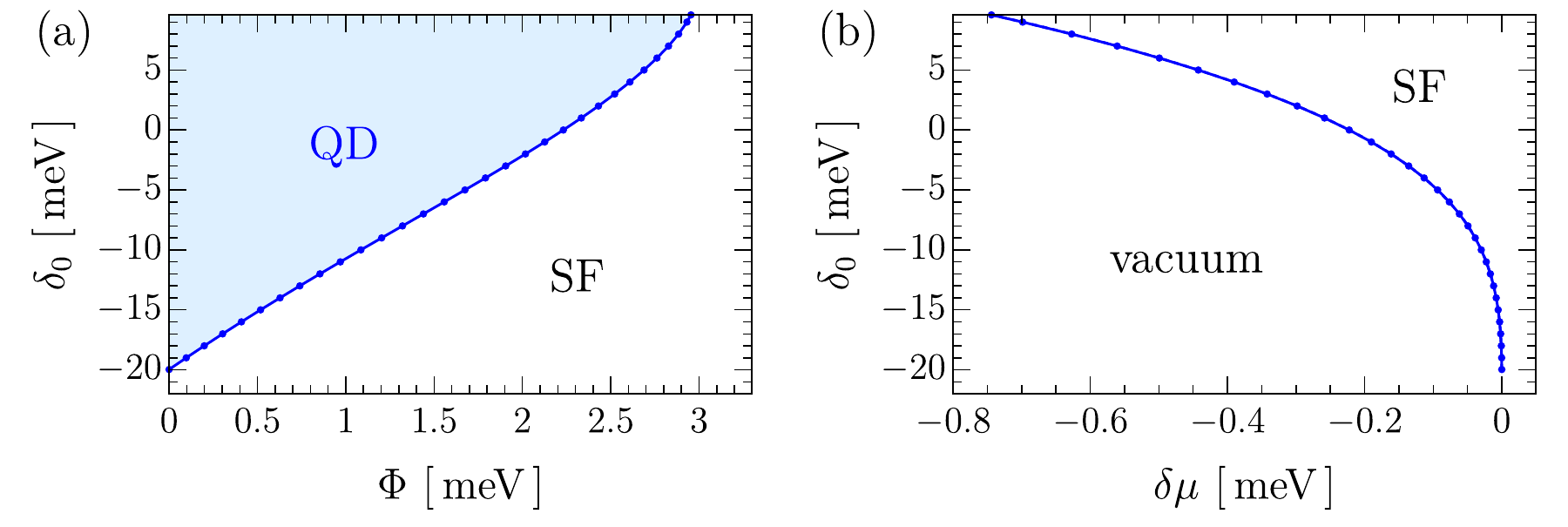}
    \end{minipage}
    \caption{(a) Phase diagram of cavity detuning \emph{vs.} pairing field. The blue solid line represents the droplet pairing fields and denotes a first order phase transition between the quantum droplet (QD) phase and a miscible superfluid (SF) phase. (b) Detuning \emph{vs.} chemical potential $\delta \mu = \mu - \epsilon_{\rm LP}$. The chemical potentials in the droplet phase ($\delta \mu_{\rm D}$) form the blue solid line which separates the vacuum from the SF phase.}
    \label{fig:phase_diags}
\end{figure}

Figure~\ref{fig:phase_diags}(a) is similar to the plot of the exciton density presented in the main text and the interpretation is the same, due to the similarity in the shape of $\Omega$ as a function of either $n_0$ or $\Phi$. Differently from $n_0$, the pairing field is a monotonic function of detuning.
Figure~\ref{fig:phase_diags}(b) shows the chemical potential in the droplet phase. First of all, it is always negative (in analogy with the $\mu<0$ condition for droplets in ultracold atomic gases) and vanishes for the detuning at which $\delta g = 0$, which denotes the onset of the miscible regime. Then we recall that the droplet chemical potential marks a first order phase transition due to the coexistence between the vacuum and the droplet a finite density. When $\delta \mu > \delta \mu_{\rm D}$ (on the right of the blue curve), the system displays an absolute minimum at negative thermodynamic potential (i.e., positive pressure) which is the SF miscible regime. For $\delta \mu < \delta \mu_{\rm D}$ (on the left of the blue curve), instead, the vacuum is the only stable phase predicted by $\Omega$ [see Fig. 2(e-f)]. We show the results for detunings up to the Feshbach resonance ($\delta_0 = 9.6$ meV).

\begin{figure}[h]
    \centering
    \begin{minipage}{0.99\textwidth}
    \centering
    \includegraphics[width=\textwidth]{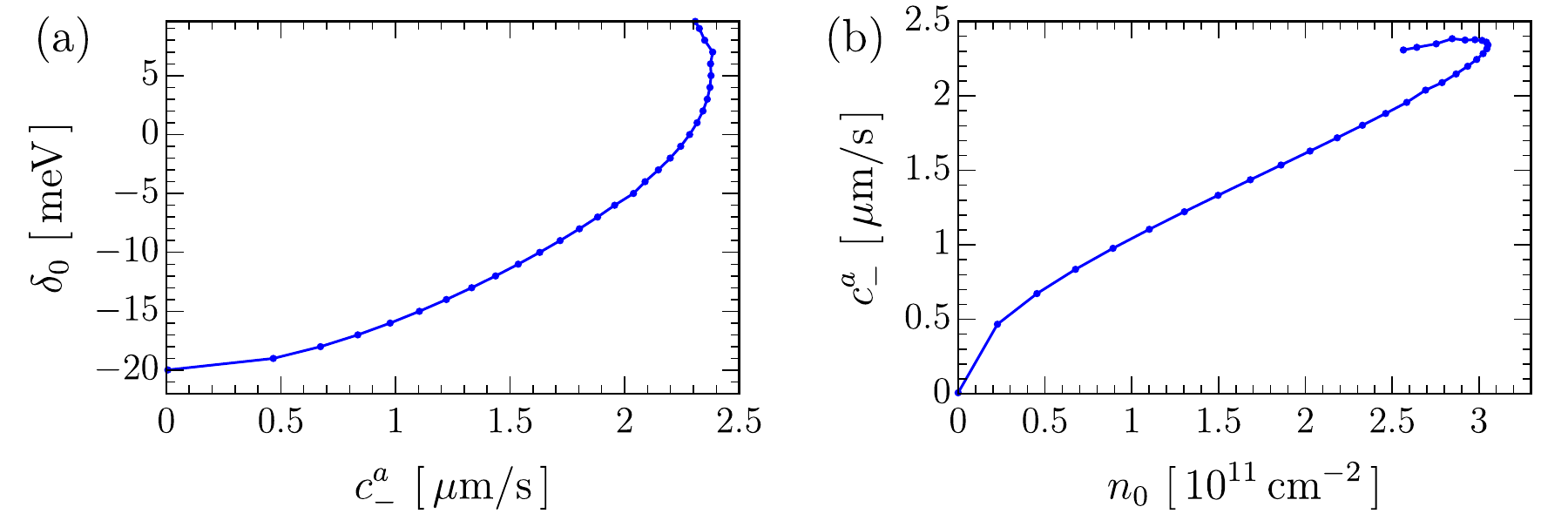}
    \end{minipage}
    \caption{Speed of sound extracted from a low-momentum expansion of the gapless branch $E_{-,k}^a$. We show this in (a) as a function of cavity detuning and in (b) as a function of the droplet saturation density.}
    \label{fig:speed_of_sound}
\end{figure}

Motivated by the experimental relevance of the Bogoliubov spectrum, we further investigate the speed of sound of the gapless mode. In fact, since the speed of sound depends on the density of the system, it will stay constant for particle numbers (per area) below the saturation density, while it will increase in the miscible superfluid phase. In particular, for any given value of cavity detuning we estimate the speed of sound from a low-momentum expansion of the lowest Bogoliubov branch, $E^{a}_{k,-} \sim c^{a}_{-} k$. The resulting Fig.~\ref{fig:speed_of_sound}(a) shows a similar qualitative behaviour as for the droplet density shown in Fig.~3(a) in the main text, in particular for the presence of a turning point at positive detuning (the irregular profile of $c^{a}_{-}(\delta_0)$ is due to small numerical errors). The curve of the speed of sound as a function of the saturation density in Fig.~\ref{fig:speed_of_sound}(b) hints a $\sim\sqrt{n_0}$ dependence in the dilute regime, as it is well--known for Bose-Bose mixtures~\cite{Petrov2015}. 

\section{Gross-Pitaevskii analysis}
In this final section we derive an effective Gross-Pitaevskii equation which describes the spatial profile of a quantum droplet in a spin mixture of exciton-polaritons. To this purpose, it is convenient to introduce the bosonic field operators $\hat{\psi}_{\rm C/X \sigma}(\mathbf{r})$ for a photon/exciton with spin $\sigma$ at position $\mathbf{r}$, and consider the system Hamiltonian in real space:
\begin{align}
    \label{eq:Hamiltonian_GP}
    \hat{H} =& \sum_{\sigma} \int d^2 r  
    \begin{pmatrix}
        \hat{\psi}_{\rm{C} \sigma}^\dagger(\mathbf{r}) & \hat{\psi}_{\rm{X} \sigma}^\dagger(\mathbf{r}) 
    \end{pmatrix}
    \begin{pmatrix}
        -\frac{\hbar^2 \nabla^2}{2m_{\rm C}} + \delta_0-\mu_{\sigma} & \Omega_{\rm R}/2 \\
        \Omega_{\rm R}/2 &  -\frac{\hbar^2 \nabla^2}{2m_{\rm X}}-\mu_{\sigma}
    \end{pmatrix}
    \begin{pmatrix}
        \hat{\psi}_{\rm{C} \sigma}(\mathbf{r}) \\
        \hat{\psi}_{\rm{X} \sigma}(\mathbf{r})
    \end{pmatrix} \notag \\
    &+ \int d^2 r \left[ \frac{g}{2} \sum_{\sigma} \hat{\psi}_{\rm{X} \sigma}^\dagger(\mathbf{r}) \hat{\psi}_{\rm{X} \sigma}^\dagger(\mathbf{r}) \hat{\psi}_{\rm{X} \sigma}(\mathbf{r}) \hat{\psi}_{\rm{X} \sigma}(\mathbf{r}) - \frac{\left| \Phi \right|^2}{g_{\uparrow \downarrow}} - \Phi \hat{\psi}_{\rm X \uparrow}^\dagger(\mathbf{r}) \hat{\psi}_{\rm X \downarrow}^\dagger(\mathbf{r}) - \Phi^* \hat{\psi}_{\rm X \downarrow}(\mathbf{r}) \hat{\psi}_{\rm X \uparrow}(\mathbf{r}) \right].  
\end{align}
Notice that we have already decoupled the attractive channel of the interaction through the pairing field at the saddle-point level $\Phi = -g_{\uparrow \downarrow} \langle \psi_{\rm X \downarrow}(\mathbf{r}) \psi_{\rm X \uparrow}(\mathbf{r}) \rangle$.
We now perform a semi-classical approximation, where we replace the field operators by complex functions of spatial coordinates, which yields the total free energy of the system
\begin{align}
    \label{eq:total_energy_GP}
    F =& \int d^2 r \Bigg\{\sum_{\sigma} \left[ \frac{\hbar^2 \left| \nabla \psi_{\rm X \sigma}(\mathbf{r}) \right|^2}{2m_{\rm X}} + \frac{\hbar^2 \left| \nabla \psi_{\rm C \sigma}(\mathbf{r}) \right|^2}{2m_{\rm C}} + \delta_0 n_{\rm C \sigma}(\mathbf{r}) + \frac{\Omega_{\rm R}}{2} \left( \psi_{\rm C \sigma}^*(\mathbf{r}) \psi_{\rm X \sigma}(\mathbf{r}) + \psi_{\rm X \sigma}^*(\mathbf{r}) \psi_{\rm C \sigma}(\mathbf{r})\right) \right] \notag  \\
    & \qquad \quad \, +\frac{g}{2} \sum_{\sigma} n^2_{\rm X \sigma}(\mathbf{r}) - \frac{\left| \Phi \right|^2}{g_{\uparrow \downarrow}} - \Phi \psi_{\rm X \uparrow}^*(\mathbf{r}) \psi_{\rm X \downarrow}^*(\mathbf{r}) - \Phi^* \psi_{\rm X \downarrow}(\mathbf{r}) \psi_{\rm X \uparrow}(\mathbf{r}) -  \sum_{\sigma} \mu_{\sigma} \left( n_{\rm X \sigma}(\mathbf{r}) + n_{\rm C \sigma}(\mathbf{r}) \right) \Bigg\} \notag \\
    =& \int d^2 r \, \mathcal{F}[n_{\rm X \sigma}(\mathbf{r}),n_{\rm C \sigma}(\mathbf{r})],
\end{align}
where the free energy density $\mathcal{F}$ is a functional of the local number densities of photons and excitons, $n_{\rm C/X \sigma}(\mathbf{r}) = \left| \psi_{\rm C/X \sigma}(\mathbf{r}) \right|^2$.

The ground state of the system is obtained through a functional minimization of the total energy, subjected to the constraint of keeping fixed the total number of particles in the system. Given the free energy in Eq.~\eqref{eq:total_energy_GP}, the Euler-Lagrange condition $\delta F = 0$ leads to the following coupled stationary Gross-Pitaevskii (GP) equations:
\begin{equation}
    \label{eq:GP_general_pairing}
    \begin{aligned}
    \left( - \frac{\hbar^2 \nabla^2}{2m_{\rm C}} + \delta_0  \right) \psi_{\rm C \sigma}(\mathbf{r}) + \frac{\Omega_{\rm R}}{2} \psi_{\rm X \sigma}(\mathbf{r}) =& \mu_{\sigma} \psi_{\rm C \sigma}(\mathbf{r}), \\
    \left( - \frac{\hbar^2 \nabla^2}{2m_{\rm X}} + g n_{\rm X \sigma}(\mathbf{r}) \right) \psi_{\rm X \sigma}(\mathbf{r}) - \Phi(\mathbf{r}) \psi_{\rm X \bar{\sigma}}^*(\mathbf{r}) + \frac{\Omega_{\rm R}}{2} \psi_{\rm C \sigma}(\mathbf{r}) =& \mu_{\sigma} \psi_{\rm X \sigma}(\mathbf{r}).
    \end{aligned}
\end{equation}

\subsection{Symmetric mixture}
For the sake of simplicity, we now restrict the discussion to the experimentally relevant case of a symmetric mixture with $\psi_{{\rm C}/{\rm X} \sigma} = \psi_{{\rm C}/{\rm X}}/\sqrt{2}$.
Moreover, within a local-density approximation spirit, we consider the mean-field relation between the photon and exciton wave functions
\begin{equation}
    \label{eq:phexrel}
    \psi_{\rm C}(\mathbf{r})=-\frac{\Omega_R}{2(\delta_0-\mu)}\psi_{\rm X}(\mathbf{r}).
\end{equation}
Substituting back into the free energy~\eqref{eq:total_energy_GP}, we get
\begin{align}
    \label{eq:free_energy_contd}
    F =& \int d^2 r \Bigg\{ \frac{\hbar^2 \left| \nabla \psi_{\rm X}(\mathbf{r}) \right|^2}{2 M_*} 
     - \frac{\left| \Phi \right|^2}{g_{\uparrow \downarrow}} - \frac{1}{2}\left( \Phi \psi_{\rm X}^*(\mathbf{r})^2 + \Phi^* \psi_{\rm X}(\mathbf{r})^2 \right) - \left(\mu + \frac{\Omega_R^2}{4(\delta_0-\mu)} \right) n_{\rm X}(\mathbf{r})+ \frac{g}{4} n^2_{\rm X}(\mathbf{r}) \Bigg\}, 
\end{align}
where we have defined the effective mass
\begin{equation}
    \label{eq:effMass}
    M_* = \left( \frac{1}{m_{\rm X}} + \frac{\Omega_R^2}{4 m_{\rm C} (\delta_0-\mu)^2} \right)^{-1}.
\end{equation}
It is worth mentioning that, in the weakly-interacting regime one gets $M_* \simeq X_0^2 M_{\rm LP}$, where $M_{\rm LP}^{-1}=C_0^2/m_{\rm C}+X_0^2/m_{\rm X}$ is the lower-polariton mass obtained from a small-k expansion of the dispersion $E_{\rm LP}(\k)$. 

\medskip
We now focus on the specific case of quantum droplets, where we can get rid of the phase degree of freedom and consider both $\psi_{\rm X}, \Phi \in \mathbb{R}$. We recall that the thermodynamic potential depends on the chemical potential only through the MF contribution, as the LHY one is evaluated at $\mu_{\rm MF}$ which is a function of the exciton density and the pairing field. Going back to the usual notation $n_{\rm X} \equiv n_0 = \alpha^2/\mathcal{A}$ and $\psi_{\rm X} \equiv \psi_0$, one can thus write
\begin{equation}
    \label{eq:OmMu}
    \Omega(\alpha,\Phi,\mu) = \mathcal{E}(\alpha,\Phi) - \left( \mu + \frac{\Omega_{\rm R}^2}{4 (\delta_0 - \mu)} \right) n_0,
\end{equation}
where the energy density $\mathcal{E}(\alpha,\Phi) = \mathcal{E}_{\rm MF}(\alpha,\Phi) + \mathcal{E}_{\rm LHY}(\alpha,\Phi)$ consists of the two parts
\begin{align}
    \mathcal{E}_{\rm MF}(\alpha,\Phi) &= - \frac{\Phi^2}{g_{\uparrow \downarrow}} - \Phi n_0 + \frac{g}{4} n_0^2, & \mathcal{E}_{\rm LHY}(\alpha,\Phi) &= \Omega_{\rm LHY}(\alpha,\Phi).
\end{align}
Notice that the light-matter coupling is responsible for an effective renormalization of the exciton chemical potential, whose meaning becomes clear upon moving into the LP basis, as we show at the end of this section. 

\begin{figure}[h]
    \centering
    \begin{minipage}{0.98\textwidth}
    \centering
    \includegraphics[width=\textwidth]{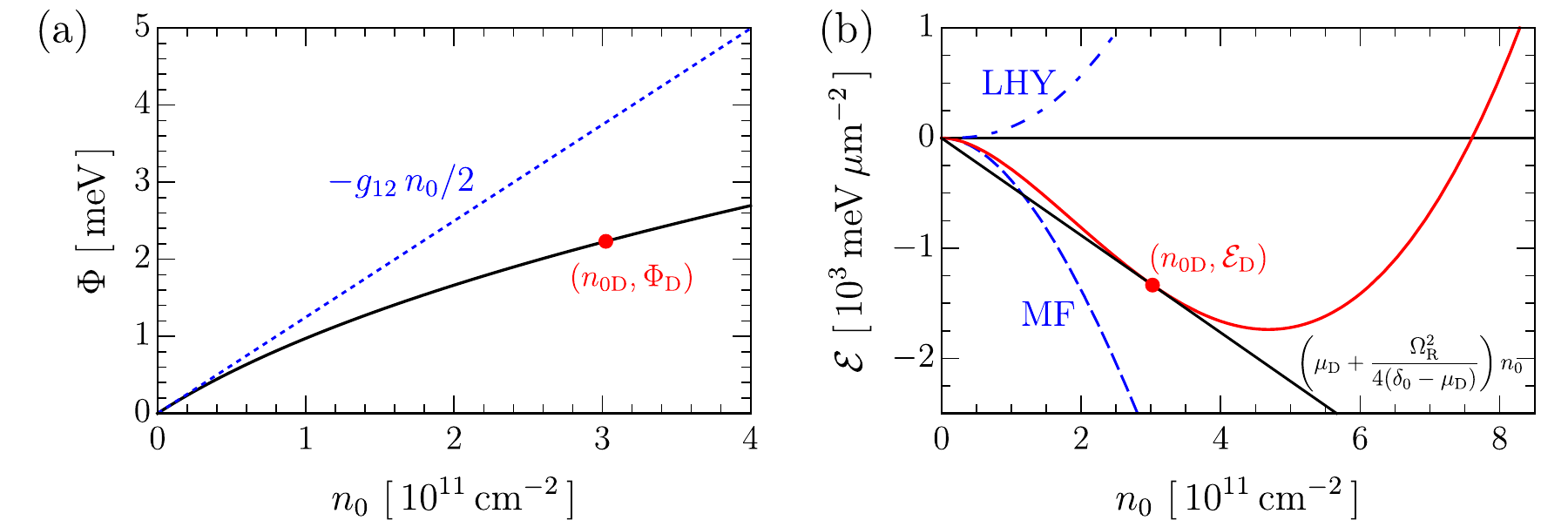}
    \end{minipage}
    \caption{(a) Pairing field as a function of density obtained from the stationary condition $\partial_{\Phi} \Omega=0$ (black line). The blue dashed line denotes the MF result. (b) Total energy density (red solid line) resulting from the competition between an attractive MF contribution (blue dashed) and a repulsive LHY one (blue dot-dashed). The tangent at the droplet phase (red dot) is given by the overall chemical potential, as predicted by the Maxwell construction. We set the cavity detuning to $\delta_0=0$ meV.}
    \label{fig:energy_densities}
\end{figure}

We now provide a procedure to derive an effective stationary Gross-Pitaevskii equation for the exciton wave function which has to be solved numerically to determine the spatial droplet profile:
\begin{itemize}
    \item[(a)] since the droplet phase is a stationary point of the thermodynamic potential, the pairing field can be related to the density through the condition $\partial_{\Phi} \Omega(\alpha,\Phi,\mu)=0$ which does not depend on the chemical potential [see Eq.~\eqref{eq:OmMu}], but only on the cavity detuning $\delta_0$.  
    The numerical solution of this implicit equation yields the function $\Phi(n_0)$ represented by the black solid curve in Fig.~\ref{fig:energy_densities}(a). The blue dashed line is the MF result obtained from $\partial_{\Phi} \Omega_{\rm MF}=0$ and deviations become larger at high densities due to the stronger effect of quantum fluctuations. Using the relation $\Phi(n_0)$ one gets the energy densities $\mathcal{E}_{\text{MF}/\text{LHY}}(n_0) \equiv \mathcal{E}_{\text{MF}/\text{LHY}}(\alpha=\sqrt{\mathcal{A} n_0},\Phi(n_0))$, which are respectively negative and positive monotonic functions of the the exciton density when $\delta g < 0$, as shown in Fig.~\ref{fig:energy_densities}(b). The droplet phase $(n_{0\rm D}, \Phi_{\rm D}, \mu_{\rm D})$ is a minimum of the energy per particle. In terms of the total energy density $\mathcal{E}(n_0)$, it corresponds to the point where the tangent is equal to $\mu_{\rm D} + \Omega_{\rm R}^2/[4(\delta_0 - \mu_{\rm D})]$, as given by the Maxwell construction;
    \item[(b)] considering a spatial dependence of the density, $n_0(\mathbf{r}) = \psi_0^2(\mathbf{r})$, we introduce the functional
    \begin{align}
        \label{eq:functionals} 
        \Omega[n_0(\mathbf{r});\mu] &= \mathcal{E}[n_0(\mathbf{r})] - \left( \mu + \frac{\Omega_{\rm R}^2}{4 (\delta_0 - \mu)} \right) n_0(\mathbf{r}),
    \end{align}
    where the notation for $\Omega$ means that it depends only parametrically on $\mu$, while $\mathcal{E}[n_0(\mathbf{r})]$ is the energy density functional. This allows us to incorporate beyond-MF effects into the free-energy~\eqref{eq:free_energy_contd}, which now takes the compact form:
    \begin{align}
        \label{eq:free_energy_new}
        F =& \int d^2 r \Bigg\{ \frac{\hbar^2 \left| \nabla \psi_0(\mathbf{r}) \right|^2}{2 M_*} + \mathcal{E}[n_0(\mathbf{r})] - \left( \mu + \frac{\Omega_{\rm R}^2}{4 (\delta_0 - \mu)} \right) n_0(\mathbf{r})
        \Bigg\}; 
    \end{align}
    \item[(c)] the Euler-Lagrange equation obtained from the condition $\delta F = 0$ gives the following effective stationary Gross-Pitaevskii equations for the exciton wave function:
    \begin{equation}
        \label{eq:effGPEX}
        \left( - \frac{\hbar^2 \nabla^2}{2 M_*} + \frac{\partial \mathcal{E}}{\partial n_0}[n_0(\mathbf{r})] \right) \psi_0(\mathbf{r}) = \left( \mu + \frac{\Omega_{\rm R}^2}{4 (\delta_0 - \mu)} \right)\psi_0(\mathbf{r}).
    \end{equation}
    First of all, we notice that the effects of light are present in both sides of Eq.~\eqref{eq:effGPEX} through the effective mass in the kinetic energy and the modified chemical potential. The droplet shape results from the interplay between the kinetic term, which is stronger on the surface, and the effective GP-single particle potential given by the density derivative of the energy density;
    \item[(d)] finally, we show how Eq.~\eqref{eq:effGPEX} can be easily recast into a GPE for lower polaritons rather than excitons. Using the parametrization of the chemical potential $\mu = \epsilon_{\rm LP}+\delta \mu$, justified by the smallness of $\delta \mu$ in the droplet phase, we perform the following expansion
    \begin{equation}
        \label{eq:expmu}
        \mu + \frac{\Omega_{\rm R}^2}{4 (\delta_0 - \mu)} \simeq \left[ 1 + \left( \frac{\Omega_{\rm R}}{2 (\delta_0 - \epsilon_{\rm LP})} \right)^2 \right] \delta \mu = \frac{\delta \mu}{X_0^2}.
    \end{equation}
    After that, according to Eq.~\eqref{eq:simpleRels} and the relative discussion, the wave function of the LP condensate is related to the one of excitons through $\psi_{\rm LP} = \psi_0/X_0$. Defining the LP density $n_{\rm LP} = \psi_{\rm LP}^2$ and recalling $M_* \simeq X_0^2 M_{\rm LP}$, Eq.~\eqref{eq:effGPEX} takes the form 
    \begin{equation}
        \label{eq:effGPELP}
        \left( - \frac{\hbar^2 \nabla^2}{2 M_{\rm LP}} +  \frac{\partial \mathcal{E}}{\partial n_{\rm LP}}[n_0(\mathbf{r})=X_0^2 n_{\rm LP}(\mathbf{r})] \right) \psi_{\rm LP}(\mathbf{r}) = \delta \mu \, \psi_{\rm LP}(\mathbf{r}).
    \end{equation}
    To give a concrete example for the form of the potential energy term, if we consider only intraspecies contact interaction such that $\mathcal{E}(n_0)=gn_0^2/4$, then we recover the expected result~\cite{Deng2010}
    \begin{align}
        \frac{\partial \mathcal{E}}{\partial n_{\rm LP}}[n_0(\mathbf{r})=X_0^2 n_{\rm LP}(\mathbf{r})] &= \frac{\partial}{\partial n_{\rm LP}} \left( \frac{g}{4} X_0^4 n_{\rm LP}^2 \right) = \frac{g_{\rm LP}}{2} n_{\rm LP}, & \text{with} \quad g_{\rm LP} = X_0^4 g;
    \end{align}
    \item[(e)] the droplet profile can be obtained from a numerical solution of either of the two Gross-Pitaevskii equations~\eqref{eq:effGPEX}, \eqref{eq:effGPELP}, where the chemical potentials on the RHS have to be determined self-consistently together with the normalization of the wave function to the total number of excitons/polaritons in the system;
    \item[(f)] Bose-Einstein condensates of polaritons are driven-dissipative systems, usually described by a complex Gross-Pitaevskii equation (cGPE). We consider a nonresonant pumping scheme as a possible way to include out-of-equilibrium effects in our theory. Following Ref.~\cite{Keeling2008}, Eq.~\eqref{eq:effGPELP} has to be augmented by a density dependent rate of gain $i (\gamma_{\rm eff} - \Gamma |\psi_\text{LP}|^2) \psi_\text{LP}$, where the effective pump rate $\gamma_{\rm eff}$ quantifies the interplay between stimulated scattering and particle decay.
    For large pumping rates, we expect the droplet saturation density to increase and at the same time its profile to be modified by the presence of supercurrents. 
\end{itemize}

\subsection{Spatial density profile in the slab-geometry configuration}
We finally present a simplified model which allows to derive an approximate yet reliable droplet profile. This model is based on two major assumptions, the first one being:
\begin{itemize}
    \item[(1)] we do not account for the finite number of particles in the system, but we fix the chemical potential to the value in the droplet phase, $\mu_{\rm D} = \epsilon_{\rm LP} + \delta \mu_{\rm D}$. It is well-known from studies in bosonic mixtures~\cite{Cabrera2018,Semeghini2018} that the atom number $N$ strongly influences not only the droplet shape, but also its stability. In fact, large values of $N$ give rise to \textquotedblleft bulk{\textquotedblright} droplets with a flat-top profile, while \textquotedblleft all-surface{\textquotedblright} droplets appear for lower $N$. There is also a critical threshold $N_c$ below which the kinetic energy dominates over the interaction energy causing the self-evaporation of the droplet. Fixing the chemical potential corresponds to considering a large number of atoms, such that droplets have a large bulk region with saturation density $n_{0 \rm D}$ and a thin surface. This is a non-restrictive assumption in light of our goal to estimate the spatial extension of quantum droplets in 2D semiconductors, i.e., it will provide an upper bound for the droplet size. 
\end{itemize}
We set $\mu= \mu_{\rm D}$ and write $\psi_0(\mathbf{r}) = \sqrt{n_0(\mathbf{r})}$ inside the free energy Eq.~\eqref{eq:free_energy_new}, which becomes
\begin{equation}
    F = \int d^2 r \mathcal{F}[ n_0(\mathbf{r})] = \int d^2 r \left\{ \frac{\hbar^2}{8 M_*} \frac{\left( \nabla n_0(\mathbf{r}) \right)^2}{n_0(\mathbf{r})} + \Omega\left[ n_0(\mathbf{r}); \mu_{\rm D} \right] \right\},
    \label{eq:freeEnDen}
\end{equation}
where we refer to Eq.~\eqref{eq:functionals} for the definition of the functional $\Omega[\dots]$. The ground state density must satisfy the condition $\delta F = 0$, therefore it is a solution of the Euler-Lagrange equations
\begin{equation}
    \frac{\partial \mathcal{F}}{\partial n_0} - \nabla \cdot \frac{\partial \mathcal{F}}{\partial (\nabla n_0)} = 0.
\end{equation}
Computing the derivatives of Eq.~\eqref{eq:freeEnDen}, one arrives at the following differential equation:
\begin{equation}
    \frac{\hbar^2}{8 M_*} \left[ - 2 \frac{\nabla^2 n_0(\mathbf{r})}{n_0(\mathbf{r})} + \left( \frac{\nabla n_0(\mathbf{r})}{n_0(\mathbf{r})} \right)^2 \right] + \frac{\partial \Omega}{\partial n_0}\left[ n_0(\mathbf{r}); \mu_{\rm D} \right] = 0.
    \label{eq:diffEqone}
\end{equation}
\begin{figure}[h]
    \centering
    \begin{minipage}{0.98\textwidth}
    \centering
    \includegraphics[width=\textwidth]{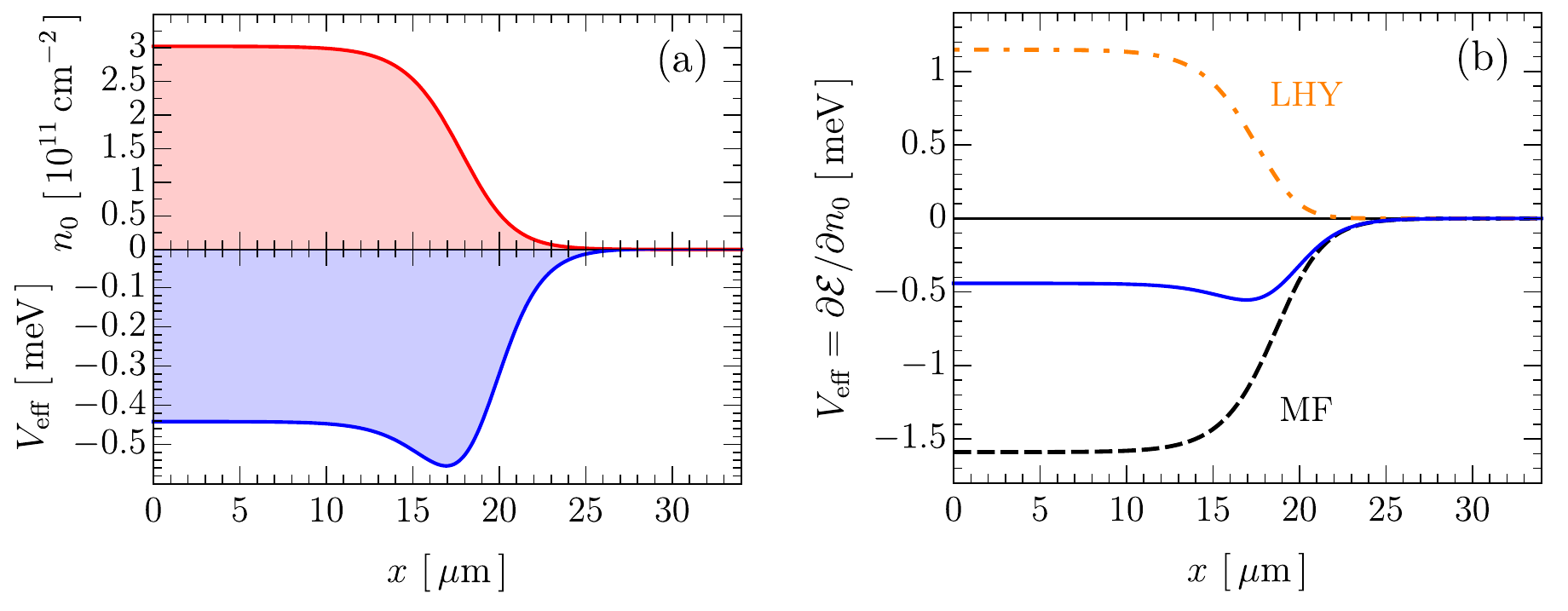}
    \end{minipage}
    \caption{(a) One-dimensional spatial droplet profile (red solid curve) in the slab-geometry configuration and the corresponding trapping potential (blue solid) as defined in the GP-like Eq.~\eqref{eq:effGPEX}. (b) This effective potential well (blue solid curve) results from the subtle interplay between the MF attraction (black dashed) and LHY repulsion (orange dot-dashed). The cavity detuning is $\delta_0 = 0$ meV.}
    \label{fig:eff_potentials}
\end{figure}

Here comes the second assumption:
\begin{enumerate}
    \item[(2)] we consider the 2D analog of a slab-geometry configuration, where we assume that the system is homogeneous along the $y$ direction, while along the $x$ direction there are two liquid-vacuum interfaces which enclose a bulk region with constant saturation density. Despite neglecting curvature effects, which may be relevant for spherical droplets, this approximation has been extensively used to characterize the spatial profile of 3D droplets in liquid He~\cite{Stringari1987}  and, more recently, in binary Bose mixtures~\cite{Cikojevic2021}. In the latter case, it has shown a remarkable agreement with the numerical solution of GPEs for droplets with a large number of atoms. 
\end{enumerate}
We are now left with a one-dimensional problem, as the exciton density only depends on the $x$ coordinate and Eq.~\eqref{eq:diffEqone} reduces to a second order ordinary differential equation (ODE). Moreover, the dependence of $\Omega$ on the spatial coordinates only through $n_0(x)$ further simplifies the treatment of this ODE. In fact, after multiplying both sides of Eq.~\eqref{eq:diffEqone} by $n_0'(x) = dn_0(x)/dx$, one obtains the first order equation
\begin{equation}
    - \frac{\hbar^2}{8 M_*} \frac{\left( n_0'(x) \right)^2}{n_0(x)} + \Omega(n_0(x)) = \text{const.} = 0,
\end{equation}
where we set the integration constant equal to zero because the equation must hold for $n_0(x)$ which approaches the vacuum region outside the droplet, where both the density and its gradient vanish [we recall that $\Omega(n_0=0)=0$]. Solving with respect to $n_0'(x)$ we get
\begin{equation}
    n_0'(x) = - \sqrt{ \frac{8 M_*}{\hbar^2} n_0(x) \Omega[n_0(x);\mu_{\rm D}]},
    \label{eq:diff_eq_profile}
\end{equation}
where we choose the minus sign because the density profile we are seeking is decreasing from the bulk to the vacuum. We use both Eqs.~(\ref{eq:diffEqone}, \ref{eq:diff_eq_profile}) to obtain the droplet density profile as follows:
\begin{enumerate}
    \item[i)] take $(x=0, n_{0}=n_D/2)$ as the starting point and use Eq.~\eqref{eq:diff_eq_profile} to determine $n_0'(0)$;
    \item[ii)] numerically solve the second order ODE~\eqref{eq:diffEqone} using the initial conditions from i) in the interval $[-x_*,+x_*]$, where $x_*>0$ is such that $n_0(-x_*)=n_{0 \rm D}$;
    \item[iii)] shift the numerical solution as $n_0(x-x_*) \mapsto n_0(x)$ to get a symmetric density profile. The result of this procedure is shown in Fig.~\ref{fig:eff_potentials}(a), as well as Fig.~3(c) in the main text. The flat-top shape (red curve) is determined by an effective potential well (blue curve) which in its turn results from the competition between the attractive MF contribution and the repulsive LHY one, see Eq.~\eqref{eq:effGPEX} and Fig.~\ref{fig:eff_potentials}(b).
\end{enumerate}


\begin{thebibliography}{107}%
\makeatletter
\providecommand \@ifxundefined [1]{%
 \@ifx{#1\undefined}
}%
\providecommand \@ifnum [1]{%
 \ifnum #1\expandafter \@firstoftwo
 \else \expandafter \@secondoftwo
 \fi
}%
\providecommand \@ifx [1]{%
 \ifx #1\expandafter \@firstoftwo
 \else \expandafter \@secondoftwo
 \fi
}%
\providecommand \natexlab [1]{#1}%
\providecommand \enquote  [1]{``#1''}%
\providecommand \bibnamefont  [1]{#1}%
\providecommand \bibfnamefont [1]{#1}%
\providecommand \citenamefont [1]{#1}%
\providecommand \href@noop [0]{\@secondoftwo}%
\providecommand \href [0]{\begingroup \@sanitize@url \@href}%
\providecommand \@href[1]{\@@startlink{#1}\@@href}%
\providecommand \@@href[1]{\endgroup#1\@@endlink}%
\providecommand \@sanitize@url [0]{\catcode `\\12\catcode `\$12\catcode `\&12\catcode `\#12\catcode `\^12\catcode `\_12\catcode `\%12\relax}%
\providecommand \@@startlink[1]{}%
\providecommand \@@endlink[0]{}%
\providecommand \url  [0]{\begingroup\@sanitize@url \@url }%
\providecommand \@url [1]{\endgroup\@href {#1}{\urlprefix }}%
\providecommand \urlprefix  [0]{URL }%
\providecommand \Eprint [0]{\href }%
\providecommand \doibase [0]{http://dx.doi.org/}%
\providecommand \selectlanguage [0]{\@gobble}%
\providecommand \bibinfo  [0]{\@secondoftwo}%
\providecommand \bibfield  [0]{\@secondoftwo}%
\providecommand \translation [1]{[#1]}%
\providecommand \BibitemOpen [0]{}%
\providecommand \bibitemStop [0]{}%
\providecommand \bibitemNoStop [0]{.\EOS\space}%
\providecommand \EOS [0]{\spacefactor3000\relax}%
\providecommand \BibitemShut  [1]{\csname bibitem#1\endcsname}%
\let\auto@bib@innerbib\@empty
\bibitem [{\citenamefont {Van~der Waals}()}]{VanDerWaals1873}%
  \BibitemOpen
  \bibfield  {author} {\bibinfo {author} {\bibfnamefont {J.~D.}\ \bibnamefont {Van~der Waals}},\ }\href@noop {} {\bibinfo {title} {\emph {{Over de Continuiteit van den Gas-en Vloeistoftoestand}}}},\ \bibinfo {note} {{PhD thesis, Leiden, The Netherlands (1873)}}\BibitemShut {NoStop}%
\bibitem [{\citenamefont {Barranco}\ \emph {et~al.}(2016)\citenamefont {Barranco}, \citenamefont {Guardiola}, \citenamefont {Hernández}, \citenamefont {Mayol}, \citenamefont {Navarro},\ and\ \citenamefont {Pi}}]{Barranco2006}%
  \BibitemOpen
  \bibfield  {author} {\bibinfo {author} {\bibfnamefont {M.}~\bibnamefont {Barranco}}, \bibinfo {author} {\bibfnamefont {R.}~\bibnamefont {Guardiola}}, \bibinfo {author} {\bibfnamefont {S.}~\bibnamefont {Hernández}}, \bibinfo {author} {\bibfnamefont {R.}~\bibnamefont {Mayol}}, \bibinfo {author} {\bibfnamefont {J.}~\bibnamefont {Navarro}}, \ and\ \bibinfo {author} {\bibfnamefont {M.}~\bibnamefont {Pi}},\ }\bibfield  {title} {\bibinfo {title} {\emph {{Helium Nanodroplets: An Overview}}},\ }\href {\doibase 10.1007/s10909-005-9267-0} {\bibfield  {journal} {\bibinfo  {journal} {Journal of Low Temperature Physics}\ }\textbf {\bibinfo {volume} {142}},\ \bibinfo {pages} {1} (\bibinfo {year} {2016})}\BibitemShut {NoStop}%
\bibitem [{\citenamefont {Leggett}(2008)}]{Leggett2008}%
  \BibitemOpen
  \bibfield  {author} {\bibinfo {author} {\bibfnamefont {A.~J.}\ \bibnamefont {Leggett}},\ }\bibfield  {title} {\bibinfo {title} {\emph {{Quantum Liquids}}},\ }\href {\doibase 10.1126/science.1152822} {\bibfield  {journal} {\bibinfo  {journal} {Science}\ }\textbf {\bibinfo {volume} {319}},\ \bibinfo {pages} {1203} (\bibinfo {year} {2008})}\BibitemShut {NoStop}%
\bibitem [{\citenamefont {Petrov}(2015)}]{Petrov2015}%
  \BibitemOpen
  \bibfield  {author} {\bibinfo {author} {\bibfnamefont {D.~S.}\ \bibnamefont {Petrov}},\ }\bibfield  {title} {\bibinfo {title} {\emph {{Quantum Mechanical Stabilization of a Collapsing Bose-Bose Mixture}}},\ }\href {\doibase 10.1103/PhysRevLett.115.155302} {\bibfield  {journal} {\bibinfo  {journal} {Phys. Rev. Lett.}\ }\textbf {\bibinfo {volume} {115}},\ \bibinfo {pages} {155302} (\bibinfo {year} {2015})}\BibitemShut {NoStop}%
\bibitem [{\citenamefont {Bulgac}(2002)}]{Bulgac2002}%
  \BibitemOpen
  \bibfield  {author} {\bibinfo {author} {\bibfnamefont {A.}~\bibnamefont {Bulgac}},\ }\bibfield  {title} {\bibinfo {title} {\emph {{Dilute Quantum Droplets}}},\ }\href {\doibase 10.1103/PhysRevLett.89.050402} {\bibfield  {journal} {\bibinfo  {journal} {Phys. Rev. Lett.}\ }\textbf {\bibinfo {volume} {89}},\ \bibinfo {pages} {050402} (\bibinfo {year} {2002})}\BibitemShut {NoStop}%
\bibitem [{\citenamefont {Lee}\ \emph {et~al.}(1957)\citenamefont {Lee}, \citenamefont {Huang},\ and\ \citenamefont {Yang}}]{Lee1957}%
  \BibitemOpen
  \bibfield  {author} {\bibinfo {author} {\bibfnamefont {T.~D.}\ \bibnamefont {Lee}}, \bibinfo {author} {\bibfnamefont {K.}~\bibnamefont {Huang}}, \ and\ \bibinfo {author} {\bibfnamefont {C.~N.}\ \bibnamefont {Yang}},\ }\bibfield  {title} {\bibinfo {title} {\emph {{Eigenvalues and Eigenfunctions of a Bose System of Hard Spheres and Its Low-Temperature Properties}}},\ }\href {\doibase 10.1103/PhysRev.106.1135} {\bibfield  {journal} {\bibinfo  {journal} {Phys. Rev.}\ }\textbf {\bibinfo {volume} {106}},\ \bibinfo {pages} {1135} (\bibinfo {year} {1957})}\BibitemShut {NoStop}%
\bibitem [{\citenamefont {Larsen}(1963)}]{Larsen1963}%
  \BibitemOpen
  \bibfield  {author} {\bibinfo {author} {\bibfnamefont {D.~M.}\ \bibnamefont {Larsen}},\ }\bibfield  {title} {\bibinfo {title} {\emph {{Binary mixtures of dilute bose gases with repulsive interactions at low temperature}}},\ }\href {\doibase https://doi.org/10.1016/0003-4916(63)90066-6} {\bibfield  {journal} {\bibinfo  {journal} {Annals of Physics}\ }\textbf {\bibinfo {volume} {24}},\ \bibinfo {pages} {89} (\bibinfo {year} {1963})}\BibitemShut {NoStop}%
\bibitem [{\citenamefont {Cabrera}\ \emph {et~al.}(2018)\citenamefont {Cabrera}, \citenamefont {Tanzi}, \citenamefont {Sanz}, \citenamefont {Naylor}, \citenamefont {Thomas}, \citenamefont {Cheiney},\ and\ \citenamefont {Tarruell}}]{Cabrera2018}%
  \BibitemOpen
  \bibfield  {author} {\bibinfo {author} {\bibfnamefont {C.~R.}\ \bibnamefont {Cabrera}}, \bibinfo {author} {\bibfnamefont {L.}~\bibnamefont {Tanzi}}, \bibinfo {author} {\bibfnamefont {J.}~\bibnamefont {Sanz}}, \bibinfo {author} {\bibfnamefont {B.}~\bibnamefont {Naylor}}, \bibinfo {author} {\bibfnamefont {P.}~\bibnamefont {Thomas}}, \bibinfo {author} {\bibfnamefont {P.}~\bibnamefont {Cheiney}}, \ and\ \bibinfo {author} {\bibfnamefont {L.}~\bibnamefont {Tarruell}},\ }\bibfield  {title} {\bibinfo {title} {\emph {{Quantum liquid droplets in a mixture of Bose-Einstein condensates}}},\ }\href {\doibase 10.1126/science.aao5686} {\bibfield  {journal} {\bibinfo  {journal} {Science}\ }\textbf {\bibinfo {volume} {359}},\ \bibinfo {pages} {301} (\bibinfo {year} {2018})}\BibitemShut {NoStop}%
\bibitem [{\citenamefont {Semeghini}\ \emph {et~al.}(2018)\citenamefont {Semeghini}, \citenamefont {Ferioli}, \citenamefont {Masi}, \citenamefont {Mazzinghi}, \citenamefont {Wolswijk}, \citenamefont {Minardi}, \citenamefont {Modugno}, \citenamefont {Modugno}, \citenamefont {Inguscio},\ and\ \citenamefont {Fattori}}]{Semeghini2018}%
  \BibitemOpen
  \bibfield  {author} {\bibinfo {author} {\bibfnamefont {G.}~\bibnamefont {Semeghini}}, \bibinfo {author} {\bibfnamefont {G.}~\bibnamefont {Ferioli}}, \bibinfo {author} {\bibfnamefont {L.}~\bibnamefont {Masi}}, \bibinfo {author} {\bibfnamefont {C.}~\bibnamefont {Mazzinghi}}, \bibinfo {author} {\bibfnamefont {L.}~\bibnamefont {Wolswijk}}, \bibinfo {author} {\bibfnamefont {F.}~\bibnamefont {Minardi}}, \bibinfo {author} {\bibfnamefont {M.}~\bibnamefont {Modugno}}, \bibinfo {author} {\bibfnamefont {G.}~\bibnamefont {Modugno}}, \bibinfo {author} {\bibfnamefont {M.}~\bibnamefont {Inguscio}}, \ and\ \bibinfo {author} {\bibfnamefont {M.}~\bibnamefont {Fattori}},\ }\bibfield  {title} {\bibinfo {title} {\emph {{Self-Bound Quantum Droplets of Atomic Mixtures in Free Space}}},\ }\href {\doibase 10.1103/PhysRevLett.120.235301} {\bibfield  {journal} {\bibinfo  {journal} {Phys. Rev. Lett.}\ }\textbf {\bibinfo {volume} {120}},\ \bibinfo {pages} {235301} (\bibinfo {year} {2018})}\BibitemShut {NoStop}%
\bibitem [{\citenamefont {D'Errico}\ \emph {et~al.}(2019)\citenamefont {D'Errico}, \citenamefont {Burchianti}, \citenamefont {Prevedelli}, \citenamefont {Salasnich}, \citenamefont {Ancilotto}, \citenamefont {Modugno}, \citenamefont {Minardi},\ and\ \citenamefont {Fort}}]{DErrico2019}%
  \BibitemOpen
  \bibfield  {author} {\bibinfo {author} {\bibfnamefont {C.}~\bibnamefont {D'Errico}}, \bibinfo {author} {\bibfnamefont {A.}~\bibnamefont {Burchianti}}, \bibinfo {author} {\bibfnamefont {M.}~\bibnamefont {Prevedelli}}, \bibinfo {author} {\bibfnamefont {L.}~\bibnamefont {Salasnich}}, \bibinfo {author} {\bibfnamefont {F.}~\bibnamefont {Ancilotto}}, \bibinfo {author} {\bibfnamefont {M.}~\bibnamefont {Modugno}}, \bibinfo {author} {\bibfnamefont {F.}~\bibnamefont {Minardi}}, \ and\ \bibinfo {author} {\bibfnamefont {C.}~\bibnamefont {Fort}},\ }\bibfield  {title} {\bibinfo {title} {\emph {{Observation of quantum droplets in a heteronuclear bosonic mixture}}},\ }\href {\doibase 10.1103/PhysRevResearch.1.033155} {\bibfield  {journal} {\bibinfo  {journal} {Phys. Rev. Res.}\ }\textbf {\bibinfo {volume} {1}},\ \bibinfo {pages} {033155} (\bibinfo {year} {2019})}\BibitemShut {NoStop}%
\bibitem [{\citenamefont {Ferrier-Barbut}\ \emph {et~al.}(2016)\citenamefont {Ferrier-Barbut}, \citenamefont {Kadau}, \citenamefont {Schmitt}, \citenamefont {Wenzel},\ and\ \citenamefont {Pfau}}]{FerrierBarbut2016}%
  \BibitemOpen
  \bibfield  {author} {\bibinfo {author} {\bibfnamefont {I.}~\bibnamefont {Ferrier-Barbut}}, \bibinfo {author} {\bibfnamefont {H.}~\bibnamefont {Kadau}}, \bibinfo {author} {\bibfnamefont {M.}~\bibnamefont {Schmitt}}, \bibinfo {author} {\bibfnamefont {M.}~\bibnamefont {Wenzel}}, \ and\ \bibinfo {author} {\bibfnamefont {T.}~\bibnamefont {Pfau}},\ }\bibfield  {title} {\bibinfo {title} {\emph {{Observation of Quantum Droplets in a Strongly Dipolar Bose Gas}}},\ }\href {\doibase 10.1103/PhysRevLett.116.215301} {\bibfield  {journal} {\bibinfo  {journal} {Phys. Rev. Lett.}\ }\textbf {\bibinfo {volume} {116}},\ \bibinfo {pages} {215301} (\bibinfo {year} {2016})}\BibitemShut {NoStop}%
\bibitem [{\citenamefont {Chomaz}\ \emph {et~al.}(2016)\citenamefont {Chomaz}, \citenamefont {Baier}, \citenamefont {Petter}, \citenamefont {Mark}, \citenamefont {W\"achtler}, \citenamefont {Santos},\ and\ \citenamefont {Ferlaino}}]{Chomaz2016}%
  \BibitemOpen
  \bibfield  {author} {\bibinfo {author} {\bibfnamefont {L.}~\bibnamefont {Chomaz}}, \bibinfo {author} {\bibfnamefont {S.}~\bibnamefont {Baier}}, \bibinfo {author} {\bibfnamefont {D.}~\bibnamefont {Petter}}, \bibinfo {author} {\bibfnamefont {M.~J.}\ \bibnamefont {Mark}}, \bibinfo {author} {\bibfnamefont {F.}~\bibnamefont {W\"achtler}}, \bibinfo {author} {\bibfnamefont {L.}~\bibnamefont {Santos}}, \ and\ \bibinfo {author} {\bibfnamefont {F.}~\bibnamefont {Ferlaino}},\ }\bibfield  {title} {\bibinfo {title} {\emph {{Quantum-Fluctuation-Driven Crossover from a Dilute Bose-Einstein Condensate to a Macrodroplet in a Dipolar Quantum Fluid}}},\ }\href {\doibase 10.1103/PhysRevX.6.041039} {\bibfield  {journal} {\bibinfo  {journal} {Phys. Rev. X}\ }\textbf {\bibinfo {volume} {6}},\ \bibinfo {pages} {041039} (\bibinfo {year} {2016})}\BibitemShut {NoStop}%
\bibitem [{\citenamefont {Petrov}()}]{Petrov2025}%
  \BibitemOpen
  \bibfield  {author} {\bibinfo {author} {\bibfnamefont {D.~S.}\ \bibnamefont {Petrov}},\ }\href@noop {} {\bibinfo {title} {\emph {{Beyond-mean-field effects in mixtures: few-body and many-body aspects}}}},\ \bibinfo {note} {in Ref.~\cite{Varenna2022book}}\BibitemShut {NoStop}%
\bibitem [{\citenamefont {Petrov}\ and\ \citenamefont {Astrakharchik}(2016)}]{Petrov2016}%
  \BibitemOpen
  \bibfield  {author} {\bibinfo {author} {\bibfnamefont {D.~S.}\ \bibnamefont {Petrov}}\ and\ \bibinfo {author} {\bibfnamefont {G.~E.}\ \bibnamefont {Astrakharchik}},\ }\bibfield  {title} {\bibinfo {title} {\emph {{Ultradilute Low-Dimensional Liquids}}},\ }\href {\doibase 10.1103/PhysRevLett.117.100401} {\bibfield  {journal} {\bibinfo  {journal} {Phys. Rev. Lett.}\ }\textbf {\bibinfo {volume} {117}},\ \bibinfo {pages} {100401} (\bibinfo {year} {2016})}\BibitemShut {NoStop}%
\bibitem [{\citenamefont {Cheiney}\ \emph {et~al.}(2018)\citenamefont {Cheiney}, \citenamefont {Cabrera}, \citenamefont {Sanz}, \citenamefont {Naylor}, \citenamefont {Tanzi},\ and\ \citenamefont {Tarruell}}]{Cheiney2018}%
  \BibitemOpen
  \bibfield  {author} {\bibinfo {author} {\bibfnamefont {P.}~\bibnamefont {Cheiney}}, \bibinfo {author} {\bibfnamefont {C.~R.}\ \bibnamefont {Cabrera}}, \bibinfo {author} {\bibfnamefont {J.}~\bibnamefont {Sanz}}, \bibinfo {author} {\bibfnamefont {B.}~\bibnamefont {Naylor}}, \bibinfo {author} {\bibfnamefont {L.}~\bibnamefont {Tanzi}}, \ and\ \bibinfo {author} {\bibfnamefont {L.}~\bibnamefont {Tarruell}},\ }\bibfield  {title} {\bibinfo {title} {\emph {{Bright Soliton to Quantum Droplet Transition in a Mixture of Bose-Einstein Condensates}}},\ }\href {\doibase 10.1103/PhysRevLett.120.135301} {\bibfield  {journal} {\bibinfo  {journal} {Phys. Rev. Lett.}\ }\textbf {\bibinfo {volume} {120}},\ \bibinfo {pages} {135301} (\bibinfo {year} {2018})}\BibitemShut {NoStop}%
\bibitem [{\citenamefont {Lavoine}\ and\ \citenamefont {Bourdel}(2021)}]{Lavoine2021b}%
  \BibitemOpen
  \bibfield  {author} {\bibinfo {author} {\bibfnamefont {L.}~\bibnamefont {Lavoine}}\ and\ \bibinfo {author} {\bibfnamefont {T.}~\bibnamefont {Bourdel}},\ }\bibfield  {title} {\bibinfo {title} {\emph {Beyond-mean-field crossover from one dimension to three dimensions in quantum droplets of binary mixtures}},\ }\href {\doibase 10.1103/PhysRevA.103.033312} {\bibfield  {journal} {\bibinfo  {journal} {Phys. Rev. A}\ }\textbf {\bibinfo {volume} {103}},\ \bibinfo {pages} {033312} (\bibinfo {year} {2021})}\BibitemShut {NoStop}%
\bibitem [{\citenamefont {Kartashov}\ \emph {et~al.}(2018)\citenamefont {Kartashov}, \citenamefont {Malomed}, \citenamefont {Tarruell},\ and\ \citenamefont {Torner}}]{Kartashov2018}%
  \BibitemOpen
  \bibfield  {author} {\bibinfo {author} {\bibfnamefont {Y.~V.}\ \bibnamefont {Kartashov}}, \bibinfo {author} {\bibfnamefont {B.~A.}\ \bibnamefont {Malomed}}, \bibinfo {author} {\bibfnamefont {L.}~\bibnamefont {Tarruell}}, \ and\ \bibinfo {author} {\bibfnamefont {L.}~\bibnamefont {Torner}},\ }\bibfield  {title} {\bibinfo {title} {\emph {{Three-dimensional droplets of swirling superfluids}}},\ }\href {\doibase 10.1103/PhysRevA.98.013612} {\bibfield  {journal} {\bibinfo  {journal} {Phys. Rev. A}\ }\textbf {\bibinfo {volume} {98}},\ \bibinfo {pages} {013612} (\bibinfo {year} {2018})}\BibitemShut {NoStop}%
\bibitem [{\citenamefont {Caldara}\ and\ \citenamefont {Ancilotto}(2022)}]{Caldara2022}%
  \BibitemOpen
  \bibfield  {author} {\bibinfo {author} {\bibfnamefont {M.}~\bibnamefont {Caldara}}\ and\ \bibinfo {author} {\bibfnamefont {F.}~\bibnamefont {Ancilotto}},\ }\bibfield  {title} {\bibinfo {title} {\emph {{Vortices in quantum droplets of heteronuclear Bose mixtures}}},\ }\href {\doibase 10.1103/PhysRevA.105.063328} {\bibfield  {journal} {\bibinfo  {journal} {Phys. Rev. A}\ }\textbf {\bibinfo {volume} {105}},\ \bibinfo {pages} {063328} (\bibinfo {year} {2022})}\BibitemShut {NoStop}%
\bibitem [{\citenamefont {Ferioli}\ \emph {et~al.}(2019)\citenamefont {Ferioli}, \citenamefont {Semeghini}, \citenamefont {Masi}, \citenamefont {Giusti}, \citenamefont {Modugno}, \citenamefont {Inguscio}, \citenamefont {Gallem\'{\i}}, \citenamefont {Recati},\ and\ \citenamefont {Fattori}}]{Ferioli2019}%
  \BibitemOpen
  \bibfield  {author} {\bibinfo {author} {\bibfnamefont {G.}~\bibnamefont {Ferioli}}, \bibinfo {author} {\bibfnamefont {G.}~\bibnamefont {Semeghini}}, \bibinfo {author} {\bibfnamefont {L.}~\bibnamefont {Masi}}, \bibinfo {author} {\bibfnamefont {G.}~\bibnamefont {Giusti}}, \bibinfo {author} {\bibfnamefont {G.}~\bibnamefont {Modugno}}, \bibinfo {author} {\bibfnamefont {M.}~\bibnamefont {Inguscio}}, \bibinfo {author} {\bibfnamefont {A.}~\bibnamefont {Gallem\'{\i}}}, \bibinfo {author} {\bibfnamefont {A.}~\bibnamefont {Recati}}, \ and\ \bibinfo {author} {\bibfnamefont {M.}~\bibnamefont {Fattori}},\ }\bibfield  {title} {\bibinfo {title} {\emph {{Collisions of Self-Bound Quantum Droplets}}},\ }\href {\doibase 10.1103/PhysRevLett.122.090401} {\bibfield  {journal} {\bibinfo  {journal} {Phys. Rev. Lett.}\ }\textbf {\bibinfo {volume} {122}},\ \bibinfo {pages} {090401} (\bibinfo {year} {2019})}\BibitemShut {NoStop}%
\bibitem [{\citenamefont {Cavicchioli}\ \emph {et~al.}(2025)\citenamefont {Cavicchioli}, \citenamefont {Fort}, \citenamefont {Ancilotto}, \citenamefont {Modugno}, \citenamefont {Minardi},\ and\ \citenamefont {Burchianti}}]{Cavicchioli2025}%
  \BibitemOpen
  \bibfield  {author} {\bibinfo {author} {\bibfnamefont {L.}~\bibnamefont {Cavicchioli}}, \bibinfo {author} {\bibfnamefont {C.}~\bibnamefont {Fort}}, \bibinfo {author} {\bibfnamefont {F.}~\bibnamefont {Ancilotto}}, \bibinfo {author} {\bibfnamefont {M.}~\bibnamefont {Modugno}}, \bibinfo {author} {\bibfnamefont {F.}~\bibnamefont {Minardi}}, \ and\ \bibinfo {author} {\bibfnamefont {A.}~\bibnamefont {Burchianti}},\ }\bibfield  {title} {\bibinfo {title} {\emph {{Dynamical Formation of Multiple Quantum Droplets in a Bose-Bose Mixture}}},\ }\href {\doibase 10.1103/PhysRevLett.134.093401} {\bibfield  {journal} {\bibinfo  {journal} {Phys. Rev. Lett.}\ }\textbf {\bibinfo {volume} {134}},\ \bibinfo {pages} {093401} (\bibinfo {year} {2025})}\BibitemShut {NoStop}%
\bibitem [{\citenamefont {Cappellaro}\ \emph {et~al.}(2017)\citenamefont {Cappellaro}, \citenamefont {Macrì}, \citenamefont {Bertacco},\ and\ \citenamefont {Salasnich}}]{Cappellaro2017}%
  \BibitemOpen
  \bibfield  {author} {\bibinfo {author} {\bibfnamefont {A.}~\bibnamefont {Cappellaro}}, \bibinfo {author} {\bibfnamefont {T.}~\bibnamefont {Macrì}}, \bibinfo {author} {\bibfnamefont {G.~F.}\ \bibnamefont {Bertacco}}, \ and\ \bibinfo {author} {\bibfnamefont {L.}~\bibnamefont {Salasnich}},\ }\bibfield  {title} {\bibinfo {title} {\emph {{Equation of state and self-bound droplet in Rabi-coupled Bose mixtures}}},\ }\href {\doibase 10.1038/s41598-017-13647-y} {\bibfield  {journal} {\bibinfo  {journal} {Scientific Reports}\ }\textbf {\bibinfo {volume} {7}},\ \bibinfo {pages} {13358} (\bibinfo {year} {2017})}\BibitemShut {NoStop}%
\bibitem [{\citenamefont {Lavoine}\ \emph {et~al.}(2021)\citenamefont {Lavoine}, \citenamefont {Hammond}, \citenamefont {Recati}, \citenamefont {Petrov},\ and\ \citenamefont {Bourdel}}]{Lavoine2021}%
  \BibitemOpen
  \bibfield  {author} {\bibinfo {author} {\bibfnamefont {L.}~\bibnamefont {Lavoine}}, \bibinfo {author} {\bibfnamefont {A.}~\bibnamefont {Hammond}}, \bibinfo {author} {\bibfnamefont {A.}~\bibnamefont {Recati}}, \bibinfo {author} {\bibfnamefont {D.~S.}\ \bibnamefont {Petrov}}, \ and\ \bibinfo {author} {\bibfnamefont {T.}~\bibnamefont {Bourdel}},\ }\bibfield  {title} {\bibinfo {title} {\emph {Beyond-Mean-Field Effects in Rabi-Coupled Two-Component Bose-Einstein Condensate}},\ }\href {\doibase 10.1103/PhysRevLett.127.203402} {\bibfield  {journal} {\bibinfo  {journal} {Phys. Rev. Lett.}\ }\textbf {\bibinfo {volume} {127}},\ \bibinfo {pages} {203402} (\bibinfo {year} {2021})}\BibitemShut {NoStop}%
\bibitem [{\citenamefont {Chiquillo}(2025)}]{Chiquillo2025}%
  \BibitemOpen
  \bibfield  {author} {\bibinfo {author} {\bibfnamefont {E.}~\bibnamefont {Chiquillo}},\ }\bibfield  {title} {\bibinfo {title} {\emph {{Nonuniversal equation of state for Rabi-coupled bosonic gases: A droplet phase}}},\ }\href {\doibase https://doi.org/10.1016/j.aop.2025.170071} {\bibfield  {journal} {\bibinfo  {journal} {Annals of Physics}\ }\textbf {\bibinfo {volume} {479}},\ \bibinfo {pages} {170071} (\bibinfo {year} {2025})}\BibitemShut {NoStop}%
\bibitem [{\citenamefont {Mixa}\ \emph {et~al.}(2025)\citenamefont {Mixa}, \citenamefont {Radonji\ifmmode~\acute{c}\else \'{c}\fi{}}, \citenamefont {Pelster},\ and\ \citenamefont {Thorwart}}]{Mixa2025}%
  \BibitemOpen
  \bibfield  {author} {\bibinfo {author} {\bibfnamefont {L.}~\bibnamefont {Mixa}}, \bibinfo {author} {\bibfnamefont {M.}~\bibnamefont {Radonji\ifmmode~\acute{c}\else \'{c}\fi{}}}, \bibinfo {author} {\bibfnamefont {A.}~\bibnamefont {Pelster}}, \ and\ \bibinfo {author} {\bibfnamefont {M.}~\bibnamefont {Thorwart}},\ }\bibfield  {title} {\bibinfo {title} {\emph {{Engineering quantum droplet formation by cavity-induced long-range interactions}}},\ }\href {\doibase 10.1103/PhysRevResearch.7.023204} {\bibfield  {journal} {\bibinfo  {journal} {Phys. Rev. Res.}\ }\textbf {\bibinfo {volume} {7}},\ \bibinfo {pages} {023204} (\bibinfo {year} {2025})}\BibitemShut {NoStop}%
\bibitem [{\citenamefont {Wang}\ \emph {et~al.}(2020)\citenamefont {Wang}, \citenamefont {Pan}, \citenamefont {Cui},\ and\ \citenamefont {Yi}}]{Wang2020}%
  \BibitemOpen
  \bibfield  {author} {\bibinfo {author} {\bibfnamefont {J.-B.}\ \bibnamefont {Wang}}, \bibinfo {author} {\bibfnamefont {J.-S.}\ \bibnamefont {Pan}}, \bibinfo {author} {\bibfnamefont {X.}~\bibnamefont {Cui}}, \ and\ \bibinfo {author} {\bibfnamefont {W.}~\bibnamefont {Yi}},\ }\bibfield  {title} {\bibinfo {title} {\emph {{Quantum Droplets in a Mixture of Bose–Fermi Superfluids}}},\ }\href {\doibase 10.1088/0256-307X/37/7/076701} {\bibfield  {journal} {\bibinfo  {journal} {Chinese Physics Letters}\ }\textbf {\bibinfo {volume} {37}},\ \bibinfo {pages} {076701} (\bibinfo {year} {2020})}\BibitemShut {NoStop}%
\bibitem [{\citenamefont {Xu}\ \emph {et~al.}(2022)\citenamefont {Xu}, \citenamefont {Lei}, \citenamefont {Du}, \citenamefont {Zhao}, \citenamefont {Hua},\ and\ \citenamefont {Zeng}}]{Xu2022}%
  \BibitemOpen
  \bibfield  {author} {\bibinfo {author} {\bibfnamefont {S.-L.}\ \bibnamefont {Xu}}, \bibinfo {author} {\bibfnamefont {Y.-B.}\ \bibnamefont {Lei}}, \bibinfo {author} {\bibfnamefont {J.-T.}\ \bibnamefont {Du}}, \bibinfo {author} {\bibfnamefont {Y.}~\bibnamefont {Zhao}}, \bibinfo {author} {\bibfnamefont {R.}~\bibnamefont {Hua}}, \ and\ \bibinfo {author} {\bibfnamefont {J.-H.}\ \bibnamefont {Zeng}},\ }\bibfield  {title} {\bibinfo {title} {\emph {{Three-dimensional quantum droplets in spin-orbit-coupled Bose-Einstein condensates}}},\ }\href {\doibase https://doi.org/10.1016/j.chaos.2022.112665} {\bibfield  {journal} {\bibinfo  {journal} {Chaos, Solitons \& Fractals}\ }\textbf {\bibinfo {volume} {164}},\ \bibinfo {pages} {112665} (\bibinfo {year} {2022})}\BibitemShut {NoStop}%
\bibitem [{\citenamefont {Deng}\ \emph {et~al.}(2010)\citenamefont {Deng}, \citenamefont {Haug},\ and\ \citenamefont {Yamamoto}}]{Deng2010}%
  \BibitemOpen
  \bibfield  {author} {\bibinfo {author} {\bibfnamefont {H.}~\bibnamefont {Deng}}, \bibinfo {author} {\bibfnamefont {H.}~\bibnamefont {Haug}}, \ and\ \bibinfo {author} {\bibfnamefont {Y.}~\bibnamefont {Yamamoto}},\ }\bibfield  {title} {\bibinfo {title} {\emph {{Exciton-polariton Bose-Einstein condensation}}},\ }\href {\doibase 10.1103/RevModPhys.82.1489} {\bibfield  {journal} {\bibinfo  {journal} {Rev. Mod. Phys.}\ }\textbf {\bibinfo {volume} {82}},\ \bibinfo {pages} {1489} (\bibinfo {year} {2010})}\BibitemShut {NoStop}%
\bibitem [{\citenamefont {Shelykh}\ \emph {et~al.}(2009)\citenamefont {Shelykh}, \citenamefont {Kavokin}, \citenamefont {Rubo}, \citenamefont {Liew},\ and\ \citenamefont {Malpuech}}]{Shelykh2010}%
  \BibitemOpen
  \bibfield  {author} {\bibinfo {author} {\bibfnamefont {I.~A.}\ \bibnamefont {Shelykh}}, \bibinfo {author} {\bibfnamefont {A.~V.}\ \bibnamefont {Kavokin}}, \bibinfo {author} {\bibfnamefont {Y.~G.}\ \bibnamefont {Rubo}}, \bibinfo {author} {\bibfnamefont {T.~C.~H.}\ \bibnamefont {Liew}}, \ and\ \bibinfo {author} {\bibfnamefont {G.}~\bibnamefont {Malpuech}},\ }\bibfield  {title} {\bibinfo {title} {\emph {{Polariton polarization-sensitive phenomena in planar semiconductor microcavities}}},\ }\href {\doibase 10.1088/0268-1242/25/1/013001} {\bibfield  {journal} {\bibinfo  {journal} {Semiconductor Science and Technology}\ }\textbf {\bibinfo {volume} {25}},\ \bibinfo {pages} {013001} (\bibinfo {year} {2009})}\BibitemShut {NoStop}%
\bibitem [{\citenamefont {Carusotto}\ and\ \citenamefont {Ciuti}(2013)}]{Carusotto2013}%
  \BibitemOpen
  \bibfield  {author} {\bibinfo {author} {\bibfnamefont {I.}~\bibnamefont {Carusotto}}\ and\ \bibinfo {author} {\bibfnamefont {C.}~\bibnamefont {Ciuti}},\ }\bibfield  {title} {\bibinfo {title} {\emph {{Quantum fluids of light}}},\ }\href {\doibase 10.1103/RevModPhys.85.299} {\bibfield  {journal} {\bibinfo  {journal} {Rev. Mod. Phys.}\ }\textbf {\bibinfo {volume} {85}},\ \bibinfo {pages} {299} (\bibinfo {year} {2013})}\BibitemShut {NoStop}%
\bibitem [{\citenamefont {Kasprzak}\ \emph {et~al.}(2006)\citenamefont {Kasprzak}, \citenamefont {Richard}, \citenamefont {Kundermann}, \citenamefont {Baas}, \citenamefont {Jeambrun}, \citenamefont {Keeling}, \citenamefont {Marchetti}, \citenamefont {Szymańska}, \citenamefont {André}, \citenamefont {Staehli}, \citenamefont {Savona}, \citenamefont {Littlewood}, \citenamefont {Deveaud},\ and\ \citenamefont {Dang}}]{Kasprzak2006}%
  \BibitemOpen
  \bibfield  {author} {\bibinfo {author} {\bibfnamefont {J.}~\bibnamefont {Kasprzak}}, \bibinfo {author} {\bibfnamefont {M.}~\bibnamefont {Richard}}, \bibinfo {author} {\bibfnamefont {S.}~\bibnamefont {Kundermann}}, \bibinfo {author} {\bibfnamefont {A.}~\bibnamefont {Baas}}, \bibinfo {author} {\bibfnamefont {P.}~\bibnamefont {Jeambrun}}, \bibinfo {author} {\bibfnamefont {J.~M.~J.}\ \bibnamefont {Keeling}}, \bibinfo {author} {\bibfnamefont {F.~M.}\ \bibnamefont {Marchetti}}, \bibinfo {author} {\bibfnamefont {M.~H.}\ \bibnamefont {Szymańska}}, \bibinfo {author} {\bibfnamefont {R.}~\bibnamefont {André}}, \bibinfo {author} {\bibfnamefont {J.~L.}\ \bibnamefont {Staehli}}, \bibinfo {author} {\bibfnamefont {V.}~\bibnamefont {Savona}}, \bibinfo {author} {\bibfnamefont {P.~B.}\ \bibnamefont {Littlewood}}, \bibinfo {author} {\bibfnamefont {B.}~\bibnamefont {Deveaud}}, \ and\ \bibinfo {author} {\bibfnamefont {L.~S.}\ \bibnamefont {Dang}},\ }\bibfield  {title} {\bibinfo {title} {\emph {{ Bose–Einstein condensation of
  exciton polaritons}}},\ }\href {\doibase 10.1038/nature05131} {\bibfield  {journal} {\bibinfo  {journal} {Nature}\ }\textbf {\bibinfo {volume} {443}},\ \bibinfo {pages} {409} (\bibinfo {year} {2006})}\BibitemShut {NoStop}%
\bibitem [{\citenamefont {Balili}\ \emph {et~al.}(2007)\citenamefont {Balili}, \citenamefont {Hartwell}, \citenamefont {Snoke}, \citenamefont {Pfeiffer},\ and\ \citenamefont {West}}]{Balili2007}%
  \BibitemOpen
  \bibfield  {author} {\bibinfo {author} {\bibfnamefont {R.}~\bibnamefont {Balili}}, \bibinfo {author} {\bibfnamefont {V.}~\bibnamefont {Hartwell}}, \bibinfo {author} {\bibfnamefont {D.}~\bibnamefont {Snoke}}, \bibinfo {author} {\bibfnamefont {L.}~\bibnamefont {Pfeiffer}}, \ and\ \bibinfo {author} {\bibfnamefont {K.}~\bibnamefont {West}},\ }\bibfield  {title} {\bibinfo {title} {\emph {{Bose-Einstein Condensation of Microcavity Polaritons in a Trap}}},\ }\href {\doibase 10.1126/science.1140990} {\bibfield  {journal} {\bibinfo  {journal} {Science}\ }\textbf {\bibinfo {volume} {316}},\ \bibinfo {pages} {1007} (\bibinfo {year} {2007})}\BibitemShut {NoStop}%
\bibitem [{\citenamefont {Amo}\ \emph {et~al.}(2009)\citenamefont {Amo}, \citenamefont {Lefrère}, \citenamefont {Pigeon}, \citenamefont {Adrados}, \citenamefont {Ciuti}, \citenamefont {Carusotto}, \citenamefont {Houdré}, \citenamefont {Giacobino},\ and\ \citenamefont {Bramati}}]{Amo2009NatPhys}%
  \BibitemOpen
  \bibfield  {author} {\bibinfo {author} {\bibfnamefont {A.}~\bibnamefont {Amo}}, \bibinfo {author} {\bibfnamefont {J.}~\bibnamefont {Lefrère}}, \bibinfo {author} {\bibfnamefont {S.}~\bibnamefont {Pigeon}}, \bibinfo {author} {\bibfnamefont {C.}~\bibnamefont {Adrados}}, \bibinfo {author} {\bibfnamefont {C.}~\bibnamefont {Ciuti}}, \bibinfo {author} {\bibfnamefont {I.}~\bibnamefont {Carusotto}}, \bibinfo {author} {\bibfnamefont {R.}~\bibnamefont {Houdré}}, \bibinfo {author} {\bibfnamefont {E.}~\bibnamefont {Giacobino}}, \ and\ \bibinfo {author} {\bibfnamefont {A.}~\bibnamefont {Bramati}},\ }\bibfield  {title} {\bibinfo {title} {\emph {{Superfluidity of polaritons in semiconductor microcavities}}},\ }\href {\doibase 10.1038/nphys1364} {\bibfield  {journal} {\bibinfo  {journal} {Nature Physics}\ }\textbf {\bibinfo {volume} {5}},\ \bibinfo {pages} {805} (\bibinfo {year} {2009})}\BibitemShut {NoStop}%
\bibitem [{\citenamefont {Lerario}\ \emph {et~al.}(2017)\citenamefont {Lerario}, \citenamefont {Fieramosca}, \citenamefont {Barachati}, \citenamefont {Ballarini}, \citenamefont {Daskalakis}, \citenamefont {Dominici}, \citenamefont {De~Giorgi}, \citenamefont {Maier}, \citenamefont {Gigli}, \citenamefont {Kéna-Cohen},\ and\ \citenamefont {Sanvitto}}]{Lerario2017}%
  \BibitemOpen
  \bibfield  {author} {\bibinfo {author} {\bibfnamefont {G.}~\bibnamefont {Lerario}}, \bibinfo {author} {\bibfnamefont {A.}~\bibnamefont {Fieramosca}}, \bibinfo {author} {\bibfnamefont {F.}~\bibnamefont {Barachati}}, \bibinfo {author} {\bibfnamefont {D.}~\bibnamefont {Ballarini}}, \bibinfo {author} {\bibfnamefont {K.~S.}\ \bibnamefont {Daskalakis}}, \bibinfo {author} {\bibfnamefont {L.}~\bibnamefont {Dominici}}, \bibinfo {author} {\bibfnamefont {M.}~\bibnamefont {De~Giorgi}}, \bibinfo {author} {\bibfnamefont {S.~A.}\ \bibnamefont {Maier}}, \bibinfo {author} {\bibfnamefont {G.}~\bibnamefont {Gigli}}, \bibinfo {author} {\bibfnamefont {S.}~\bibnamefont {Kéna-Cohen}}, \ and\ \bibinfo {author} {\bibfnamefont {D.}~\bibnamefont {Sanvitto}},\ }\bibfield  {title} {\bibinfo {title} {\emph {{Room-temperature superfluidity in a polariton condensate}}},\ }\href {\doibase 10.1038/nphys4147} {\bibfield  {journal} {\bibinfo  {journal} {Nature Physics}\ }\textbf {\bibinfo {volume} {13}},\ \bibinfo {pages} {837} (\bibinfo
  {year} {2017})}\BibitemShut {NoStop}%
\bibitem [{\citenamefont {Lagoudakis}\ \emph {et~al.}(2008)\citenamefont {Lagoudakis}, \citenamefont {Wouters}, \citenamefont {Richard}, \citenamefont {Baas}, \citenamefont {Carusotto}, \citenamefont {André}, \citenamefont {Dang},\ and\ \citenamefont {Deveaud-Plédran}}]{Lagoudakis2008}%
  \BibitemOpen
  \bibfield  {author} {\bibinfo {author} {\bibfnamefont {K.~G.}\ \bibnamefont {Lagoudakis}}, \bibinfo {author} {\bibfnamefont {M.}~\bibnamefont {Wouters}}, \bibinfo {author} {\bibfnamefont {M.}~\bibnamefont {Richard}}, \bibinfo {author} {\bibfnamefont {A.}~\bibnamefont {Baas}}, \bibinfo {author} {\bibfnamefont {I.}~\bibnamefont {Carusotto}}, \bibinfo {author} {\bibfnamefont {R.}~\bibnamefont {André}}, \bibinfo {author} {\bibfnamefont {L.~S.}\ \bibnamefont {Dang}}, \ and\ \bibinfo {author} {\bibfnamefont {B.}~\bibnamefont {Deveaud-Plédran}},\ }\bibfield  {title} {\bibinfo {title} {\emph {{Quantized vortices in an exciton–polariton condensate}}},\ }\href {\doibase 10.1038/nphys1051} {\bibfield  {journal} {\bibinfo  {journal} {Nature Physics}\ }\textbf {\bibinfo {volume} {4}},\ \bibinfo {pages} {706} (\bibinfo {year} {2008})}\BibitemShut {NoStop}%
\bibitem [{\citenamefont {Lagoudakis}\ \emph {et~al.}(2009)\citenamefont {Lagoudakis}, \citenamefont {Ostatnický}, \citenamefont {Kavokin}, \citenamefont {Rubo}, \citenamefont {André},\ and\ \citenamefont {Deveaud-Plédran}}]{Lagoudakis2009}%
  \BibitemOpen
  \bibfield  {author} {\bibinfo {author} {\bibfnamefont {K.~G.}\ \bibnamefont {Lagoudakis}}, \bibinfo {author} {\bibfnamefont {T.}~\bibnamefont {Ostatnický}}, \bibinfo {author} {\bibfnamefont {A.~V.}\ \bibnamefont {Kavokin}}, \bibinfo {author} {\bibfnamefont {Y.~G.}\ \bibnamefont {Rubo}}, \bibinfo {author} {\bibfnamefont {R.}~\bibnamefont {André}}, \ and\ \bibinfo {author} {\bibfnamefont {B.}~\bibnamefont {Deveaud-Plédran}},\ }\bibfield  {title} {\bibinfo {title} {\emph {{Observation of Half-Quantum Vortices in an Exciton-Polariton Condensate}}},\ }\href {\doibase 10.1126/science.1177980} {\bibfield  {journal} {\bibinfo  {journal} {Science}\ }\textbf {\bibinfo {volume} {326}},\ \bibinfo {pages} {974} (\bibinfo {year} {2009})}\BibitemShut {NoStop}%
\bibitem [{\citenamefont {Sanvitto}\ \emph {et~al.}(2010)\citenamefont {Sanvitto}, \citenamefont {Marchetti}, \citenamefont {Szymańska}, \citenamefont {Tosi}, \citenamefont {Baudisch}, \citenamefont {Laussy}, \citenamefont {Krizhanovskii}, \citenamefont {Skolnick}, \citenamefont {Marrucci}, \citenamefont {Lemaître}, \citenamefont {Bloch}, \citenamefont {Tejedor},\ and\ \citenamefont {Viña}}]{Sanvitto2010}%
  \BibitemOpen
  \bibfield  {author} {\bibinfo {author} {\bibfnamefont {D.}~\bibnamefont {Sanvitto}}, \bibinfo {author} {\bibfnamefont {F.~M.}\ \bibnamefont {Marchetti}}, \bibinfo {author} {\bibfnamefont {M.~H.}\ \bibnamefont {Szymańska}}, \bibinfo {author} {\bibfnamefont {G.}~\bibnamefont {Tosi}}, \bibinfo {author} {\bibfnamefont {M.}~\bibnamefont {Baudisch}}, \bibinfo {author} {\bibfnamefont {F.~P.}\ \bibnamefont {Laussy}}, \bibinfo {author} {\bibfnamefont {D.~N.}\ \bibnamefont {Krizhanovskii}}, \bibinfo {author} {\bibfnamefont {M.~S.}\ \bibnamefont {Skolnick}}, \bibinfo {author} {\bibfnamefont {L.}~\bibnamefont {Marrucci}}, \bibinfo {author} {\bibfnamefont {A.}~\bibnamefont {Lemaître}}, \bibinfo {author} {\bibfnamefont {J.}~\bibnamefont {Bloch}}, \bibinfo {author} {\bibfnamefont {C.}~\bibnamefont {Tejedor}}, \ and\ \bibinfo {author} {\bibfnamefont {L.}~\bibnamefont {Viña}},\ }\bibfield  {title} {\bibinfo {title} {\emph {{Persistent currents and quantized vortices in a polariton superfluid}}},\ }\href {\doibase
  10.1038/nphys1668} {\bibfield  {journal} {\bibinfo  {journal} {Nature Physics}\ }\textbf {\bibinfo {volume} {6}},\ \bibinfo {pages} {527} (\bibinfo {year} {2010})}\BibitemShut {NoStop}%
\bibitem [{\citenamefont {Gerace}\ \emph {et~al.}(2009)\citenamefont {Gerace}, \citenamefont {Türeci}, \citenamefont {Imamoglu}, \citenamefont {Giovannetti},\ and\ \citenamefont {Fazio}}]{Gerace2009}%
  \BibitemOpen
  \bibfield  {author} {\bibinfo {author} {\bibfnamefont {D.}~\bibnamefont {Gerace}}, \bibinfo {author} {\bibfnamefont {H.~E.}\ \bibnamefont {Türeci}}, \bibinfo {author} {\bibfnamefont {A.}~\bibnamefont {Imamoglu}}, \bibinfo {author} {\bibfnamefont {V.}~\bibnamefont {Giovannetti}}, \ and\ \bibinfo {author} {\bibfnamefont {R.}~\bibnamefont {Fazio}},\ }\bibfield  {title} {\bibinfo {title} {\emph {{The quantum-optical Josephson interferometer}}},\ }\href {\doibase 10.1038/nphys1223} {\bibfield  {journal} {\bibinfo  {journal} {Nature Physics}\ }\textbf {\bibinfo {volume} {5}},\ \bibinfo {pages} {281} (\bibinfo {year} {2009})}\BibitemShut {NoStop}%
\bibitem [{\citenamefont {Ballarini}\ \emph {et~al.}(2013)\citenamefont {Ballarini}, \citenamefont {De~Giorgi}, \citenamefont {Cancellieri}, \citenamefont {Houdré}, \citenamefont {Giacobino}, \citenamefont {Cingolani}, \citenamefont {Bramati}, \citenamefont {Gigli},\ and\ \citenamefont {Sanvitto}}]{Ballarini2013}%
  \BibitemOpen
  \bibfield  {author} {\bibinfo {author} {\bibfnamefont {D.}~\bibnamefont {Ballarini}}, \bibinfo {author} {\bibfnamefont {M.}~\bibnamefont {De~Giorgi}}, \bibinfo {author} {\bibfnamefont {E.}~\bibnamefont {Cancellieri}}, \bibinfo {author} {\bibfnamefont {R.}~\bibnamefont {Houdré}}, \bibinfo {author} {\bibfnamefont {E.}~\bibnamefont {Giacobino}}, \bibinfo {author} {\bibfnamefont {R.}~\bibnamefont {Cingolani}}, \bibinfo {author} {\bibfnamefont {A.}~\bibnamefont {Bramati}}, \bibinfo {author} {\bibfnamefont {G.}~\bibnamefont {Gigli}}, \ and\ \bibinfo {author} {\bibfnamefont {D.}~\bibnamefont {Sanvitto}},\ }\bibfield  {title} {\bibinfo {title} {\emph {{All-optical polariton transistor}}},\ }\href {\doibase 10.1038/ncomms2734} {\bibfield  {journal} {\bibinfo  {journal} {Nature Communications}\ }\textbf {\bibinfo {volume} {4}},\ \bibinfo {pages} {1778} (\bibinfo {year} {2013})}\BibitemShut {NoStop}%
\bibitem [{\citenamefont {Sanvitto}\ and\ \citenamefont {Kéna-Cohen}(2016)}]{Sanvitto2016}%
  \BibitemOpen
  \bibfield  {author} {\bibinfo {author} {\bibfnamefont {D.}~\bibnamefont {Sanvitto}}\ and\ \bibinfo {author} {\bibfnamefont {S.}~\bibnamefont {Kéna-Cohen}},\ }\bibfield  {title} {\bibinfo {title} {\emph {{The road towards polaritonic devices}}},\ }\href {\doibase 10.1038/nmat4668} {\bibfield  {journal} {\bibinfo  {journal} {Nature Materials}\ }\textbf {\bibinfo {volume} {15}},\ \bibinfo {pages} {1061} (\bibinfo {year} {2016})}\BibitemShut {NoStop}%
\bibitem [{\citenamefont {Gerace}\ \emph {et~al.}(2019)\citenamefont {Gerace}, \citenamefont {Laussy},\ and\ \citenamefont {Sanvitto}}]{Gerace2019}%
  \BibitemOpen
  \bibfield  {author} {\bibinfo {author} {\bibfnamefont {D.}~\bibnamefont {Gerace}}, \bibinfo {author} {\bibfnamefont {F.}~\bibnamefont {Laussy}}, \ and\ \bibinfo {author} {\bibfnamefont {D.}~\bibnamefont {Sanvitto}},\ }\bibfield  {title} {\bibinfo {title} {\emph {{Quantum nonlinearities at the single-particle level}}},\ }\href {\doibase 10.1038/s41563-019-0298-3} {\bibfield  {journal} {\bibinfo  {journal} {Nature Materials}\ }\textbf {\bibinfo {volume} {18}},\ \bibinfo {pages} {200} (\bibinfo {year} {2019})}\BibitemShut {NoStop}%
\bibitem [{\citenamefont {Liew}(2023)}]{Liew2023}%
  \BibitemOpen
  \bibfield  {author} {\bibinfo {author} {\bibfnamefont {T.~C.~H.}\ \bibnamefont {Liew}},\ }\bibfield  {title} {\bibinfo {title} {\emph {{The future of quantum in polariton systems: opinion}}},\ }\href {\doibase 10.1364/OME.492503} {\bibfield  {journal} {\bibinfo  {journal} {Opt. Mater. Express}\ }\textbf {\bibinfo {volume} {13}},\ \bibinfo {pages} {1938} (\bibinfo {year} {2023})}\BibitemShut {NoStop}%
\bibitem [{\citenamefont {Rivera}\ \emph {et~al.}(2015)\citenamefont {Rivera}, \citenamefont {Schaibley}, \citenamefont {Jones}, \citenamefont {Ross}, \citenamefont {Wu}, \citenamefont {Aivazian}, \citenamefont {Klement}, \citenamefont {Seyler}, \citenamefont {Clark}, \citenamefont {Ghimire}, \citenamefont {Yan}, \citenamefont {Mandrus}, \citenamefont {Yao},\ and\ \citenamefont {Xu}}]{Rivera2015}%
  \BibitemOpen
  \bibfield  {author} {\bibinfo {author} {\bibfnamefont {P.}~\bibnamefont {Rivera}}, \bibinfo {author} {\bibfnamefont {J.~R.}\ \bibnamefont {Schaibley}}, \bibinfo {author} {\bibfnamefont {A.~M.}\ \bibnamefont {Jones}}, \bibinfo {author} {\bibfnamefont {J.~S.}\ \bibnamefont {Ross}}, \bibinfo {author} {\bibfnamefont {S.}~\bibnamefont {Wu}}, \bibinfo {author} {\bibfnamefont {G.}~\bibnamefont {Aivazian}}, \bibinfo {author} {\bibfnamefont {P.}~\bibnamefont {Klement}}, \bibinfo {author} {\bibfnamefont {K.}~\bibnamefont {Seyler}}, \bibinfo {author} {\bibfnamefont {G.}~\bibnamefont {Clark}}, \bibinfo {author} {\bibfnamefont {N.~J.}\ \bibnamefont {Ghimire}}, \bibinfo {author} {\bibfnamefont {J.}~\bibnamefont {Yan}}, \bibinfo {author} {\bibfnamefont {D.~G.}\ \bibnamefont {Mandrus}}, \bibinfo {author} {\bibfnamefont {W.}~\bibnamefont {Yao}}, \ and\ \bibinfo {author} {\bibfnamefont {X.}~\bibnamefont {Xu}},\ }\bibfield  {title} {\bibinfo {title} {\emph {{Observation of long-lived interlayer excitons in monolayer
  ${\mathrm{MoSe}}_{2}$–${\mathrm{WSe}}_{2}$ heterostructures}}},\ }\href {\doibase 10.1038/ncomms7242} {\bibfield  {journal} {\bibinfo  {journal} {Nature Communications}\ }\textbf {\bibinfo {volume} {6}},\ \bibinfo {pages} {6242} (\bibinfo {year} {2015})}\BibitemShut {NoStop}%
\bibitem [{\citenamefont {Arora}\ \emph {et~al.}(2017)\citenamefont {Arora}, \citenamefont {Drüppel}, \citenamefont {Schmidt}, \citenamefont {Deilmann}, \citenamefont {Schneider}, \citenamefont {Molas}, \citenamefont {Marauhn}, \citenamefont {Michaelis~de Vasconcellos}, \citenamefont {Potemski}, \citenamefont {Rohlfing},\ and\ \citenamefont {Bratschitsch}}]{Arora2017}%
  \BibitemOpen
  \bibfield  {author} {\bibinfo {author} {\bibfnamefont {A.}~\bibnamefont {Arora}}, \bibinfo {author} {\bibfnamefont {M.}~\bibnamefont {Drüppel}}, \bibinfo {author} {\bibfnamefont {R.}~\bibnamefont {Schmidt}}, \bibinfo {author} {\bibfnamefont {T.}~\bibnamefont {Deilmann}}, \bibinfo {author} {\bibfnamefont {R.}~\bibnamefont {Schneider}}, \bibinfo {author} {\bibfnamefont {M.~R.}\ \bibnamefont {Molas}}, \bibinfo {author} {\bibfnamefont {P.}~\bibnamefont {Marauhn}}, \bibinfo {author} {\bibfnamefont {S.}~\bibnamefont {Michaelis~de Vasconcellos}}, \bibinfo {author} {\bibfnamefont {M.}~\bibnamefont {Potemski}}, \bibinfo {author} {\bibfnamefont {M.}~\bibnamefont {Rohlfing}}, \ and\ \bibinfo {author} {\bibfnamefont {R.}~\bibnamefont {Bratschitsch}},\ }\bibfield  {title} {\bibinfo {title} {\emph {{Interlayer excitons in a bulk van der Waals semiconductor}}},\ }\href {\doibase 10.1038/s41467-017-00691-5} {\bibfield  {journal} {\bibinfo  {journal} {Nature Communications}\ }\textbf {\bibinfo {volume} {8}},\ \bibinfo
  {pages} {639} (\bibinfo {year} {2017})}\BibitemShut {NoStop}%
\bibitem [{\citenamefont {Calman}\ \emph {et~al.}(2018)\citenamefont {Calman}, \citenamefont {Fogler}, \citenamefont {Butov}, \citenamefont {Hu}, \citenamefont {Mishchenko},\ and\ \citenamefont {Geim}}]{Calman2018}%
  \BibitemOpen
  \bibfield  {author} {\bibinfo {author} {\bibfnamefont {E.~V.}\ \bibnamefont {Calman}}, \bibinfo {author} {\bibfnamefont {M.~M.}\ \bibnamefont {Fogler}}, \bibinfo {author} {\bibfnamefont {L.~V.}\ \bibnamefont {Butov}}, \bibinfo {author} {\bibfnamefont {S.}~\bibnamefont {Hu}}, \bibinfo {author} {\bibfnamefont {A.}~\bibnamefont {Mishchenko}}, \ and\ \bibinfo {author} {\bibfnamefont {A.~K.}\ \bibnamefont {Geim}},\ }\bibfield  {title} {\bibinfo {title} {\emph {{Indirect excitons in van der Waals heterostructures at room temperature}}},\ }\href {\doibase 10.1038/s41467-018-04293-7} {\bibfield  {journal} {\bibinfo  {journal} {Nature Communications}\ }\textbf {\bibinfo {volume} {9}},\ \bibinfo {pages} {1895} (\bibinfo {year} {2018})}\BibitemShut {NoStop}%
\bibitem [{\citenamefont {Horng}\ \emph {et~al.}(2018)\citenamefont {Horng}, \citenamefont {Stroucken}, \citenamefont {Zhang}, \citenamefont {Paik}, \citenamefont {Deng},\ and\ \citenamefont {Koch}}]{Horng2018}%
  \BibitemOpen
  \bibfield  {author} {\bibinfo {author} {\bibfnamefont {J.}~\bibnamefont {Horng}}, \bibinfo {author} {\bibfnamefont {T.}~\bibnamefont {Stroucken}}, \bibinfo {author} {\bibfnamefont {L.}~\bibnamefont {Zhang}}, \bibinfo {author} {\bibfnamefont {E.~Y.}\ \bibnamefont {Paik}}, \bibinfo {author} {\bibfnamefont {H.}~\bibnamefont {Deng}}, \ and\ \bibinfo {author} {\bibfnamefont {S.~W.}\ \bibnamefont {Koch}},\ }\bibfield  {title} {\bibinfo {title} {\emph {{Observation of interlayer excitons in ${\mathrm{MoSe}}_{2}$ single crystals}}},\ }\href {\doibase 10.1103/PhysRevB.97.241404} {\bibfield  {journal} {\bibinfo  {journal} {Phys. Rev. B}\ }\textbf {\bibinfo {volume} {97}},\ \bibinfo {pages} {241404} (\bibinfo {year} {2018})}\BibitemShut {NoStop}%
\bibitem [{\citenamefont {Niehues}\ \emph {et~al.}(2019)\citenamefont {Niehues}, \citenamefont {Blob}, \citenamefont {Stiehm}, \citenamefont {Michaelis~de Vasconcellos},\ and\ \citenamefont {Bratschitsch}}]{Niehues2019}%
  \BibitemOpen
  \bibfield  {author} {\bibinfo {author} {\bibfnamefont {I.}~\bibnamefont {Niehues}}, \bibinfo {author} {\bibfnamefont {A.}~\bibnamefont {Blob}}, \bibinfo {author} {\bibfnamefont {T.}~\bibnamefont {Stiehm}}, \bibinfo {author} {\bibfnamefont {S.}~\bibnamefont {Michaelis~de Vasconcellos}}, \ and\ \bibinfo {author} {\bibfnamefont {R.}~\bibnamefont {Bratschitsch}},\ }\bibfield  {title} {\bibinfo {title} {\emph {{Interlayer excitons in bilayer ${\mathrm{MoS}}_{2}$ under uniaxial tensile strain}}},\ }\href {\doibase 10.1039/C9NR03332G} {\bibfield  {journal} {\bibinfo  {journal} {Nanoscale}\ }\textbf {\bibinfo {volume} {11}},\ \bibinfo {pages} {12788} (\bibinfo {year} {2019})}\BibitemShut {NoStop}%
\bibitem [{\citenamefont {Gu}\ \emph {et~al.}(2021)\citenamefont {Gu}, \citenamefont {Walther}, \citenamefont {Waldecker}, \citenamefont {Rhodes}, \citenamefont {Raja}, \citenamefont {Hone}, \citenamefont {Heinz}, \citenamefont {K{\'e}na-Cohen}, \citenamefont {Pohl},\ and\ \citenamefont {Menon}}]{Gu_NatComm2021}%
  \BibitemOpen
  \bibfield  {author} {\bibinfo {author} {\bibfnamefont {J.}~\bibnamefont {Gu}}, \bibinfo {author} {\bibfnamefont {V.}~\bibnamefont {Walther}}, \bibinfo {author} {\bibfnamefont {L.}~\bibnamefont {Waldecker}}, \bibinfo {author} {\bibfnamefont {D.}~\bibnamefont {Rhodes}}, \bibinfo {author} {\bibfnamefont {A.}~\bibnamefont {Raja}}, \bibinfo {author} {\bibfnamefont {J.~C.}\ \bibnamefont {Hone}}, \bibinfo {author} {\bibfnamefont {T.~F.}\ \bibnamefont {Heinz}}, \bibinfo {author} {\bibfnamefont {S.}~\bibnamefont {K{\'e}na-Cohen}}, \bibinfo {author} {\bibfnamefont {T.}~\bibnamefont {Pohl}}, \ and\ \bibinfo {author} {\bibfnamefont {V.~M.}\ \bibnamefont {Menon}},\ }\bibfield  {title} {\bibinfo {title} {\emph {{Enhanced nonlinear interaction of polaritons via excitonic Rydberg states in monolayer ${\mathrm{WSe}}_{2}$}}},\ }\href {\doibase 10.1038/s41467-021-22537-x} {\bibfield  {journal} {\bibinfo  {journal} {Nature Communications}\ }\textbf {\bibinfo {volume} {12}},\ \bibinfo {pages} {2269} (\bibinfo {year}
  {2021})}\BibitemShut {NoStop}%
\bibitem [{\citenamefont {Orfanakis}\ \emph {et~al.}(2022)\citenamefont {Orfanakis}, \citenamefont {Rajendran}, \citenamefont {Walther}, \citenamefont {Volz}, \citenamefont {Pohl},\ and\ \citenamefont {Ohadi}}]{Orfanakis2022}%
  \BibitemOpen
  \bibfield  {author} {\bibinfo {author} {\bibfnamefont {K.}~\bibnamefont {Orfanakis}}, \bibinfo {author} {\bibfnamefont {S.~K.}\ \bibnamefont {Rajendran}}, \bibinfo {author} {\bibfnamefont {V.}~\bibnamefont {Walther}}, \bibinfo {author} {\bibfnamefont {T.}~\bibnamefont {Volz}}, \bibinfo {author} {\bibfnamefont {T.}~\bibnamefont {Pohl}}, \ and\ \bibinfo {author} {\bibfnamefont {H.}~\bibnamefont {Ohadi}},\ }\bibfield  {title} {\bibinfo {title} {\emph {{Rydberg exciton–polaritons in a Cu$_2$O microcavity}}},\ }\href {\doibase 10.1038/s41563-022-01230-4} {\bibfield  {journal} {\bibinfo  {journal} {Nature Materials}\ }\textbf {\bibinfo {volume} {21}},\ \bibinfo {pages} {767} (\bibinfo {year} {2022})}\BibitemShut {NoStop}%
\bibitem [{\citenamefont {Makhonin}\ \emph {et~al.}(2024)\citenamefont {Makhonin}, \citenamefont {Delphan}, \citenamefont {Song}, \citenamefont {Walker}, \citenamefont {Isoniemi}, \citenamefont {Claronino}, \citenamefont {Orfanakis}, \citenamefont {Rajendran}, \citenamefont {Ohadi}, \citenamefont {Heck{\"o}tter}, \citenamefont {Assmann}, \citenamefont {Bayer}, \citenamefont {Tartakovskii}, \citenamefont {Skolnick}, \citenamefont {Kyriienko},\ and\ \citenamefont {Krizhanovskii}}]{Makhonin_LSA2024}%
  \BibitemOpen
  \bibfield  {author} {\bibinfo {author} {\bibfnamefont {M.}~\bibnamefont {Makhonin}}, \bibinfo {author} {\bibfnamefont {A.}~\bibnamefont {Delphan}}, \bibinfo {author} {\bibfnamefont {K.~W.}\ \bibnamefont {Song}}, \bibinfo {author} {\bibfnamefont {P.}~\bibnamefont {Walker}}, \bibinfo {author} {\bibfnamefont {T.}~\bibnamefont {Isoniemi}}, \bibinfo {author} {\bibfnamefont {P.}~\bibnamefont {Claronino}}, \bibinfo {author} {\bibfnamefont {K.}~\bibnamefont {Orfanakis}}, \bibinfo {author} {\bibfnamefont {S.~K.}\ \bibnamefont {Rajendran}}, \bibinfo {author} {\bibfnamefont {H.}~\bibnamefont {Ohadi}}, \bibinfo {author} {\bibfnamefont {J.}~\bibnamefont {Heck{\"o}tter}}, \bibinfo {author} {\bibfnamefont {M.}~\bibnamefont {Assmann}}, \bibinfo {author} {\bibfnamefont {M.}~\bibnamefont {Bayer}}, \bibinfo {author} {\bibfnamefont {A.}~\bibnamefont {Tartakovskii}}, \bibinfo {author} {\bibfnamefont {M.}~\bibnamefont {Skolnick}}, \bibinfo {author} {\bibfnamefont {O.}~\bibnamefont {Kyriienko}}, \ and\ \bibinfo {author}
  {\bibfnamefont {D.}~\bibnamefont {Krizhanovskii}},\ }\bibfield  {title} {\bibinfo {title} {\emph {{Nonlinear Rydberg exciton-polaritons in ${\mathrm{Cu}}_2{\mathrm{O}}$ microcavities}}},\ }\href {\doibase 10.1038/s41377-024-01382-9} {\bibfield  {journal} {\bibinfo  {journal} {Light: Science {\&} Applications}\ }\textbf {\bibinfo {volume} {13}},\ \bibinfo {pages} {47} (\bibinfo {year} {2024})}\BibitemShut {NoStop}%
\bibitem [{\citenamefont {Muñoz-Matutano}\ \emph {et~al.}(2019)\citenamefont {Muñoz-Matutano}, \citenamefont {Wood}, \citenamefont {Johnsson}, \citenamefont {Vidal}, \citenamefont {Baragiola}, \citenamefont {Reinhard}, \citenamefont {Lemaître}, \citenamefont {Bloch}, \citenamefont {Amo}, \citenamefont {Nogues}, \citenamefont {Besga}, \citenamefont {Richard},\ and\ \citenamefont {Volz}}]{MunozMatutano2019}%
  \BibitemOpen
  \bibfield  {author} {\bibinfo {author} {\bibfnamefont {G.}~\bibnamefont {Muñoz-Matutano}}, \bibinfo {author} {\bibfnamefont {A.}~\bibnamefont {Wood}}, \bibinfo {author} {\bibfnamefont {M.}~\bibnamefont {Johnsson}}, \bibinfo {author} {\bibfnamefont {X.}~\bibnamefont {Vidal}}, \bibinfo {author} {\bibfnamefont {B.~Q.}\ \bibnamefont {Baragiola}}, \bibinfo {author} {\bibfnamefont {A.}~\bibnamefont {Reinhard}}, \bibinfo {author} {\bibfnamefont {A.}~\bibnamefont {Lemaître}}, \bibinfo {author} {\bibfnamefont {J.}~\bibnamefont {Bloch}}, \bibinfo {author} {\bibfnamefont {A.}~\bibnamefont {Amo}}, \bibinfo {author} {\bibfnamefont {G.}~\bibnamefont {Nogues}}, \bibinfo {author} {\bibfnamefont {B.}~\bibnamefont {Besga}}, \bibinfo {author} {\bibfnamefont {M.}~\bibnamefont {Richard}}, \ and\ \bibinfo {author} {\bibfnamefont {T.}~\bibnamefont {Volz}},\ }\bibfield  {title} {\bibinfo {title} {\emph {{Emergence of quantum correlations from interacting fibre-cavity polaritons}}},\ }\href {\doibase 10.1038/s41563-019-0281-z}
  {\bibfield  {journal} {\bibinfo  {journal} {Nature Materials}\ }\textbf {\bibinfo {volume} {18}},\ \bibinfo {pages} {213} (\bibinfo {year} {2019})}\BibitemShut {NoStop}%
\bibitem [{\citenamefont {Delteil}\ \emph {et~al.}(2019)\citenamefont {Delteil}, \citenamefont {Fink}, \citenamefont {Schade}, \citenamefont {Höfling}, \citenamefont {Schneider},\ and\ \citenamefont {İmamoğlu}}]{Delteil2019}%
  \BibitemOpen
  \bibfield  {author} {\bibinfo {author} {\bibfnamefont {A.}~\bibnamefont {Delteil}}, \bibinfo {author} {\bibfnamefont {T.}~\bibnamefont {Fink}}, \bibinfo {author} {\bibfnamefont {A.}~\bibnamefont {Schade}}, \bibinfo {author} {\bibfnamefont {S.}~\bibnamefont {Höfling}}, \bibinfo {author} {\bibfnamefont {C.}~\bibnamefont {Schneider}}, \ and\ \bibinfo {author} {\bibfnamefont {A.}~\bibnamefont {İmamoğlu}},\ }\bibfield  {title} {\bibinfo {title} {\emph {{Towards polariton blockade of confined exciton–polaritons}}},\ }\href {\doibase 10.1038/s41563-019-0282-y} {\bibfield  {journal} {\bibinfo  {journal} {Nature Materials}\ }\textbf {\bibinfo {volume} {18}},\ \bibinfo {pages} {219} (\bibinfo {year} {2019})}\BibitemShut {NoStop}%
\bibitem [{\citenamefont {Kuriakose}\ \emph {et~al.}(2022)\citenamefont {Kuriakose}, \citenamefont {Walker}, \citenamefont {Dowling}, \citenamefont {Kyriienko}, \citenamefont {Shelykh}, \citenamefont {St-Jean}, \citenamefont {Zambon}, \citenamefont {Lemaître}, \citenamefont {Sagnes}, \citenamefont {Legratiet}, \citenamefont {Harouri}, \citenamefont {Ravets}, \citenamefont {Skolnick}, \citenamefont {Amo}, \citenamefont {Bloch},\ and\ \citenamefont {Krizhanovskii}}]{Kuriakose2022}%
  \BibitemOpen
  \bibfield  {author} {\bibinfo {author} {\bibfnamefont {T.}~\bibnamefont {Kuriakose}}, \bibinfo {author} {\bibfnamefont {P.~M.}\ \bibnamefont {Walker}}, \bibinfo {author} {\bibfnamefont {T.}~\bibnamefont {Dowling}}, \bibinfo {author} {\bibfnamefont {O.}~\bibnamefont {Kyriienko}}, \bibinfo {author} {\bibfnamefont {I.~A.}\ \bibnamefont {Shelykh}}, \bibinfo {author} {\bibfnamefont {P.}~\bibnamefont {St-Jean}}, \bibinfo {author} {\bibfnamefont {N.~C.}\ \bibnamefont {Zambon}}, \bibinfo {author} {\bibfnamefont {A.}~\bibnamefont {Lemaître}}, \bibinfo {author} {\bibfnamefont {I.}~\bibnamefont {Sagnes}}, \bibinfo {author} {\bibfnamefont {L.}~\bibnamefont {Legratiet}}, \bibinfo {author} {\bibfnamefont {A.}~\bibnamefont {Harouri}}, \bibinfo {author} {\bibfnamefont {S.}~\bibnamefont {Ravets}}, \bibinfo {author} {\bibfnamefont {M.~S.}\ \bibnamefont {Skolnick}}, \bibinfo {author} {\bibfnamefont {A.}~\bibnamefont {Amo}}, \bibinfo {author} {\bibfnamefont {J.}~\bibnamefont {Bloch}}, \ and\ \bibinfo {author} {\bibfnamefont
  {D.~N.}\ \bibnamefont {Krizhanovskii}},\ }\bibfield  {title} {\bibinfo {title} {\emph {{Few-photon all-optical phase rotation in a quantum-well micropillar cavity}}},\ }\href {\doibase 10.1038/s41566-022-01019-6} {\bibfield  {journal} {\bibinfo  {journal} {Nature Photonics}\ }\textbf {\bibinfo {volume} {16}},\ \bibinfo {pages} {566} (\bibinfo {year} {2022})}\BibitemShut {NoStop}%
\bibitem [{\citenamefont {Saba}\ \emph {et~al.}(2000)\citenamefont {Saba}, \citenamefont {Quochi}, \citenamefont {Ciuti}, \citenamefont {Oesterle}, \citenamefont {Staehli}, \citenamefont {Deveaud}, \citenamefont {Bongiovanni},\ and\ \citenamefont {Mura}}]{Saba_PRL2000}%
  \BibitemOpen
  \bibfield  {author} {\bibinfo {author} {\bibfnamefont {M.}~\bibnamefont {Saba}}, \bibinfo {author} {\bibfnamefont {F.}~\bibnamefont {Quochi}}, \bibinfo {author} {\bibfnamefont {C.}~\bibnamefont {Ciuti}}, \bibinfo {author} {\bibfnamefont {U.}~\bibnamefont {Oesterle}}, \bibinfo {author} {\bibfnamefont {J.~L.}\ \bibnamefont {Staehli}}, \bibinfo {author} {\bibfnamefont {B.}~\bibnamefont {Deveaud}}, \bibinfo {author} {\bibfnamefont {G.}~\bibnamefont {Bongiovanni}}, \ and\ \bibinfo {author} {\bibfnamefont {A.}~\bibnamefont {Mura}},\ }\bibfield  {title} {\bibinfo {title} {\emph {{Crossover from Exciton to Biexciton Polaritons in Semiconductor Microcavities}}},\ }\href {\doibase 10.1103/PhysRevLett.85.385} {\bibfield  {journal} {\bibinfo  {journal} {Phys. Rev. Lett.}\ }\textbf {\bibinfo {volume} {85}},\ \bibinfo {pages} {385} (\bibinfo {year} {2000})}\BibitemShut {NoStop}%
\bibitem [{\citenamefont {Borri}\ \emph {et~al.}(2000)\citenamefont {Borri}, \citenamefont {Langbein}, \citenamefont {Woggon}, \citenamefont {Jensen},\ and\ \citenamefont {Hvam}}]{Borri2000}%
  \BibitemOpen
  \bibfield  {author} {\bibinfo {author} {\bibfnamefont {P.}~\bibnamefont {Borri}}, \bibinfo {author} {\bibfnamefont {W.}~\bibnamefont {Langbein}}, \bibinfo {author} {\bibfnamefont {U.}~\bibnamefont {Woggon}}, \bibinfo {author} {\bibfnamefont {J.~R.}\ \bibnamefont {Jensen}}, \ and\ \bibinfo {author} {\bibfnamefont {J.~M.}\ \bibnamefont {Hvam}},\ }\bibfield  {title} {\bibinfo {title} {\emph {{Biexcitons or bipolaritons in a semiconductor microcavity}}},\ }\href {\doibase 10.1103/PhysRevB.62.R7763} {\bibfield  {journal} {\bibinfo  {journal} {Phys. Rev. B}\ }\textbf {\bibinfo {volume} {62}},\ \bibinfo {pages} {R7763} (\bibinfo {year} {2000})}\BibitemShut {NoStop}%
\bibitem [{\citenamefont {Wen}\ \emph {et~al.}(2013)\citenamefont {Wen}, \citenamefont {Christmann}, \citenamefont {Baumberg},\ and\ \citenamefont {Nelson}}]{Wen_NJP2013}%
  \BibitemOpen
  \bibfield  {author} {\bibinfo {author} {\bibfnamefont {P.}~\bibnamefont {Wen}}, \bibinfo {author} {\bibfnamefont {G.}~\bibnamefont {Christmann}}, \bibinfo {author} {\bibfnamefont {J.~J.}\ \bibnamefont {Baumberg}}, \ and\ \bibinfo {author} {\bibfnamefont {K.~A.}\ \bibnamefont {Nelson}},\ }\bibfield  {title} {\bibinfo {title} {\emph {{Influence of multi-exciton correlations on nonlinear polariton dynamics in semiconductor microcavities}}},\ }\href {\doibase 10.1088/1367-2630/15/2/025005} {\bibfield  {journal} {\bibinfo  {journal} {New Journal of Physics}\ }\textbf {\bibinfo {volume} {15}},\ \bibinfo {pages} {025005} (\bibinfo {year} {2013})}\BibitemShut {NoStop}%
\bibitem [{\citenamefont {Wouters}(2007)}]{Wouters2007}%
  \BibitemOpen
  \bibfield  {author} {\bibinfo {author} {\bibfnamefont {M.}~\bibnamefont {Wouters}},\ }\bibfield  {title} {\bibinfo {title} {\emph {{Resonant polariton-polariton scattering in semiconductor microcavities}}},\ }\href {\doibase 10.1103/PhysRevB.76.045319} {\bibfield  {journal} {\bibinfo  {journal} {Phys. Rev. B}\ }\textbf {\bibinfo {volume} {76}},\ \bibinfo {pages} {045319} (\bibinfo {year} {2007})}\BibitemShut {NoStop}%
\bibitem [{\citenamefont {Carusotto}\ \emph {et~al.}(2010)\citenamefont {Carusotto}, \citenamefont {Volz},\ and\ \citenamefont {Imamoğlu}}]{Carusotto2010}%
  \BibitemOpen
  \bibfield  {author} {\bibinfo {author} {\bibfnamefont {I.}~\bibnamefont {Carusotto}}, \bibinfo {author} {\bibfnamefont {T.}~\bibnamefont {Volz}}, \ and\ \bibinfo {author} {\bibfnamefont {A.}~\bibnamefont {Imamoğlu}},\ }\bibfield  {title} {\bibinfo {title} {\emph {{Feshbach blockade: Single-photon nonlinear optics using resonantly enhanced cavity polariton scattering from biexciton states}}},\ }\href {\doibase 10.1209/0295-5075/90/37001} {\bibfield  {journal} {\bibinfo  {journal} {Europhysics Letters}\ }\textbf {\bibinfo {volume} {90}},\ \bibinfo {pages} {37001} (\bibinfo {year} {2010})}\BibitemShut {NoStop}%
\bibitem [{\citenamefont {Bleu}\ \emph {et~al.}(2020)\citenamefont {Bleu}, \citenamefont {Li}, \citenamefont {Levinsen},\ and\ \citenamefont {Parish}}]{Bleu2020}%
  \BibitemOpen
  \bibfield  {author} {\bibinfo {author} {\bibfnamefont {O.}~\bibnamefont {Bleu}}, \bibinfo {author} {\bibfnamefont {G.}~\bibnamefont {Li}}, \bibinfo {author} {\bibfnamefont {J.}~\bibnamefont {Levinsen}}, \ and\ \bibinfo {author} {\bibfnamefont {M.~M.}\ \bibnamefont {Parish}},\ }\bibfield  {title} {\bibinfo {title} {\emph {{Polariton interactions in microcavities with atomically thin semiconductor layers}}},\ }\href {\doibase 10.1103/PhysRevResearch.2.043185} {\bibfield  {journal} {\bibinfo  {journal} {Phys. Rev. Res.}\ }\textbf {\bibinfo {volume} {2}},\ \bibinfo {pages} {043185} (\bibinfo {year} {2020})}\BibitemShut {NoStop}%
\bibitem [{\citenamefont {Miller}\ \emph {et~al.}(1982)\citenamefont {Miller}, \citenamefont {Kleinman}, \citenamefont {Gossard},\ and\ \citenamefont {Munteanu}}]{Miller_PRB1982}%
  \BibitemOpen
  \bibfield  {author} {\bibinfo {author} {\bibfnamefont {R.~C.}\ \bibnamefont {Miller}}, \bibinfo {author} {\bibfnamefont {D.~A.}\ \bibnamefont {Kleinman}}, \bibinfo {author} {\bibfnamefont {A.~C.}\ \bibnamefont {Gossard}}, \ and\ \bibinfo {author} {\bibfnamefont {O.}~\bibnamefont {Munteanu}},\ }\bibfield  {title} {\bibinfo {title} {\emph {{Biexcitons in ${\mathrm{GaAs}}$ quantum wells}}},\ }\href {\doibase 10.1103/PhysRevB.25.6545} {\bibfield  {journal} {\bibinfo  {journal} {Phys. Rev. B}\ }\textbf {\bibinfo {volume} {25}},\ \bibinfo {pages} {6545} (\bibinfo {year} {1982})}\BibitemShut {NoStop}%
\bibitem [{\citenamefont {Lovering}\ \emph {et~al.}(1992)\citenamefont {Lovering}, \citenamefont {Phillips}, \citenamefont {Denton},\ and\ \citenamefont {Smith}}]{Lovering_PRL1992}%
  \BibitemOpen
  \bibfield  {author} {\bibinfo {author} {\bibfnamefont {D.~J.}\ \bibnamefont {Lovering}}, \bibinfo {author} {\bibfnamefont {R.~T.}\ \bibnamefont {Phillips}}, \bibinfo {author} {\bibfnamefont {G.~J.}\ \bibnamefont {Denton}}, \ and\ \bibinfo {author} {\bibfnamefont {G.~W.}\ \bibnamefont {Smith}},\ }\bibfield  {title} {\bibinfo {title} {\emph {{Resonant generation of biexcitons in a ${\mathrm{GaAs}}$ quantum well}}},\ }\href {\doibase 10.1103/PhysRevLett.68.1880} {\bibfield  {journal} {\bibinfo  {journal} {Phys. Rev. Lett.}\ }\textbf {\bibinfo {volume} {68}},\ \bibinfo {pages} {1880} (\bibinfo {year} {1992})}\BibitemShut {NoStop}%
\bibitem [{\citenamefont {Borri}\ \emph {et~al.}(2003)\citenamefont {Borri}, \citenamefont {Langbein}, \citenamefont {Woggon}, \citenamefont {Esser}, \citenamefont {Jensen},\ and\ \citenamefont {Hvam}}]{Borri_SST2003}%
  \BibitemOpen
  \bibfield  {author} {\bibinfo {author} {\bibfnamefont {P.}~\bibnamefont {Borri}}, \bibinfo {author} {\bibfnamefont {W.}~\bibnamefont {Langbein}}, \bibinfo {author} {\bibfnamefont {U.}~\bibnamefont {Woggon}}, \bibinfo {author} {\bibfnamefont {A.}~\bibnamefont {Esser}}, \bibinfo {author} {\bibfnamefont {J.~R.}\ \bibnamefont {Jensen}}, \ and\ \bibinfo {author} {\bibfnamefont {J.~M.}\ \bibnamefont {Hvam}},\ }\bibfield  {title} {\bibinfo {title} {\emph {{Biexcitons in semiconductor microcavities}}},\ }\href {\doibase 10.1088/0268-1242/18/10/309} {\bibfield  {journal} {\bibinfo  {journal} {Semiconductor Science and Technology}\ }\textbf {\bibinfo {volume} {18}},\ \bibinfo {pages} {S351} (\bibinfo {year} {2003})}\BibitemShut {NoStop}%
\bibitem [{\citenamefont {Takemura}\ \emph {et~al.}(2014)\citenamefont {Takemura}, \citenamefont {Trebaol}, \citenamefont {Wouters}, \citenamefont {Portella-Oberli},\ and\ \citenamefont {Deveaud}}]{TakemuraNatPhys2014}%
  \BibitemOpen
  \bibfield  {author} {\bibinfo {author} {\bibfnamefont {N.}~\bibnamefont {Takemura}}, \bibinfo {author} {\bibfnamefont {S.}~\bibnamefont {Trebaol}}, \bibinfo {author} {\bibfnamefont {M.}~\bibnamefont {Wouters}}, \bibinfo {author} {\bibfnamefont {M.~T.}\ \bibnamefont {Portella-Oberli}}, \ and\ \bibinfo {author} {\bibfnamefont {B.}~\bibnamefont {Deveaud}},\ }\bibfield  {title} {\bibinfo {title} {\emph {{Polaritonic Feshbach resonance}}},\ }\href {\doibase 10.1038/nphys2999} {\bibfield  {journal} {\bibinfo  {journal} {Nature Physics}\ }\textbf {\bibinfo {volume} {10}},\ \bibinfo {pages} {500} (\bibinfo {year} {2014})}\BibitemShut {NoStop}%
\bibitem [{\citenamefont {Takemura}\ \emph {et~al.}(2017)\citenamefont {Takemura}, \citenamefont {Anderson}, \citenamefont {Navadeh-Toupchi}, \citenamefont {Oberli}, \citenamefont {Portella-Oberli},\ and\ \citenamefont {Deveaud}}]{Takemura2017}%
  \BibitemOpen
  \bibfield  {author} {\bibinfo {author} {\bibfnamefont {N.}~\bibnamefont {Takemura}}, \bibinfo {author} {\bibfnamefont {M.~D.}\ \bibnamefont {Anderson}}, \bibinfo {author} {\bibfnamefont {M.}~\bibnamefont {Navadeh-Toupchi}}, \bibinfo {author} {\bibfnamefont {D.~Y.}\ \bibnamefont {Oberli}}, \bibinfo {author} {\bibfnamefont {M.~T.}\ \bibnamefont {Portella-Oberli}}, \ and\ \bibinfo {author} {\bibfnamefont {B.}~\bibnamefont {Deveaud}},\ }\bibfield  {title} {\bibinfo {title} {\emph {{Spin anisotropic interactions of lower polaritons in the vicinity of polaritonic Feshbach resonance}}},\ }\href {\doibase 10.1103/PhysRevB.95.205303} {\bibfield  {journal} {\bibinfo  {journal} {Phys. Rev. B}\ }\textbf {\bibinfo {volume} {95}},\ \bibinfo {pages} {205303} (\bibinfo {year} {2017})}\BibitemShut {NoStop}%
\bibitem [{\citenamefont {Navadeh-Toupchi}\ \emph {et~al.}(2019)\citenamefont {Navadeh-Toupchi}, \citenamefont {Takemura}, \citenamefont {Anderson}, \citenamefont {Oberli},\ and\ \citenamefont {Portella-Oberli}}]{NavadehToupchi2019}%
  \BibitemOpen
  \bibfield  {author} {\bibinfo {author} {\bibfnamefont {M.}~\bibnamefont {Navadeh-Toupchi}}, \bibinfo {author} {\bibfnamefont {N.}~\bibnamefont {Takemura}}, \bibinfo {author} {\bibfnamefont {M.~D.}\ \bibnamefont {Anderson}}, \bibinfo {author} {\bibfnamefont {D.~Y.}\ \bibnamefont {Oberli}}, \ and\ \bibinfo {author} {\bibfnamefont {M.~T.}\ \bibnamefont {Portella-Oberli}},\ }\bibfield  {title} {\bibinfo {title} {\emph {{Polaritonic Cross Feshbach Resonance}}},\ }\href {\doibase 10.1103/PhysRevLett.122.047402} {\bibfield  {journal} {\bibinfo  {journal} {Phys. Rev. Lett.}\ }\textbf {\bibinfo {volume} {122}},\ \bibinfo {pages} {047402} (\bibinfo {year} {2019})}\BibitemShut {NoStop}%
\bibitem [{\citenamefont {Tan}\ \emph {et~al.}(2023)\citenamefont {Tan}, \citenamefont {Diessel}, \citenamefont {Popert}, \citenamefont {Schmidt}, \citenamefont {Imamoglu},\ and\ \citenamefont {Kroner}}]{BingTan2023}%
  \BibitemOpen
  \bibfield  {author} {\bibinfo {author} {\bibfnamefont {L.~B.}\ \bibnamefont {Tan}}, \bibinfo {author} {\bibfnamefont {O.~K.}\ \bibnamefont {Diessel}}, \bibinfo {author} {\bibfnamefont {A.}~\bibnamefont {Popert}}, \bibinfo {author} {\bibfnamefont {R.}~\bibnamefont {Schmidt}}, \bibinfo {author} {\bibfnamefont {A.}~\bibnamefont {Imamoglu}}, \ and\ \bibinfo {author} {\bibfnamefont {M.}~\bibnamefont {Kroner}},\ }\bibfield  {title} {\bibinfo {title} {\emph {{Bose Polaron Interactions in a Cavity-Coupled Monolayer Semiconductor}}},\ }\href {\doibase 10.1103/PhysRevX.13.031036} {\bibfield  {journal} {\bibinfo  {journal} {Phys. Rev. X}\ }\textbf {\bibinfo {volume} {13}},\ \bibinfo {pages} {031036} (\bibinfo {year} {2023})}\BibitemShut {NoStop}%
\bibitem [{\citenamefont {Zhao}\ \emph {et~al.}(2024)\citenamefont {Zhao}, \citenamefont {Fieramosca}, \citenamefont {Dini}, \citenamefont {Shang}, \citenamefont {Bao}, \citenamefont {Luo}, \citenamefont {Shen}, \citenamefont {Zhao}, \citenamefont {Su}, \citenamefont {Perez}, \citenamefont {Gao}, \citenamefont {Ardizzone}, \citenamefont {Sanvitto}, \citenamefont {Xiong},\ and\ \citenamefont {Liew}}]{Zhao2024}%
  \BibitemOpen
  \bibfield  {author} {\bibinfo {author} {\bibfnamefont {J.}~\bibnamefont {Zhao}}, \bibinfo {author} {\bibfnamefont {A.}~\bibnamefont {Fieramosca}}, \bibinfo {author} {\bibfnamefont {K.}~\bibnamefont {Dini}}, \bibinfo {author} {\bibfnamefont {Q.}~\bibnamefont {Shang}}, \bibinfo {author} {\bibfnamefont {R.}~\bibnamefont {Bao}}, \bibinfo {author} {\bibfnamefont {Y.}~\bibnamefont {Luo}}, \bibinfo {author} {\bibfnamefont {K.}~\bibnamefont {Shen}}, \bibinfo {author} {\bibfnamefont {Y.}~\bibnamefont {Zhao}}, \bibinfo {author} {\bibfnamefont {R.}~\bibnamefont {Su}}, \bibinfo {author} {\bibfnamefont {J.~Z.}\ \bibnamefont {Perez}}, \bibinfo {author} {\bibfnamefont {W.}~\bibnamefont {Gao}}, \bibinfo {author} {\bibfnamefont {V.}~\bibnamefont {Ardizzone}}, \bibinfo {author} {\bibfnamefont {D.}~\bibnamefont {Sanvitto}}, \bibinfo {author} {\bibfnamefont {Q.}~\bibnamefont {Xiong}}, \ and\ \bibinfo {author} {\bibfnamefont {T.~C.~H.}\ \bibnamefont {Liew}},\ }\bibfield  {title} {\bibinfo {title} {\emph {{Room temperature
  spin-layer locking of exciton-polariton nonlinearities}}},\ }\href {https://arxiv.org/abs/2410.18474} {\bibfield  {journal} {\bibinfo  {journal} {arXiv:2410.18474}\ } (\bibinfo {year} {2024})}\BibitemShut {NoStop}%
\bibitem [{\citenamefont {Levinsen}\ \emph {et~al.}(2019)\citenamefont {Levinsen}, \citenamefont {Marchetti}, \citenamefont {Keeling},\ and\ \citenamefont {Parish}}]{Levinsen2019}%
  \BibitemOpen
  \bibfield  {author} {\bibinfo {author} {\bibfnamefont {J.}~\bibnamefont {Levinsen}}, \bibinfo {author} {\bibfnamefont {F.~M.}\ \bibnamefont {Marchetti}}, \bibinfo {author} {\bibfnamefont {J.}~\bibnamefont {Keeling}}, \ and\ \bibinfo {author} {\bibfnamefont {M.~M.}\ \bibnamefont {Parish}},\ }\bibfield  {title} {\bibinfo {title} {\emph {{Spectroscopic Signatures of Quantum Many-Body Correlations in Polariton Microcavities}}},\ }\href {\doibase 10.1103/PhysRevLett.123.266401} {\bibfield  {journal} {\bibinfo  {journal} {Phys. Rev. Lett.}\ }\textbf {\bibinfo {volume} {123}},\ \bibinfo {pages} {266401} (\bibinfo {year} {2019})}\BibitemShut {NoStop}%
\bibitem [{\citenamefont {Bastarrachea-Magnani}\ \emph {et~al.}(2019)\citenamefont {Bastarrachea-Magnani}, \citenamefont {Camacho-Guardian}, \citenamefont {Wouters},\ and\ \citenamefont {Bruun}}]{BastarracheaMagnani2019}%
  \BibitemOpen
  \bibfield  {author} {\bibinfo {author} {\bibfnamefont {M.~A.}\ \bibnamefont {Bastarrachea-Magnani}}, \bibinfo {author} {\bibfnamefont {A.}~\bibnamefont {Camacho-Guardian}}, \bibinfo {author} {\bibfnamefont {M.}~\bibnamefont {Wouters}}, \ and\ \bibinfo {author} {\bibfnamefont {G.~M.}\ \bibnamefont {Bruun}},\ }\bibfield  {title} {\bibinfo {title} {\emph {{Strong interactions and biexcitons in a polariton mixture}}},\ }\href {\doibase 10.1103/PhysRevB.100.195301} {\bibfield  {journal} {\bibinfo  {journal} {Phys. Rev. B}\ }\textbf {\bibinfo {volume} {100}},\ \bibinfo {pages} {195301} (\bibinfo {year} {2019})}\BibitemShut {NoStop}%
\bibitem [{\citenamefont {Choo}\ \emph {et~al.}(2024)\citenamefont {Choo}, \citenamefont {Bleu}, \citenamefont {Levinsen},\ and\ \citenamefont {Parish}}]{Choo2024}%
  \BibitemOpen
  \bibfield  {author} {\bibinfo {author} {\bibfnamefont {K.}~\bibnamefont {Choo}}, \bibinfo {author} {\bibfnamefont {O.}~\bibnamefont {Bleu}}, \bibinfo {author} {\bibfnamefont {J.}~\bibnamefont {Levinsen}}, \ and\ \bibinfo {author} {\bibfnamefont {M.~M.}\ \bibnamefont {Parish}},\ }\bibfield  {title} {\bibinfo {title} {\emph {{Polaronic polariton quasiparticles in a dark excitonic medium}}},\ }\href {\doibase 10.1103/PhysRevB.109.195432} {\bibfield  {journal} {\bibinfo  {journal} {Phys. Rev. B}\ }\textbf {\bibinfo {volume} {109}},\ \bibinfo {pages} {195432} (\bibinfo {year} {2024})}\BibitemShut {NoStop}%
\bibitem [{\citenamefont {Camacho-Guardian}\ \emph {et~al.}(2021)\citenamefont {Camacho-Guardian}, \citenamefont {Bastarrachea-Magnani},\ and\ \citenamefont {Bruun}}]{CamachoGuardian2021}%
  \BibitemOpen
  \bibfield  {author} {\bibinfo {author} {\bibfnamefont {A.}~\bibnamefont {Camacho-Guardian}}, \bibinfo {author} {\bibfnamefont {M.~A.}\ \bibnamefont {Bastarrachea-Magnani}}, \ and\ \bibinfo {author} {\bibfnamefont {G.~M.}\ \bibnamefont {Bruun}},\ }\bibfield  {title} {\bibinfo {title} {\emph {{Mediated Interactions and Photon Bound States in an Exciton-Polariton Mixture}}},\ }\href {\doibase 10.1103/PhysRevLett.126.017401} {\bibfield  {journal} {\bibinfo  {journal} {Phys. Rev. Lett.}\ }\textbf {\bibinfo {volume} {126}},\ \bibinfo {pages} {017401} (\bibinfo {year} {2021})}\BibitemShut {NoStop}%
\bibitem [{\citenamefont {Marchetti}\ and\ \citenamefont {Keeling}(2014)}]{Marchetti2014}%
  \BibitemOpen
  \bibfield  {author} {\bibinfo {author} {\bibfnamefont {F.~M.}\ \bibnamefont {Marchetti}}\ and\ \bibinfo {author} {\bibfnamefont {J.}~\bibnamefont {Keeling}},\ }\bibfield  {title} {\bibinfo {title} {\emph {{Collective Pairing of Resonantly Coupled Microcavity Polaritons}}},\ }\href {\doibase 10.1103/PhysRevLett.113.216405} {\bibfield  {journal} {\bibinfo  {journal} {Phys. Rev. Lett.}\ }\textbf {\bibinfo {volume} {113}},\ \bibinfo {pages} {216405} (\bibinfo {year} {2014})}\BibitemShut {NoStop}%
\bibitem [{\citenamefont {Vermilyea}\ and\ \citenamefont {Fogler}(2024)}]{Vermileya2024}%
  \BibitemOpen
  \bibfield  {author} {\bibinfo {author} {\bibfnamefont {B.}~\bibnamefont {Vermilyea}}\ and\ \bibinfo {author} {\bibfnamefont {M.~M.}\ \bibnamefont {Fogler}},\ }\bibfield  {title} {\bibinfo {title} {\emph {{Feshbach resonance of heavy exciton-polaritons}}},\ }\href {\doibase 10.1103/PhysRevB.109.115401} {\bibfield  {journal} {\bibinfo  {journal} {Phys. Rev. B}\ }\textbf {\bibinfo {volume} {109}},\ \bibinfo {pages} {115401} (\bibinfo {year} {2024})}\BibitemShut {NoStop}%
\bibitem [{\citenamefont {Nozières}\ and\ \citenamefont {Saint~James}(1982)}]{Nozieres1982}%
  \BibitemOpen
  \bibfield  {author} {\bibinfo {author} {\bibfnamefont {P.}~\bibnamefont {Nozières}}\ and\ \bibinfo {author} {\bibfnamefont {D.}~\bibnamefont {Saint~James}},\ }\bibfield  {title} {\bibinfo {title} {\emph {{Particle vs. pair condensation in attractive Bose liquids}}},\ }\href {\doibase 10.1051/jphys:019820043070113300} {\bibfield  {journal} {\bibinfo  {journal} {J. Phys. France}\ }\textbf {\bibinfo {volume} {43}},\ \bibinfo {pages} {1133} (\bibinfo {year} {1982})}\BibitemShut {NoStop}%
\bibitem [{\citenamefont {Hu}\ and\ \citenamefont {Liu}(2020)}]{HuPRL2020}%
  \BibitemOpen
  \bibfield  {author} {\bibinfo {author} {\bibfnamefont {H.}~\bibnamefont {Hu}}\ and\ \bibinfo {author} {\bibfnamefont {X.-J.}\ \bibnamefont {Liu}},\ }\bibfield  {title} {\bibinfo {title} {\emph {{Consistent Theory of Self-Bound Quantum Droplets with Bosonic Pairing}}},\ }\href {\doibase 10.1103/PhysRevLett.125.195302} {\bibfield  {journal} {\bibinfo  {journal} {Phys. Rev. Lett.}\ }\textbf {\bibinfo {volume} {125}},\ \bibinfo {pages} {195302} (\bibinfo {year} {2020})}\BibitemShut {NoStop}%
\bibitem [{\citenamefont {Hu}\ \emph {et~al.}(2020)\citenamefont {Hu}, \citenamefont {Wang},\ and\ \citenamefont {Liu}}]{HuPRA2020}%
  \BibitemOpen
  \bibfield  {author} {\bibinfo {author} {\bibfnamefont {H.}~\bibnamefont {Hu}}, \bibinfo {author} {\bibfnamefont {J.}~\bibnamefont {Wang}}, \ and\ \bibinfo {author} {\bibfnamefont {X.-J.}\ \bibnamefont {Liu}},\ }\bibfield  {title} {\bibinfo {title} {\emph {{Microscopic pairing theory of a binary Bose mixture with interspecies attractions: Bosonic BEC-BCS crossover and ultradilute low-dimensional quantum droplets}}},\ }\href {\doibase 10.1103/PhysRevA.102.043301} {\bibfield  {journal} {\bibinfo  {journal} {Phys. Rev. A}\ }\textbf {\bibinfo {volume} {102}},\ \bibinfo {pages} {043301} (\bibinfo {year} {2020})}\BibitemShut {NoStop}%
\bibitem [{\citenamefont {Dufferwiel}\ \emph {et~al.}(2015)\citenamefont {Dufferwiel}, \citenamefont {Schwarz}, \citenamefont {Withers}, \citenamefont {Trichet}, \citenamefont {Li}, \citenamefont {Sich}, \citenamefont {Del Pozo-Zamudio}, \citenamefont {Clark}, \citenamefont {Nalitov}, \citenamefont {Solnyshkov}, \citenamefont {Malpuech}, \citenamefont {Novoselov}, \citenamefont {Smith}, \citenamefont {Skolnick}, \citenamefont {Krizhanovskii},\ and\ \citenamefont {Tartakovskii}}]{Dufferwiel2015}%
  \BibitemOpen
  \bibfield  {author} {\bibinfo {author} {\bibfnamefont {S.}~\bibnamefont {Dufferwiel}}, \bibinfo {author} {\bibfnamefont {S.}~\bibnamefont {Schwarz}}, \bibinfo {author} {\bibfnamefont {F.}~\bibnamefont {Withers}}, \bibinfo {author} {\bibfnamefont {A.~A.~P.}\ \bibnamefont {Trichet}}, \bibinfo {author} {\bibfnamefont {F.}~\bibnamefont {Li}}, \bibinfo {author} {\bibfnamefont {M.}~\bibnamefont {Sich}}, \bibinfo {author} {\bibfnamefont {O.}~\bibnamefont {Del Pozo-Zamudio}}, \bibinfo {author} {\bibfnamefont {C.}~\bibnamefont {Clark}}, \bibinfo {author} {\bibfnamefont {A.}~\bibnamefont {Nalitov}}, \bibinfo {author} {\bibfnamefont {D.~D.}\ \bibnamefont {Solnyshkov}}, \bibinfo {author} {\bibfnamefont {G.}~\bibnamefont {Malpuech}}, \bibinfo {author} {\bibfnamefont {K.~S.}\ \bibnamefont {Novoselov}}, \bibinfo {author} {\bibfnamefont {J.~M.}\ \bibnamefont {Smith}}, \bibinfo {author} {\bibfnamefont {M.~S.}\ \bibnamefont {Skolnick}}, \bibinfo {author} {\bibfnamefont {D.~N.}\ \bibnamefont {Krizhanovskii}}, \ and\ \bibinfo
  {author} {\bibfnamefont {A.~I.}\ \bibnamefont {Tartakovskii}},\ }\bibfield  {title} {\bibinfo {title} {\emph {{Exciton–polaritons in van der Waals heterostructures embedded in tunable microcavities}}},\ }\href {\doibase 10.1038/ncomms9579} {\bibfield  {journal} {\bibinfo  {journal} {Nature Communications}\ }\textbf {\bibinfo {volume} {6}},\ \bibinfo {pages} {8579} (\bibinfo {year} {2015})}\BibitemShut {NoStop}%
\bibitem [{\citenamefont {Hao}\ \emph {et~al.}(2017)\citenamefont {Hao}, \citenamefont {Specht}, \citenamefont {Nagler}, \citenamefont {Xu}, \citenamefont {Tran}, \citenamefont {Singh}, \citenamefont {Dass}, \citenamefont {Schüller}, \citenamefont {Korn}, \citenamefont {Richter}, \citenamefont {Knorr}, \citenamefont {Li},\ and\ \citenamefont {Moody}}]{Hao2017}%
  \BibitemOpen
  \bibfield  {author} {\bibinfo {author} {\bibfnamefont {K.}~\bibnamefont {Hao}}, \bibinfo {author} {\bibfnamefont {J.~F.}\ \bibnamefont {Specht}}, \bibinfo {author} {\bibfnamefont {P.}~\bibnamefont {Nagler}}, \bibinfo {author} {\bibfnamefont {L.}~\bibnamefont {Xu}}, \bibinfo {author} {\bibfnamefont {K.}~\bibnamefont {Tran}}, \bibinfo {author} {\bibfnamefont {A.}~\bibnamefont {Singh}}, \bibinfo {author} {\bibfnamefont {C.~K.}\ \bibnamefont {Dass}}, \bibinfo {author} {\bibfnamefont {C.}~\bibnamefont {Schüller}}, \bibinfo {author} {\bibfnamefont {T.}~\bibnamefont {Korn}}, \bibinfo {author} {\bibfnamefont {M.}~\bibnamefont {Richter}}, \bibinfo {author} {\bibfnamefont {A.}~\bibnamefont {Knorr}}, \bibinfo {author} {\bibfnamefont {X.}~\bibnamefont {Li}}, \ and\ \bibinfo {author} {\bibfnamefont {G.}~\bibnamefont {Moody}},\ }\bibfield  {title} {\bibinfo {title} {\emph {{Neutral and charged inter-valley biexcitons in monolayer ${\mathrm{MoSe}}_{2}$}}},\ }\href {\doibase 10.1038/ncomms15552} {\bibfield  {journal}
  {\bibinfo  {journal} {Nature Communications}\ }\textbf {\bibinfo {volume} {8}},\ \bibinfo {pages} {15552} (\bibinfo {year} {2017})}\BibitemShut {NoStop}%
\bibitem [{\citenamefont {Kyl\"anp\"a\"a}\ and\ \citenamefont {Komsa}(2015)}]{Kylanpaa2015}%
  \BibitemOpen
  \bibfield  {author} {\bibinfo {author} {\bibfnamefont {I.}~\bibnamefont {Kyl\"anp\"a\"a}}\ and\ \bibinfo {author} {\bibfnamefont {H.-P.}\ \bibnamefont {Komsa}},\ }\bibfield  {title} {\bibinfo {title} {\emph {{Binding energies of exciton complexes in transition metal dichalcogenide monolayers and effect of dielectric environment}}},\ }\href {\doibase 10.1103/PhysRevB.92.205418} {\bibfield  {journal} {\bibinfo  {journal} {Phys. Rev. B}\ }\textbf {\bibinfo {volume} {92}},\ \bibinfo {pages} {205418} (\bibinfo {year} {2015})}\BibitemShut {NoStop}%
\bibitem [{SM()}]{SM}%
  \BibitemOpen
  \href@noop {} {}\bibinfo {note} {See Supplemental Material at [URL] for details on the interaction strength and polariton $T$ matrix, the Bogoliubov and bosonic pairing theory, and the Gross-Pitaevskii analysis of the droplet shape. This includes references to \cite{LandauLifshitzStat2, Levinsen2015, Hu2022}}\BibitemShut {NoStop}%
\bibitem [{\citenamefont {Ciuti}\ \emph {et~al.}(1998)\citenamefont {Ciuti}, \citenamefont {Savona}, \citenamefont {Piermarocchi}, \citenamefont {Quattropani},\ and\ \citenamefont {Schwendimann}}]{Ciuti1998}%
  \BibitemOpen
  \bibfield  {author} {\bibinfo {author} {\bibfnamefont {C.}~\bibnamefont {Ciuti}}, \bibinfo {author} {\bibfnamefont {V.}~\bibnamefont {Savona}}, \bibinfo {author} {\bibfnamefont {C.}~\bibnamefont {Piermarocchi}}, \bibinfo {author} {\bibfnamefont {A.}~\bibnamefont {Quattropani}}, \ and\ \bibinfo {author} {\bibfnamefont {P.}~\bibnamefont {Schwendimann}},\ }\bibfield  {title} {\bibinfo {title} {\emph {{Role of the exchange of carriers in elastic exciton-exciton scattering in quantum wells}}},\ }\href {\doibase 10.1103/PhysRevB.58.7926} {\bibfield  {journal} {\bibinfo  {journal} {Phys. Rev. B}\ }\textbf {\bibinfo {volume} {58}},\ \bibinfo {pages} {7926} (\bibinfo {year} {1998})}\BibitemShut {NoStop}%
\bibitem [{\citenamefont {Tassone}\ and\ \citenamefont {Yamamoto}(1999)}]{Tassone1999}%
  \BibitemOpen
  \bibfield  {author} {\bibinfo {author} {\bibfnamefont {F.}~\bibnamefont {Tassone}}\ and\ \bibinfo {author} {\bibfnamefont {Y.}~\bibnamefont {Yamamoto}},\ }\bibfield  {title} {\bibinfo {title} {\emph {{Exciton-exciton scattering dynamics in a semiconductor microcavity and stimulated scattering into polaritons}}},\ }\href {\doibase 10.1103/PhysRevB.59.10830} {\bibfield  {journal} {\bibinfo  {journal} {Phys. Rev. B}\ }\textbf {\bibinfo {volume} {59}},\ \bibinfo {pages} {10830} (\bibinfo {year} {1999})}\BibitemShut {NoStop}%
\bibitem [{\citenamefont {de~la Fuente~Pico}\ \emph {et~al.}(2025)\citenamefont {de~la Fuente~Pico}, \citenamefont {Levinsen}, \citenamefont {Laird}, \citenamefont {Parish},\ and\ \citenamefont {Marchetti}}]{deLaFuente2025}%
  \BibitemOpen
  \bibfield  {author} {\bibinfo {author} {\bibfnamefont {D.}~\bibnamefont {de~la Fuente~Pico}}, \bibinfo {author} {\bibfnamefont {J.}~\bibnamefont {Levinsen}}, \bibinfo {author} {\bibfnamefont {E.}~\bibnamefont {Laird}}, \bibinfo {author} {\bibfnamefont {M.~M.}\ \bibnamefont {Parish}}, \ and\ \bibinfo {author} {\bibfnamefont {F.~M.}\ \bibnamefont {Marchetti}},\ }\bibfield  {title} {\bibinfo {title} {\emph {{Rydberg excitons and polaritons in monolayer transition metal dichalcogenides in a magnetic field}}},\ }\href {\doibase 10.1103/PhysRevB.111.035432} {\bibfield  {journal} {\bibinfo  {journal} {Phys. Rev. B}\ }\textbf {\bibinfo {volume} {111}},\ \bibinfo {pages} {035432} (\bibinfo {year} {2025})}\BibitemShut {NoStop}%
\bibitem [{Note1()}]{Note1}%
  \BibitemOpen
  \bibinfo {note} {Note that the biexciton resonance essentially coincides with the bipolariton resonance due to the smallness of the cavity photon mass~\cite {Borri_SST2003}.}\BibitemShut {Stop}%
\bibitem [{\citenamefont {Vladimirova}\ \emph {et~al.}(2010)\citenamefont {Vladimirova}, \citenamefont {Cronenberger}, \citenamefont {Scalbert}, \citenamefont {Kavokin}, \citenamefont {Miard}, \citenamefont {Lema\^{\i}tre}, \citenamefont {Bloch}, \citenamefont {Solnyshkov}, \citenamefont {Malpuech},\ and\ \citenamefont {Kavokin}}]{Vladimirova2010}%
  \BibitemOpen
  \bibfield  {author} {\bibinfo {author} {\bibfnamefont {M.}~\bibnamefont {Vladimirova}}, \bibinfo {author} {\bibfnamefont {S.}~\bibnamefont {Cronenberger}}, \bibinfo {author} {\bibfnamefont {D.}~\bibnamefont {Scalbert}}, \bibinfo {author} {\bibfnamefont {K.~V.}\ \bibnamefont {Kavokin}}, \bibinfo {author} {\bibfnamefont {A.}~\bibnamefont {Miard}}, \bibinfo {author} {\bibfnamefont {A.}~\bibnamefont {Lema\^{\i}tre}}, \bibinfo {author} {\bibfnamefont {J.}~\bibnamefont {Bloch}}, \bibinfo {author} {\bibfnamefont {D.}~\bibnamefont {Solnyshkov}}, \bibinfo {author} {\bibfnamefont {G.}~\bibnamefont {Malpuech}}, \ and\ \bibinfo {author} {\bibfnamefont {A.~V.}\ \bibnamefont {Kavokin}},\ }\bibfield  {title} {\bibinfo {title} {\emph {{Polariton-polariton interaction constants in microcavities}}},\ }\href {\doibase 10.1103/PhysRevB.82.075301} {\bibfield  {journal} {\bibinfo  {journal} {Phys. Rev. B}\ }\textbf {\bibinfo {volume} {82}},\ \bibinfo {pages} {075301} (\bibinfo {year} {2010})}\BibitemShut {NoStop}%
\bibitem [{\citenamefont {Li}\ \emph {et~al.}(2021)\citenamefont {Li}, \citenamefont {Parish},\ and\ \citenamefont {Levinsen}}]{Li2021}%
  \BibitemOpen
  \bibfield  {author} {\bibinfo {author} {\bibfnamefont {G.}~\bibnamefont {Li}}, \bibinfo {author} {\bibfnamefont {M.~M.}\ \bibnamefont {Parish}}, \ and\ \bibinfo {author} {\bibfnamefont {J.}~\bibnamefont {Levinsen}},\ }\bibfield  {title} {\bibinfo {title} {\emph {{Microscopic calculation of polariton scattering in semiconductor microcavities}}},\ }\href {\doibase 10.1103/PhysRevB.104.245404} {\bibfield  {journal} {\bibinfo  {journal} {Phys. Rev. B}\ }\textbf {\bibinfo {volume} {104}},\ \bibinfo {pages} {245404} (\bibinfo {year} {2021})}\BibitemShut {NoStop}%
\bibitem [{\citenamefont {Marchetti}\ \emph {et~al.}(2008)\citenamefont {Marchetti}, \citenamefont {Mathy}, \citenamefont {Huse},\ and\ \citenamefont {Parish}}]{Marchetti2008}%
  \BibitemOpen
  \bibfield  {author} {\bibinfo {author} {\bibfnamefont {F.~M.}\ \bibnamefont {Marchetti}}, \bibinfo {author} {\bibfnamefont {C.~J.~M.}\ \bibnamefont {Mathy}}, \bibinfo {author} {\bibfnamefont {D.~A.}\ \bibnamefont {Huse}}, \ and\ \bibinfo {author} {\bibfnamefont {M.~M.}\ \bibnamefont {Parish}},\ }\bibfield  {title} {\bibinfo {title} {\emph {{Phase separation and collapse in Bose-Fermi mixtures with a Feshbach resonance}}},\ }\href {\doibase 10.1103/PhysRevB.78.134517} {\bibfield  {journal} {\bibinfo  {journal} {Phys. Rev. B}\ }\textbf {\bibinfo {volume} {78}},\ \bibinfo {pages} {134517} (\bibinfo {year} {2008})}\BibitemShut {NoStop}%
\bibitem [{\citenamefont {Pieczarka}\ \emph {et~al.}(2020)\citenamefont {Pieczarka}, \citenamefont {Estrecho}, \citenamefont {Boozarjmehr}, \citenamefont {Bleu}, \citenamefont {Steger}, \citenamefont {West}, \citenamefont {Pfeiffer}, \citenamefont {Snoke}, \citenamefont {Levinsen}, \citenamefont {Parish}, \citenamefont {Truscott},\ and\ \citenamefont {Ostrovskaya}}]{Pieczarka2020}%
  \BibitemOpen
  \bibfield  {author} {\bibinfo {author} {\bibfnamefont {M.}~\bibnamefont {Pieczarka}}, \bibinfo {author} {\bibfnamefont {E.}~\bibnamefont {Estrecho}}, \bibinfo {author} {\bibfnamefont {M.}~\bibnamefont {Boozarjmehr}}, \bibinfo {author} {\bibfnamefont {O.}~\bibnamefont {Bleu}}, \bibinfo {author} {\bibfnamefont {M.}~\bibnamefont {Steger}}, \bibinfo {author} {\bibfnamefont {K.}~\bibnamefont {West}}, \bibinfo {author} {\bibfnamefont {L.~N.}\ \bibnamefont {Pfeiffer}}, \bibinfo {author} {\bibfnamefont {D.~W.}\ \bibnamefont {Snoke}}, \bibinfo {author} {\bibfnamefont {J.}~\bibnamefont {Levinsen}}, \bibinfo {author} {\bibfnamefont {M.~M.}\ \bibnamefont {Parish}}, \bibinfo {author} {\bibfnamefont {A.~G.}\ \bibnamefont {Truscott}}, \ and\ \bibinfo {author} {\bibfnamefont {E.~A.}\ \bibnamefont {Ostrovskaya}},\ }\bibfield  {title} {\bibinfo {title} {\emph {{Observation of quantum depletion in a non-equilibrium exciton–polariton condensate}}},\ }\href {\doibase 10.1038/s41467-019-14243-6} {\bibfield  {journal} {\bibinfo
   {journal} {Nature Communications}\ }\textbf {\bibinfo {volume} {11}},\ \bibinfo {pages} {429} (\bibinfo {year} {2020})}\BibitemShut {NoStop}%
\bibitem [{\citenamefont {Ballarini}\ \emph {et~al.}(2020)\citenamefont {Ballarini}, \citenamefont {Caputo}, \citenamefont {Dagvadorj}, \citenamefont {Juggins}, \citenamefont {De~Giorgi}, \citenamefont {Dominici}, \citenamefont {West}, \citenamefont {Pfeiffer}, \citenamefont {Gigli}, \citenamefont {Szymańska},\ and\ \citenamefont {Sanvitto}}]{Ballarini2020}%
  \BibitemOpen
  \bibfield  {author} {\bibinfo {author} {\bibfnamefont {D.}~\bibnamefont {Ballarini}}, \bibinfo {author} {\bibfnamefont {D.}~\bibnamefont {Caputo}}, \bibinfo {author} {\bibfnamefont {G.}~\bibnamefont {Dagvadorj}}, \bibinfo {author} {\bibfnamefont {R.}~\bibnamefont {Juggins}}, \bibinfo {author} {\bibfnamefont {M.}~\bibnamefont {De~Giorgi}}, \bibinfo {author} {\bibfnamefont {L.}~\bibnamefont {Dominici}}, \bibinfo {author} {\bibfnamefont {K.}~\bibnamefont {West}}, \bibinfo {author} {\bibfnamefont {L.~N.}\ \bibnamefont {Pfeiffer}}, \bibinfo {author} {\bibfnamefont {G.}~\bibnamefont {Gigli}}, \bibinfo {author} {\bibfnamefont {M.~H.}\ \bibnamefont {Szymańska}}, \ and\ \bibinfo {author} {\bibfnamefont {D.}~\bibnamefont {Sanvitto}},\ }\bibfield  {title} {\bibinfo {title} {\emph {{Directional Goldstone waves in polariton condensates close to equilibrium}}},\ }\href {\doibase 10.1038/s41467-019-13733-x} {\bibfield  {journal} {\bibinfo  {journal} {Nature Communications}\ }\textbf {\bibinfo {volume} {11}},\ \bibinfo
  {pages} {217} (\bibinfo {year} {2020})}\BibitemShut {NoStop}%
\bibitem [{\citenamefont {Pieczarka}\ \emph {et~al.}(2022)\citenamefont {Pieczarka}, \citenamefont {Bleu}, \citenamefont {Estrecho}, \citenamefont {Wurdack}, \citenamefont {Steger}, \citenamefont {Snoke}, \citenamefont {West}, \citenamefont {Pfeiffer}, \citenamefont {Truscott}, \citenamefont {Ostrovskaya}, \citenamefont {Levinsen},\ and\ \citenamefont {Parish}}]{Pieczarka2022}%
  \BibitemOpen
  \bibfield  {author} {\bibinfo {author} {\bibfnamefont {M.}~\bibnamefont {Pieczarka}}, \bibinfo {author} {\bibfnamefont {O.}~\bibnamefont {Bleu}}, \bibinfo {author} {\bibfnamefont {E.}~\bibnamefont {Estrecho}}, \bibinfo {author} {\bibfnamefont {M.}~\bibnamefont {Wurdack}}, \bibinfo {author} {\bibfnamefont {M.}~\bibnamefont {Steger}}, \bibinfo {author} {\bibfnamefont {D.~W.}\ \bibnamefont {Snoke}}, \bibinfo {author} {\bibfnamefont {K.}~\bibnamefont {West}}, \bibinfo {author} {\bibfnamefont {L.~N.}\ \bibnamefont {Pfeiffer}}, \bibinfo {author} {\bibfnamefont {A.~G.}\ \bibnamefont {Truscott}}, \bibinfo {author} {\bibfnamefont {E.~A.}\ \bibnamefont {Ostrovskaya}}, \bibinfo {author} {\bibfnamefont {J.}~\bibnamefont {Levinsen}}, \ and\ \bibinfo {author} {\bibfnamefont {M.~M.}\ \bibnamefont {Parish}},\ }\bibfield  {title} {\bibinfo {title} {\emph {{Bogoliubov excitations of a polariton condensate in dynamical equilibrium with an incoherent reservoir}}},\ }\href {\doibase 10.1103/PhysRevB.105.224515} {\bibfield
  {journal} {\bibinfo  {journal} {Phys. Rev. B}\ }\textbf {\bibinfo {volume} {105}},\ \bibinfo {pages} {224515} (\bibinfo {year} {2022})}\BibitemShut {NoStop}%
\bibitem [{\citenamefont {Claude}\ \emph {et~al.}(2023)\citenamefont {Claude}, \citenamefont {Jacquet}, \citenamefont {Carusotto}, \citenamefont {Glorieux}, \citenamefont {Giacobino},\ and\ \citenamefont {Bramati}}]{Claude2023}%
  \BibitemOpen
  \bibfield  {author} {\bibinfo {author} {\bibfnamefont {F.}~\bibnamefont {Claude}}, \bibinfo {author} {\bibfnamefont {M.~J.}\ \bibnamefont {Jacquet}}, \bibinfo {author} {\bibfnamefont {I.}~\bibnamefont {Carusotto}}, \bibinfo {author} {\bibfnamefont {Q.}~\bibnamefont {Glorieux}}, \bibinfo {author} {\bibfnamefont {E.}~\bibnamefont {Giacobino}}, \ and\ \bibinfo {author} {\bibfnamefont {A.}~\bibnamefont {Bramati}},\ }\bibfield  {title} {\bibinfo {title} {\emph {{Spectrum of collective excitations of a quantum fluid of polaritons}}},\ }\href {\doibase 10.1103/PhysRevB.107.174507} {\bibfield  {journal} {\bibinfo  {journal} {Phys. Rev. B}\ }\textbf {\bibinfo {volume} {107}},\ \bibinfo {pages} {174507} (\bibinfo {year} {2023})}\BibitemShut {NoStop}%
\bibitem [{\citenamefont {Fr\'erot}\ \emph {et~al.}(2023)\citenamefont {Fr\'erot}, \citenamefont {Vashisht}, \citenamefont {Morassi}, \citenamefont {Lema\^{\i}tre}, \citenamefont {Ravets}, \citenamefont {Bloch}, \citenamefont {Minguzzi},\ and\ \citenamefont {Richard}}]{Frerot2023}%
  \BibitemOpen
  \bibfield  {author} {\bibinfo {author} {\bibfnamefont {I.}~\bibnamefont {Fr\'erot}}, \bibinfo {author} {\bibfnamefont {A.}~\bibnamefont {Vashisht}}, \bibinfo {author} {\bibfnamefont {M.}~\bibnamefont {Morassi}}, \bibinfo {author} {\bibfnamefont {A.}~\bibnamefont {Lema\^{\i}tre}}, \bibinfo {author} {\bibfnamefont {S.}~\bibnamefont {Ravets}}, \bibinfo {author} {\bibfnamefont {J.}~\bibnamefont {Bloch}}, \bibinfo {author} {\bibfnamefont {A.}~\bibnamefont {Minguzzi}}, \ and\ \bibinfo {author} {\bibfnamefont {M.}~\bibnamefont {Richard}},\ }\bibfield  {title} {\bibinfo {title} {\emph {{Bogoliubov Excitations Driven by Thermal Lattice Phonons in a Quantum Fluid of Light}}},\ }\href {\doibase 10.1103/PhysRevX.13.041058} {\bibfield  {journal} {\bibinfo  {journal} {Phys. Rev. X}\ }\textbf {\bibinfo {volume} {13}},\ \bibinfo {pages} {041058} (\bibinfo {year} {2023})}\BibitemShut {NoStop}%
\bibitem [{\citenamefont {\ifmmode~\acute{C}\else \'{C}\fi{}wik}\ \emph {et~al.}(2016)\citenamefont {\ifmmode~\acute{C}\else \'{C}\fi{}wik}, \citenamefont {Kirton}, \citenamefont {De~Liberato},\ and\ \citenamefont {Keeling}}]{Cwik2016}%
  \BibitemOpen
  \bibfield  {author} {\bibinfo {author} {\bibfnamefont {J.~A.}\ \bibnamefont {\ifmmode~\acute{C}\else \'{C}\fi{}wik}}, \bibinfo {author} {\bibfnamefont {P.}~\bibnamefont {Kirton}}, \bibinfo {author} {\bibfnamefont {S.}~\bibnamefont {De~Liberato}}, \ and\ \bibinfo {author} {\bibfnamefont {J.}~\bibnamefont {Keeling}},\ }\bibfield  {title} {\bibinfo {title} {\emph {{Excitonic spectral features in strongly coupled organic polaritons}}},\ }\href {\doibase 10.1103/PhysRevA.93.033840} {\bibfield  {journal} {\bibinfo  {journal} {Phys. Rev. A}\ }\textbf {\bibinfo {volume} {93}},\ \bibinfo {pages} {033840} (\bibinfo {year} {2016})}\BibitemShut {NoStop}%
\bibitem [{\citenamefont {Stringari}\ and\ \citenamefont {Treiner}(1987)}]{Stringari1987}%
  \BibitemOpen
  \bibfield  {author} {\bibinfo {author} {\bibfnamefont {S.}~\bibnamefont {Stringari}}\ and\ \bibinfo {author} {\bibfnamefont {J.}~\bibnamefont {Treiner}},\ }\bibfield  {title} {\bibinfo {title} {\emph {{Surface properties of liquid $^{3}{\mathrm{He}}$ and $^{4}{\mathrm{He}}$: A density-functional approach}}},\ }\href {\doibase 10.1103/PhysRevB.36.8369} {\bibfield  {journal} {\bibinfo  {journal} {Phys. Rev. B}\ }\textbf {\bibinfo {volume} {36}},\ \bibinfo {pages} {8369} (\bibinfo {year} {1987})}\BibitemShut {NoStop}%
\bibitem [{\citenamefont {Cikojevi\ifmmode~\acute{c}\else \'{c}\fi{}}\ \emph {et~al.}(2021)\citenamefont {Cikojevi\ifmmode~\acute{c}\else \'{c}\fi{}}, \citenamefont {Poli}, \citenamefont {Ancilotto}, \citenamefont {Vranje\ifmmode \check{s}\else \v{s}\fi{}-Marki\ifmmode~\acute{c}\else \'{c}\fi{}},\ and\ \citenamefont {Boronat}}]{Cikojevic2021}%
  \BibitemOpen
  \bibfield  {author} {\bibinfo {author} {\bibfnamefont {V.}~\bibnamefont {Cikojevi\ifmmode~\acute{c}\else \'{c}\fi{}}}, \bibinfo {author} {\bibfnamefont {E.}~\bibnamefont {Poli}}, \bibinfo {author} {\bibfnamefont {F.}~\bibnamefont {Ancilotto}}, \bibinfo {author} {\bibfnamefont {L.}~\bibnamefont {Vranje\ifmmode \check{s}\else \v{s}\fi{}-Marki\ifmmode~\acute{c}\else \'{c}\fi{}}}, \ and\ \bibinfo {author} {\bibfnamefont {J.}~\bibnamefont {Boronat}},\ }\bibfield  {title} {\bibinfo {title} {\emph {{Dilute quantum liquid in a K-Rb Bose mixture}}},\ }\href {\doibase 10.1103/PhysRevA.104.033319} {\bibfield  {journal} {\bibinfo  {journal} {Phys. Rev. A}\ }\textbf {\bibinfo {volume} {104}},\ \bibinfo {pages} {033319} (\bibinfo {year} {2021})}\BibitemShut {NoStop}%
\bibitem [{\citenamefont {Michinel}\ \emph {et~al.}(2002)\citenamefont {Michinel}, \citenamefont {Campo-T\'aboas}, \citenamefont {Garc\'{\i}a-Fern\'andez}, \citenamefont {Salgueiro},\ and\ \citenamefont {Quiroga-Teixeiro}}]{Michinel2002}%
  \BibitemOpen
  \bibfield  {author} {\bibinfo {author} {\bibfnamefont {H.}~\bibnamefont {Michinel}}, \bibinfo {author} {\bibfnamefont {J.}~\bibnamefont {Campo-T\'aboas}}, \bibinfo {author} {\bibfnamefont {R.}~\bibnamefont {Garc\'{\i}a-Fern\'andez}}, \bibinfo {author} {\bibfnamefont {J.~R.}\ \bibnamefont {Salgueiro}}, \ and\ \bibinfo {author} {\bibfnamefont {M.~L.}\ \bibnamefont {Quiroga-Teixeiro}},\ }\bibfield  {title} {\bibinfo {title} {\emph {{Liquid light condensates}}},\ }\href {\doibase 10.1103/PhysRevE.65.066604} {\bibfield  {journal} {\bibinfo  {journal} {Phys. Rev. E}\ }\textbf {\bibinfo {volume} {65}},\ \bibinfo {pages} {066604} (\bibinfo {year} {2002})}\BibitemShut {NoStop}%
\bibitem [{\citenamefont {Michinel}\ \emph {et~al.}(2006)\citenamefont {Michinel}, \citenamefont {Paz-Alonso},\ and\ \citenamefont {P\'erez-Garc\'{\i}a}}]{Michinel2006}%
  \BibitemOpen
  \bibfield  {author} {\bibinfo {author} {\bibfnamefont {H.}~\bibnamefont {Michinel}}, \bibinfo {author} {\bibfnamefont {M.~J.}\ \bibnamefont {Paz-Alonso}}, \ and\ \bibinfo {author} {\bibfnamefont {V.~M.}\ \bibnamefont {P\'erez-Garc\'{\i}a}},\ }\bibfield  {title} {\bibinfo {title} {\emph {{Turning Light into a Liquid via Atomic Coherence}}},\ }\href {\doibase 10.1103/PhysRevLett.96.023903} {\bibfield  {journal} {\bibinfo  {journal} {Phys. Rev. Lett.}\ }\textbf {\bibinfo {volume} {96}},\ \bibinfo {pages} {023903} (\bibinfo {year} {2006})}\BibitemShut {NoStop}%
\bibitem [{\citenamefont {Novoa}\ \emph {et~al.}(2009)\citenamefont {Novoa}, \citenamefont {Michinel},\ and\ \citenamefont {Tommasini}}]{Novoa2009}%
  \BibitemOpen
  \bibfield  {author} {\bibinfo {author} {\bibfnamefont {D.}~\bibnamefont {Novoa}}, \bibinfo {author} {\bibfnamefont {H.}~\bibnamefont {Michinel}}, \ and\ \bibinfo {author} {\bibfnamefont {D.}~\bibnamefont {Tommasini}},\ }\bibfield  {title} {\bibinfo {title} {\emph {Pressure, Surface Tension, and Dripping of Self-Trapped Laser Beams}},\ }\href {\doibase 10.1103/PhysRevLett.103.023903} {\bibfield  {journal} {\bibinfo  {journal} {Phys. Rev. Lett.}\ }\textbf {\bibinfo {volume} {103}},\ \bibinfo {pages} {023903} (\bibinfo {year} {2009})}\BibitemShut {NoStop}%
\bibitem [{\citenamefont {Wu}\ \emph {et~al.}(2013)\citenamefont {Wu}, \citenamefont {Zhang}, \citenamefont {Yuan}, \citenamefont {Wen}, \citenamefont {Zheng}, \citenamefont {Zhang},\ and\ \citenamefont {Xiao}}]{Wu2013}%
  \BibitemOpen
  \bibfield  {author} {\bibinfo {author} {\bibfnamefont {Z.}~\bibnamefont {Wu}}, \bibinfo {author} {\bibfnamefont {Y.}~\bibnamefont {Zhang}}, \bibinfo {author} {\bibfnamefont {C.}~\bibnamefont {Yuan}}, \bibinfo {author} {\bibfnamefont {F.}~\bibnamefont {Wen}}, \bibinfo {author} {\bibfnamefont {H.}~\bibnamefont {Zheng}}, \bibinfo {author} {\bibfnamefont {Y.}~\bibnamefont {Zhang}}, \ and\ \bibinfo {author} {\bibfnamefont {M.}~\bibnamefont {Xiao}},\ }\bibfield  {title} {\bibinfo {title} {\emph {{Cubic-quintic condensate solitons in four-wave mixing}}},\ }\href {\doibase 10.1103/PhysRevA.88.063828} {\bibfield  {journal} {\bibinfo  {journal} {Phys. Rev. A}\ }\textbf {\bibinfo {volume} {88}},\ \bibinfo {pages} {063828} (\bibinfo {year} {2013})}\BibitemShut {NoStop}%
\bibitem [{\citenamefont {Figueiredo}\ \emph {et~al.}(2024)\citenamefont {Figueiredo}, \citenamefont {Mendon\ifmmode~\mbox{\c{c}}\else \c{c}\fi{}a},\ and\ \citenamefont {Ter\ifmmode~\mbox{\c{c}}\else \c{c}\fi{}as}}]{Figueiredo2024}%
  \BibitemOpen
  \bibfield  {author} {\bibinfo {author} {\bibfnamefont {J.~L.}\ \bibnamefont {Figueiredo}}, \bibinfo {author} {\bibfnamefont {J.~T.}\ \bibnamefont {Mendon\ifmmode~\mbox{\c{c}}\else \c{c}\fi{}a}}, \ and\ \bibinfo {author} {\bibfnamefont {H.}~\bibnamefont {Ter\ifmmode~\mbox{\c{c}}\else \c{c}\fi{}as}},\ }\bibfield  {title} {\bibinfo {title} {\emph {{Quantum kinetic theory of light-matter interactions in degenerate plasmas}}},\ }\href {\doibase 10.1103/PhysRevA.110.063519} {\bibfield  {journal} {\bibinfo  {journal} {Phys. Rev. A}\ }\textbf {\bibinfo {volume} {110}},\ \bibinfo {pages} {063519} (\bibinfo {year} {2024})}\BibitemShut {NoStop}%
\bibitem [{\citenamefont {Braaten}\ and\ \citenamefont {Hammer}(2006)}]{Braaten2006}%
  \BibitemOpen
  \bibfield  {author} {\bibinfo {author} {\bibfnamefont {E.}~\bibnamefont {Braaten}}\ and\ \bibinfo {author} {\bibfnamefont {H.-W.}\ \bibnamefont {Hammer}},\ }\bibfield  {title} {\bibinfo {title} {\emph {{Universality in few-body systems with large scattering length}}},\ }\href {\doibase https://doi.org/10.1016/j.physrep.2006.03.001} {\bibfield  {journal} {\bibinfo  {journal} {Physics Reports}\ }\textbf {\bibinfo {volume} {428}},\ \bibinfo {pages} {259} (\bibinfo {year} {2006})}\BibitemShut {NoStop}%
\bibitem [{\citenamefont {Sun}\ \emph {et~al.}(2017)\citenamefont {Sun}, \citenamefont {Wen}, \citenamefont {Yoon}, \citenamefont {Liu}, \citenamefont {Steger}, \citenamefont {Pfeiffer}, \citenamefont {West}, \citenamefont {Snoke},\ and\ \citenamefont {Nelson}}]{Sun2017}%
  \BibitemOpen
  \bibfield  {author} {\bibinfo {author} {\bibfnamefont {Y.}~\bibnamefont {Sun}}, \bibinfo {author} {\bibfnamefont {P.}~\bibnamefont {Wen}}, \bibinfo {author} {\bibfnamefont {Y.}~\bibnamefont {Yoon}}, \bibinfo {author} {\bibfnamefont {G.}~\bibnamefont {Liu}}, \bibinfo {author} {\bibfnamefont {M.}~\bibnamefont {Steger}}, \bibinfo {author} {\bibfnamefont {L.~N.}\ \bibnamefont {Pfeiffer}}, \bibinfo {author} {\bibfnamefont {K.}~\bibnamefont {West}}, \bibinfo {author} {\bibfnamefont {D.~W.}\ \bibnamefont {Snoke}}, \ and\ \bibinfo {author} {\bibfnamefont {K.~A.}\ \bibnamefont {Nelson}},\ }\bibfield  {title} {\bibinfo {title} {\emph {{Bose-Einstein Condensation of Long-Lifetime Polaritons in Thermal Equilibrium}}},\ }\href {\doibase 10.1103/PhysRevLett.118.016602} {\bibfield  {journal} {\bibinfo  {journal} {Phys. Rev. Lett.}\ }\textbf {\bibinfo {volume} {118}},\ \bibinfo {pages} {016602} (\bibinfo {year} {2017})}\BibitemShut {NoStop}%
\bibitem [{\citenamefont {Keeling}\ and\ \citenamefont {Berloff}(2008)}]{Keeling2008}%
  \BibitemOpen
  \bibfield  {author} {\bibinfo {author} {\bibfnamefont {J.}~\bibnamefont {Keeling}}\ and\ \bibinfo {author} {\bibfnamefont {N.~G.}\ \bibnamefont {Berloff}},\ }\bibfield  {title} {\bibinfo {title} {\emph {{Spontaneous Rotating Vortex Lattices in a Pumped Decaying Condensate}}},\ }\href {\doibase 10.1103/PhysRevLett.100.250401} {\bibfield  {journal} {\bibinfo  {journal} {Phys. Rev. Lett.}\ }\textbf {\bibinfo {volume} {100}},\ \bibinfo {pages} {250401} (\bibinfo {year} {2008})}\BibitemShut {NoStop}%
\bibitem [{\citenamefont {Caldara}\ \emph {et~al.}()\citenamefont {Caldara}, \citenamefont {Bleu}, \citenamefont {Marchetti}, \citenamefont {Levinsen},\ and\ \citenamefont {Parish}}]{dataav}%
  \BibitemOpen
  \bibfield  {author} {\bibinfo {author} {\bibfnamefont {M.}~\bibnamefont {Caldara}}, \bibinfo {author} {\bibfnamefont {O.}~\bibnamefont {Bleu}}, \bibinfo {author} {\bibfnamefont {F.~M.}\ \bibnamefont {Marchetti}}, \bibinfo {author} {\bibfnamefont {J.}~\bibnamefont {Levinsen}}, \ and\ \bibinfo {author} {\bibfnamefont {M.~M.}\ \bibnamefont {Parish}},\ }\href@noop {} {}\bibinfo {note} {Quantum Droplets of Light in Semiconductor Microcavities (2025), \href{https://bridges.monash.edu/articles/dataset/Quantum_Droplets_of_Light_in_Semiconductor_Microcavities/30843377}{10.26180/30843377}}\BibitemShut {NoStop}%
\bibitem [{\citenamefont {Grimm}\ \emph {et~al.}()\citenamefont {Grimm}, \citenamefont {Inguscio}, \citenamefont {Stringari},\ and\ \citenamefont {Lamporesi}}]{Varenna2022book}%
  \BibitemOpen
  \bibinfo {editor} {\bibfnamefont {R.}~\bibnamefont {Grimm}}, \bibinfo {editor} {\bibfnamefont {M.}~\bibnamefont {Inguscio}}, \bibinfo {editor} {\bibfnamefont {S.}~\bibnamefont {Stringari}}, \ and\ \bibinfo {editor} {\bibfnamefont {G.}~\bibnamefont {Lamporesi}},\ eds.,\ \href {https://ebooks.iospress.nl/volume/quantum-mixtures-with-ultra-cold-atoms} {\emph {\bibinfo {title} {{Quantum Mixtures with Ultra-cold Atoms}}}}\ (\bibinfo  {publisher} {IOS Press, Amsterdam, 2025})\ \bibinfo {note} {{P}roceedings of the International School of Physics ``Enrico Fermi'', Course CCXI, Varenna, 18-23 July 2022}\BibitemShut {NoStop}%
\bibitem [{\citenamefont {Lifshitz}\ and\ \citenamefont {Pitaevskii}(1995)}]{LandauLifshitzStat2}%
  \BibitemOpen
  \bibfield  {author} {\bibinfo {author} {\bibfnamefont {E.~M.}\ \bibnamefont {Lifshitz}}\ and\ \bibinfo {author} {\bibfnamefont {L.~P.}\ \bibnamefont {Pitaevskii}},\ }\href@noop {} {\emph {\bibinfo {title} {Statistical physics. Part 2, Theory of the condensed state}}},\ \bibinfo {edition} {3rd}\ ed.,\ Course of Theoretical Physics; Volume 9\ (\bibinfo  {publisher} {Butterworth-Heinemann},\ \bibinfo {address} {Oxford, England},\ \bibinfo {year} {1995})\BibitemShut {NoStop}%
\bibitem [{\citenamefont {Levinsen}\ and\ \citenamefont {Parish}(2015)}]{Levinsen2015}%
  \BibitemOpen
  \bibfield  {author} {\bibinfo {author} {\bibfnamefont {J.}~\bibnamefont {Levinsen}}\ and\ \bibinfo {author} {\bibfnamefont {M.~M.}\ \bibnamefont {Parish}},\ }\bibfield  {title} {\bibinfo {title} {\emph {{Strongly interacting two-dimensional Fermi gases}}},\ }\href {\doibase 10.1142/9789814667746_0001} {\bibfield  {journal} {\bibinfo  {journal} {Annual Review of Cold Atoms and Molecules}\ }\textbf {\bibinfo {volume} {3}},\ \bibinfo {pages} {1} (\bibinfo {year} {2015})}\BibitemShut {NoStop}%
\bibitem [{\citenamefont {Hu}\ \emph {et~al.}(2022)\citenamefont {Hu}, \citenamefont {Deng},\ and\ \citenamefont {Liu}}]{Hu2022}%
  \BibitemOpen
  \bibfield  {author} {\bibinfo {author} {\bibfnamefont {H.}~\bibnamefont {Hu}}, \bibinfo {author} {\bibfnamefont {H.}~\bibnamefont {Deng}}, \ and\ \bibinfo {author} {\bibfnamefont {X.-J.}\ \bibnamefont {Liu}},\ }\bibfield  {title} {\bibinfo {title} {\emph {{Two-dimensional exciton-polariton interactions beyond the Born approximation}}},\ }\href {\doibase 10.1103/PhysRevA.106.063303} {\bibfield  {journal} {\bibinfo  {journal} {Phys. Rev. A}\ }\textbf {\bibinfo {volume} {106}},\ \bibinfo {pages} {063303} (\bibinfo {year} {2022})}\BibitemShut {NoStop}%
\end{thebibliography}
\end{document}